\RequirePackage{ifpdf}
\documentclass[11pt]{JHEP3}
\pdfoutput=1
\usepackage{graphicx}
\usepackage{cite}
 \usepackage{ae}
 \usepackage{aecompl}
\usepackage{amsmath,epsfig}
\usepackage{amssymb,amsfonts}
\usepackage{latexsym}
\usepackage{epsfig}
\usepackage[caption=false]{subfig}
\usepackage{empheq}
\usepackage{float}
\usepackage{comment}
\usepackage{ulem}

\allowdisplaybreaks[1]

\relax
\renewcommand{\theequation}{\arabic{section}.\arabic{equation}}

\def\D{\mathrm{d}}
\def\be{\begin{equation}}
\def\ee{\end{equation}}

\newcommand{\bear}{\begin{eqnarray}}
\newcommand{\bea}{\begin{eqnarray}}
\newcommand{\eear}{\end{eqnarray}}
\newcommand{\eea}{\end{eqnarray}}
\newbox\pippobox

\def\II{\relax{\rm I\kern-.18em I}}

\title{% Fermi surfaces in holographic Bose-Fermi systems
Polarized solutions and Fermi surfaces in holographic Bose-Fermi systems
}

\author{Francesco~Nitti$^{a}$, Giuseppe~Policastro$^b$, Thomas~Vanel$^{c,d}$
~\\ \\
$^a$ APC, Universit\'e Paris 7, CNRS/IN2P3, CEA/IRFU, Obs. de Paris, Sorbonne Paris Cit\'e,
B\^atiment Condorcet, F-75205, Paris Cedex 13, France (UMR du CNRS 7164)
~\\ \\
$^b$ Laboratoire de Physique Th\'eorique, Ecole Normale Sup\'erieure, 24 rue Lhomond, 75231
Paris Cedex 05, France (UMR du CNRS 8549)
~\\ \\
$^c$ Sorbonne Universit\'es, UPMC Univ Paris 06, UMR 7589, LPTHE, F-75005, Paris, France
~\\ \\
$^d$ CNRS, UMR 7589, LPTHE, F-75005, Paris, France 
~\\
}

\preprint{}

\abstract{We use holography to study the ground state of a system with interacting bosonic and fermionic degrees of freedom at finite density. The gravitational model consists of Einstein-Maxwell gravity coupled to a perfect fluid of charged fermions and to a charged scalar field which interact through a current-current interaction. When the scalar field is non-trivial, in addition to compact electron stars, the screening of the fermion electric charge by the scalar condensate allows the formation of solutions where the fermion fluid is made of antiparticles, as well as solutions with coexisting, separated regions of particle-like and antiparticle-like fermion fluids. We show that, when the latter solutions exist, they are thermodynamically favored. By computing the two-point Green function of the boundary fermionic operator we show that, in addition to the charged scalar condensate, the dual field theory state exhibits electron-like and/or hole-like Fermi surfaces. Compared to fluid-only solutions, the presence of the scalar condensate destroys the Fermi surfaces with lowest Fermi momenta.  We interpret this as a signal of the onset of superconductivity.}

\begin{document}

%\maketitle %%%%%%%%%% THIS IS IGNORED %%%%%%%%%%%

%%%%%%%%%%%%%%%%%%%%%%%%%%%%%%%%%%%%%%%%%%%%%%%%%%%%%%%%%%%%%%%%%%%%%%%%%%%%%%%%%%%%%%%%%%%%%%%
%%%%%%%%%%%%%%%%%%%%%%%%%%%%%%%%%%%%%%%%%%%%%%%%%%%%%%%%%%%%%%%%%%%%%%%%%%%%%%%%%%%%%%%%%%%%%%%
\section{Introduction}
\label{sec:intro}
%%%%%%%%%%%%%%%%%%%%%%%%%%%%%%%%%%%%%%%%%%%%%%%%%%%%%%%%%%%%%%%%%%%%%%%%%%%%%%%%%%%%%%%%%%%%%%%
%%%%%%%%%%%%%%%%%%%%%%%%%%%%%%%%%%%%%%%%%%%%%%%%%%%%%%%%%%%%%%%%%%%%%%%%%%%%%%%%%%%%%%%%%%%%%%%

The holographic correspondence between field theories in $d$ dimensions and gravitational
theories in $d+1$ dimension has been extensively used to study properties 
of strongly coupled systems, obtaining information that is not easily accessible by ordinary
methods. In particular, fermionic systems at finite density pose a particularly difficult
problem, as there are no theoretical models that can be studied reliably in a controlled
approximation and lattice simulations are marred by the \lq \lq sign problem''.  In this
context, the holographic method has proved useful by  offering a number of insights into
possible exotic phases of matter that are not described by Landau's theory of Fermi liquids or
other weakly-coupled descriptions~\cite{Hartnoll:2009sz,Herzog:2009xv,McGreevy:2009xe,Sachdev:2012tj}. 

It was suggested in~\cite{Sachdev:2010um} that the presence of a charged horizon in the simplest gravity
solutions dual to finite-density states admits an interpretation in terms of {\it
fractionalization} of the fundamental charged degrees of freedom. When the charge is sourced by
matter fields in the bulk, it corresponds to fermionic operators in the boundary. But when the 
charge emanates from the horizon, it cannot be associated to any gauge-invariant observable.
This leads to a violation of the Luttinger's theorem: the charge contained in the Fermi surface
(or surfaces) does not account for the total charge of the system. The phase transition
corresponding to the onset of fractionalization was studied in~\cite{Hartnoll:2011pp,Adam:2012mw,Gouteraux:2012yr}. 
When there are charged scalars in the theory, they can also become excited and condense,
carrying part of the charge. In this case however the $U(1)$ symmetry is broken, and 
the interpretation is different: the system undergoes a transition to a superconducting
state~\cite{Adam:2012mw}. 
If both components are allowed to be present, there can be a competition between the bosons and the fermions \footnote{For other examples of competing orders in holography see \cite{Basu:2010fa,Donos:2012yu,Musso:2013ija,Cai:2013wma,Amado:2013lia,Donos:2013woa,Li:2014wca}}. 
We have considered this situation in a previous paper~\cite{Nitti:2013xaa} where we studied gravitational solutions in which the
charged matter in the bulk can be both fermionic and bosonic, in the regime where the fermions
can be treated as a perfect fluid. We have found phase transitions between the electron star,
the holographic superconductor, and a new class of solutions in which the electron star is
dressed by a scalar condensate. We called them {\it compact stars} since they extend only
across a finite range in the radial direction of the bulk geometry: the fermionic fluid  is confined in  a radial shell, and the infrared geometry is the same as for the zero-temperature holographic superconductor of \cite{Horowitz:2009ij}. A similar system was studied in \cite{Liu:2013yaa}, but with a different potential that also allows for an unbounded fluid with Lifshitz asymptotics in the IR. 
% We determined the phase diagram of the system in a certain region of
%the parameter space. 

In the present paper, we continue the study of the system analyzed in \cite{Nitti:2013xaa}, adding a new ingredient, namely a
direct coupling between the scalar and the fermions. The reason for introducing this coupling
is the following: in the holographic superconductor, the boundary bosonic operator that
develops a vacuum expectation value is interpreted as a strongly coupled 
analogue of the Cooper pair that condenses in the superconducting state in the usual BCS
theory. In our model fermions are also present at the same time, so if the boson has to be
interpreted as a bound state of fermions, it seems unnatural that they should interact only
from the exchange of gauge fields. A natural interaction term would be a Yukawa coupling, as it
was considered in~\cite{Hartman:2010fk}. 
This requires dealing with microscopic fermions and going away from the fluid approximation;
moreover the coupling can only be there in the case where the boson has twice the charge of the fermions.  

In \cite{Liu:2014mva}, an interacting description in the fluid approximation was given that treats the scalar as a BCS-like fermion bilinear. In the fluid approximation, the presence of the condensate modifies  the fermionic fluid  equation of state. At the macroscopic level, this system  too may arise from a Yukawa interaction of a scalar whose charge is twice that of the elementary fermions. 

In a strongly coupled system we cannot discard the possibility that the
fermions that condense are not the fundamental electrons but some other excitation, perhaps
fractionalized, therefore we want to leave the ratio of the charges arbitrary.

\begin{figure}[h!]
\centering
\includegraphics[width=0.6\textwidth]{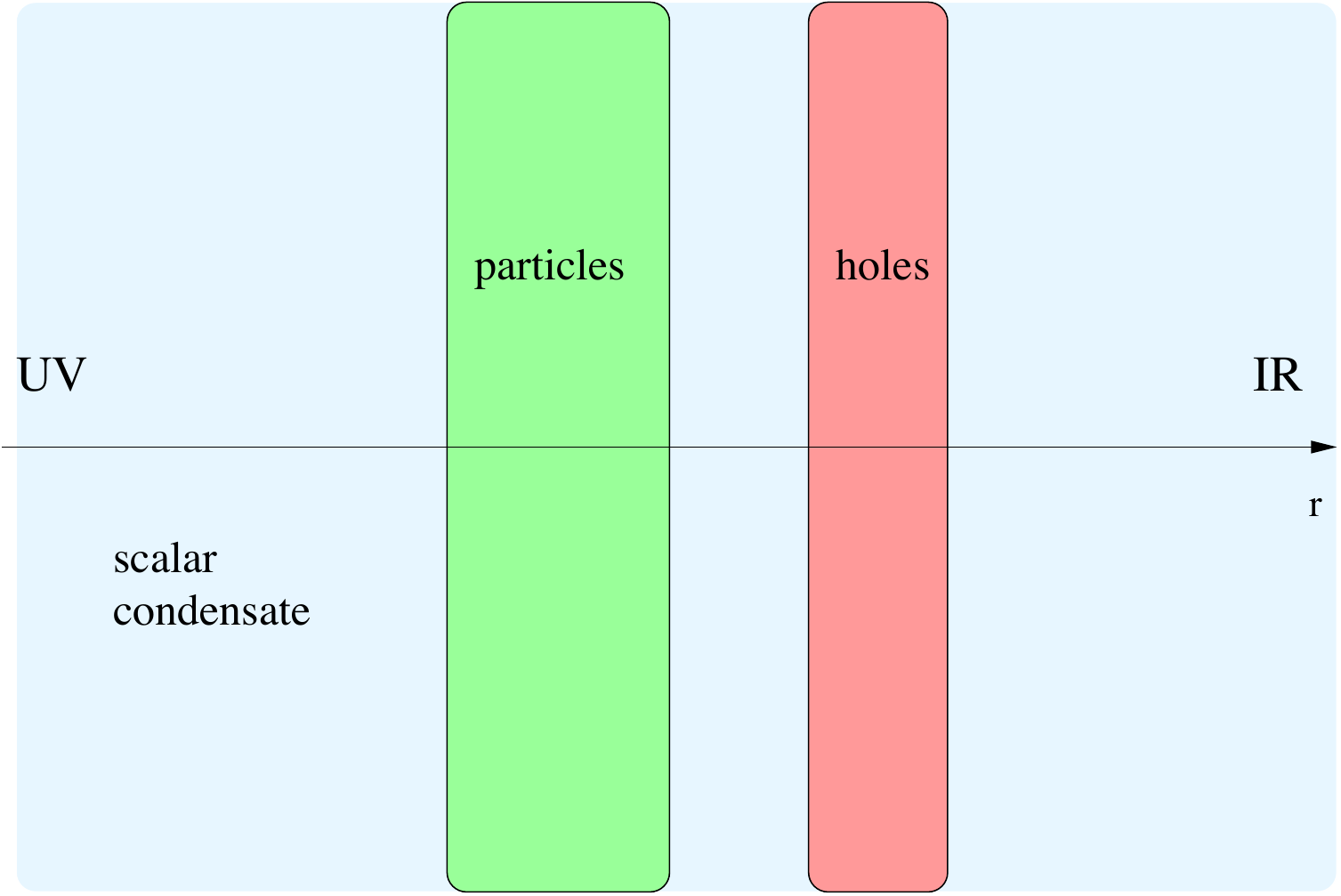}
\caption{Structure of compact electron-positron star solutions. The bulk fermion fluid  is confined within two spherical shells: the outer shell (green) is made out of particles (electrons) and the inner one (red) is made of holes (positrons). The whole configuration is  immersed in the non-zero bulk scalar condensate, which overscreens the hole-like fluid making it effectively repel the particle shell. This combination of electromagnetic repulsion and  gravitational attraction renders the system stable.}
\label{fig:polarized}
\end{figure}

If we want to 
work directly with  the fluid approximation, and at the same time leave the scalar and fermion charges generic,   the simplest interaction we can write is a current-current
coupling. 
We found that in the presence of this interaction, a surprising phenomenon takes place: in the
bulk we can have a {\it polarized} charged system, which is constituted of radially separated shells of positively and negatively charged components of the fluid, immersed in a non-zero scalar condensate. We call these solutions {\it electron-positron stars}, and they are  illustrated schematically in Figure~\ref{fig:polarized}.

The reason these solutions may arise is that, due to the current-current interaction term,
the local chemical potential can have opposite signs  in different regions of the bulk geometry. This leads  to both fermionic particles and
antiparticles being populated in separate regions. The solution is stable because the scalar condensate effectively
screens the negative  charge of the positrons so  that these  do not feel the electric attraction of the electrons
but rather they are repelled. The system is kept together by the gravitational
attraction, which balances the electromagnetic repulsion.  

\begin{figure}[ht!]
\centering
\includegraphics[width=0.6\textwidth]{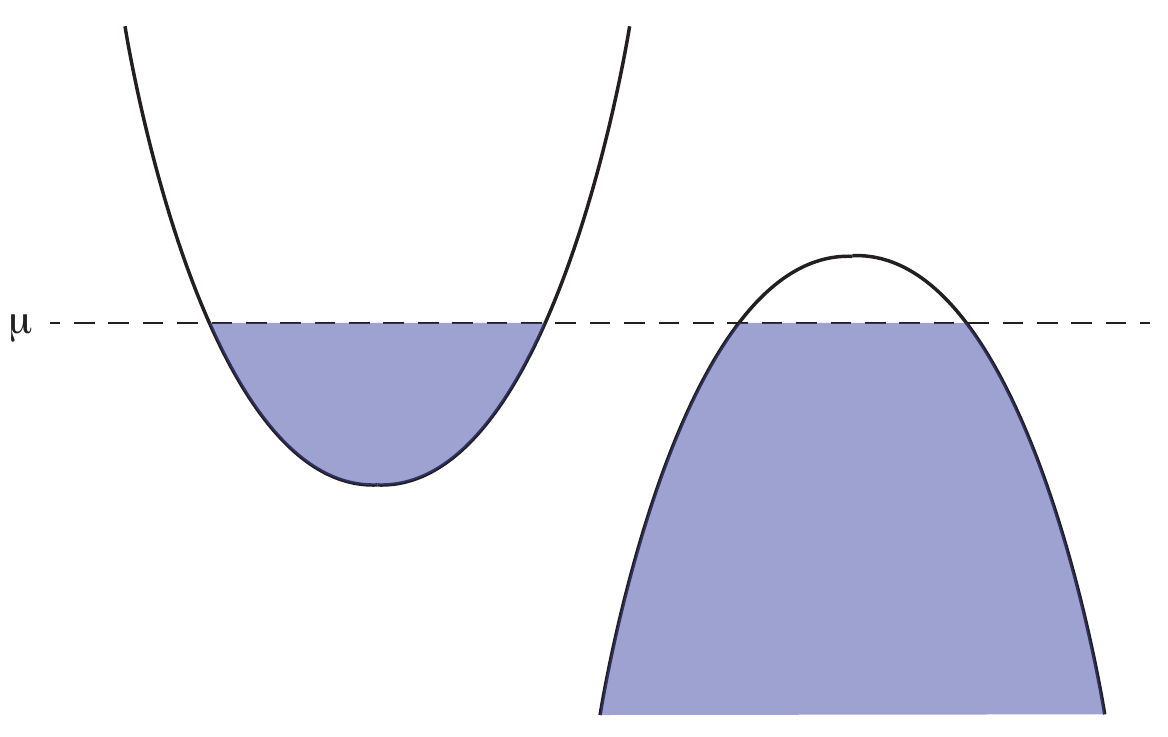}
\caption{Schematic band structure of (dual of) the compact polarized stars solutions.}
\label{fig:bands}
\end{figure}

The simplest boundary interpretation of these solutions appears to be in terms of different flavors
of fermions, each of them having a certain band structure but with the zero energy level having
a different offset for different flavors, so that a given chemical potential intersects the
conductance band for some fermions and the valence band for others (see Figure~\ref{fig:bands} for  a schematic representation of this phenomenon). We
do not know of any realistic system that displays such features, it would be interesting to 
find some real-world realization of this situation.

To shed more light on the physics of the system, we study the spectrum of
low-energy fermionic excitations. We consider a probe fermion in the geometry, that is also
interacting with the Maxwell field and with the scalar current, since it is supposed to be one
of the fermions making up the fluid. We solve the Dirac equation in the WKB approximation and
find the normal and quasi-normal modes, that correspond to poles of the boundary fermionic
Green's function. The analysis of the poles  at zero frequency and finite momentum reveals the  presence of a finite number of Fermi surfaces.

 One can compare this situation  with that of  the unbounded electron star in the absence of the scalar field \cite{Hartnoll:2011dm}. In
the latter case, there are infinitely many Fermi surfaces, with Fermi momenta accumulating
exponentially close to zero. As discussed in \cite{Cubrovic:2011xm}, the dual field theory interpretation is that of a Fermi system in a  limit where  the number of constituent fermions is infinite.  In compact stars (both in the electron and the
electron-positron version) there are only {\it finitely many} Fermi surfaces, meaning that all except for a finite number of Fermi surfaces become gapped due to  the scalar condensate.
 For small
frequency, in the electron star the modes have a small dissipation 
due to the possibility of tunneling into the IR  Lifshitz part of the geometry.  This was interpreted as the effect of the  interaction of the fermions  with bosonic critical modes. In compact star solutions, 
on the other hand, we find that for a certain range of energies above zero, the excitations are stable, so
there is no residual interaction at low energies. The comparison with the electron star also
reveals that the most \lq\lq shallow" modes, corresponding to Fermi surfaces with smallest momenta,
are the ones that disappear from the spectrum; we interpret this as a signal that the system
has become gapped due to the superconductivity; however the gap concerns only part of the
system, as other Fermi surfaces remain gapless. Naively one would expect that the smallest
Fermi surfaces should be more robust, since they have a lower temperature for the
superconducting transition, as predicted by BCS~\cite{Hartman:2010fk}. A possible explanation
would be if the mechanism for superconductivity is not described by BCS. In fact, our findings are consistent with the UV/IR duality displayed by holographic Fermionic fluids discussed in \cite{Cubrovic:2011xm}, where it was argued that the states corresponding to the smallest Fermi momenta are the last to be filled. 

The plan of the paper is as follows: in Section~\ref{sec:matter in AdS} we review the results
of our previous work, in particular the compact star solutions. In Section~\ref{sec:mixture}
we study the system in presence of the current-current coupling and describe the new solutions,
find the free energy and determine the phase diagram. In Section~\ref{sec:spectrum} we analyze
the probe fermions and determine the low-energy spectrum. We conclude in
Section~\ref{sec:conclusion} summarizing the results and indicating the open questions and
future directions of investigation.

%%%%%%%%%%%%%%%%%%%%%%%%%%%%%%%%%%%%%%%%%%%%%%%%%%%%%%%%%%%%%%%%%%%%%%%%%%%%%%%%%%%%%%%%%%%%%%%
%%%%%%%%%%%%%%%%%%%%%%%%%%%%%%%%%%%%%%%%%%%%%%%%%%%%%%%%%%%%%%%%%%%%%%%%%%%%%%%%%%%%%%%%%%%%%%%
\section{Bosonic and fermionic matter in asymptotically AdS spacetimes}
\label{sec:matter in AdS}
%%%%%%%%%%%%%%%%%%%%%%%%%%%%%%%%%%%%%%%%%%%%%%%%%%%%%%%%%%%%%%%%%%%%%%%%%%%%%%%%%%%%%%%%%%%%%%%
%%%%%%%%%%%%%%%%%%%%%%%%%%%%%%%%%%%%%%%%%%%%%%%%%%%%%%%%%%%%%%%%%%%%%%%%%%%%%%%%%%%%%%%%%%%%%%%

We will consider four-dimensional gravitational solutions which are asymptotically $AdS_4$ and
have zero temperature and finite charge density.
These solutions arise from models which have an action of the form
\begin{align}
\label{eq:action}
 S = \int\D^4 x \sqrt{-g} \left[ \frac{1}{2\kappa^2}\left(R+\frac{6}{L^2}\right)
-\frac{1}{4e^2}F_{ab}F^{ab} \right] + S_\textnormal{matter} + S_\textnormal{bdry}
\end{align}
where $F=\D A$ is the field strength of the gauge field $A$, $\kappa$ is Newton's constant,
$L$ is the asymptotic $AdS_4$ length and $e$ is the $U(1)$ coupling.
The term $S_\textnormal{bdry}$ represents the Gibbons-Hawking term and the counterterms
necessary
for the holographic renormalization and $S_\textnormal{matter}$ is the action for the matter
fields.

% We will study  solutions which are static, homogeneous and  isotropic  with respect to the field theory coordinates $(t,x^i)$.
% \begin{align}
% \label{eq:ansatz}
% \D s^2 = L^2\left[-f(r)\D t^2 + g(r)\D r^2 + \frac{1}{r^2} \left(\D x^2 + \D
% y^2\right)\right] \, , \qquad A = \frac{eL}{\kappa} h(r) \D t \, ,
% \end{align}
% %

As in \cite{Nitti:2013xaa,Liu:2013yaa}, we will consider the simultaneous presence of two types of bulk matter:

\begin{enumerate} 
\item A charged scalar field $\psi$ with action:  
\begin{align}
\label{eq:action-scalar}
 S_\textnormal{scalar} = -\frac{1}{2} \int\D^4 x \sqrt{-g} \left( \left| \nabla\psi-iqA\psi
\right|^2 + m_s^2 |\psi|^2 \right) \, ,
\end{align}
where $q$ is the scalar field $U(1)$ charge and $m_s^2$ is  negative and 
satisfies the Breitenlohner-Freedman (BF) bound, $-9/4<L^2m_s^2<0$. 

\item 
A fermionic component described effectively by a fluid stress tensor and electromagnetic
current:
\begin{equation}
\label{eq:stress-tensor-fluid}
 T_{ab}^\textnormal{fluid} = (\rho+p)u_a u_b + p g_{ab}  \, , \qquad J^a_\textnormal{fluid} = \sigma u^a \, .
\end{equation}
The form of the fluid energy density $\rho$, pressure $p$ and charge density $\sigma$ will be given shortly. 
\end{enumerate}

%
% and its electromagnetic current
% %
% \begin{align}
% \label{eq:current-fluid}
%  J^a_\textnormal{fluid} = \sigma u^a \, ,
% \end{align}
% %
We will restrict to static, homogeneous and isotropic solutions, which  by performing diffeomorphisms and
gauge transformations can always be written as,
\begin{equation}
\begin{aligned}
\label{eq:ansatz-full}
\D s^2 &= L^2\left[-f(r)\D t^2 + g(r)\D r^2 + \frac{1}{r^2} \left(\D x^2 + \D
y^2\right)\right] \, , \\
A &= \frac{eL}{\kappa} h(r) \D t \ , \qquad \psi = \psi(r) \, , \qquad u^a = (u^t(r),0,0,0) \, ,
\end{aligned}
\end{equation}
in which $r=0$ is the $AdS$ boundary.
As discussed in detail in  Appendix~\ref{app:action}, we rescale all quantities by suitable powers of $e$, $L$ and
$\kappa$ and denote the rescaled quantities with hats.
In the UV region we have:
\begin{align}
\label{eq:f-g-h-uv}
 r\to0 : \qquad f(r)% {c^2} \sim g(r)
 \sim \frac{1}{r^2} \, , \qquad h(r) \sim
% c
\hat{\mu}-\hat{Q}r \, ,
\end{align}
where $\hat{\mu}$ is the chemical potential of the boundary quantum field theory and
$\hat{Q}$ the total charge of the system.
% The constant $c>0$ is in general different from unity.
% It can be eliminated by rescaling the time coordinate, leading to the field theory Minkowski
% metric on the boundary.
We will assume that the UV asymptotics of the scalar field corresponds to zero source term for the dual operator, i.e.:
 \begin{align}
\label{eq:psi-uv}
 \psi \sim \psi_+ \, r^\Delta \, , \qquad \Delta =
\frac{3}{2}+\frac{3}{2}\sqrt{1+\frac{4\hat{m}_s^2}{9}} \, ,
\end{align}
where $\Delta$ is the conformal dimension of the field theory scalar operator $\mathcal{O}$.

The energy density $\rho$,  pressure $p$, charge density $\sigma$  and fluid velocity $u^a$
are assumed to locally satisfy the chemical equilibrium equation of state of a free Fermi gas, as in  \cite{Hartnoll:2010gu}, with a density of states given by:
\begin{align}
 g(\epsilon) =
\left\{
\begin{array}{lr}
\beta \, \epsilon \, \sqrt{\epsilon^2-m_f^2} &\qquad \epsilon > m_f \, , \\
0 &\qquad 0 < \epsilon < m_f \, ,
\end{array}
\right.
\end{align}
where $m_f$ is the constituent fermion mass and $\beta$ is a phenomenological parameter related to the spin of the fermions. 

It is
convenient to describe the fluid using rescaled energy density, pressure and charge density 
(defined in Appendix \ref{app:ansatz} and denoted by a hat), which under the local chemical equilibrium condition are given by: 
%The (rescaled) pressure, energy density and charge density are then given by
%
\begin{align}
\label{eq:fluid-quantities}
\hat{\rho}(r) &=  \hat{\beta} \int_{\hat{m}_f}^{|\hat{\mu}_l(r)|}\D \epsilon \
\Theta\left(\epsilon-\hat{m}_f\right) \epsilon^2 \, \sqrt{\epsilon^2-\hat{m}_f^2}
\, , \\
\qquad \hat{\sigma}(r) &= \textnormal{sign}(\hat{\mu}_l) \, \hat{\beta} \int_{\hat{m}_f}^{|\hat{\mu}_l(r)|}\D \epsilon \
\Theta\left(\epsilon-\hat{m}_f\right) \epsilon \, \sqrt{\epsilon^2-\hat{m}_f^2}\, ,\\
 -\hat{p}(r) & = \hat{\rho}(r) -\hat{\mu}_l(r) \hat{\sigma}(r)\, ,
\end{align}
where $\hat{\mu}_l(r)$ is the local chemical potential in the bulk, defined  in Appendix \ref{app:action}, and 
\begin{align}\label{mf}
 \hat{m}_f = \frac{\kappa}{e}\frac{m_f}{|q_f|} \, , \qquad \hat{\beta} =
\frac{e^4L^2}{\kappa^2}\beta  q_f^4 \, ,
\end{align}
where  $q_f$ is the elementary charge of the
constituent fermions. 

With respect to the original construction in  \cite{Hartnoll:2010gu}, we allow for both signs of $q_f$, i.e. we allow for the possibility of the fluid to be made up of particles\footnote{We follow the conventions of  \cite{Hartnoll:2010gu}, where   ``electrons'' denotes  positively charged particles.} (electrons, $q_f>0$) and antiparticles (holes, or positrons, $q_f<0$). This leads to a slight generalization of equations (\ref{eq:fluid-quantities}) with respect to the original formulae of \cite{Hartnoll:2010gu}. The details  of how the signs and absolute values arise in (\ref{eq:fluid-quantities}) can be found in Appendix \ref{app:ansatz}. It is important to  note that the sign of  $\hat{\mu}_l(r)$ is the same as the sign of   $q_f$, thus a negative chemical potential in the bulk will lead to the formation of a fluid made out of  negatively charged constituents.  

For $|\hat{\mu}_l|<m_f$, $\hat{\rho}=\hat{\sigma}=\hat{p}=0$ due to the Heaviside step
functions, and there is no fluid.

% Equations (\ref{eq:fluid-quantities})  mean that we are assuming that the fluid obeys a  flat
%space free  Fermi fluid equation of state at each radial position. This requires  bulk
%gravitational effects  to be negligible on the fermion equation of state, and that flat-space
%Thomas-Fermi approximation be valid at each $r$. This is the case if the following combination
%of parameters is large \cite{Hartnoll:2010gu}
% \be
% \gamma \equiv {|q_f| e L \over \kappa} \gg 1. 
% \ee

The explicit expression of the local chemical potential  depends on the way the fermions couple to the Maxwell and scalar field in the bulk. 
If there is no direct interaction between the scalar and fermionic components, it is given by
\begin{align}
\label{eq:local-mu-0}
\hat{\mu}_l(r) = \frac{h(r)}{\sqrt{f(r)}} \, ,
\end{align}
and it has the same sign of the electric potential $h(r)$ and of the boundary chemical potential $\hat{\mu}$ (Eq.~(\ref{eq:f-g-h-uv})). For this reason, without direct interactions between the fermionic fluid  and the scalar field, only positively charged solutions are possible for positive boundary chemical potential. This is the case covered in \cite{Hartnoll:2010gu} and \cite{Nitti:2013xaa}, and the  possible bulk solutions are reviewed in the next subsection. 

As we will see in Section \ref{sec:mixture}, a direct current-current interaction between the fermion fluid and the scalar field will allow $\hat{\mu}_l(r)$ to have both signs, or even to change sign in the bulk, 
and solutions with both signs of the constituent fermion charge will be possible.

% The expressions~(\ref{eq:fluid-quantities}) are obtained by applying the
% rescaling~(\ref{eq:rescalings}) to the parameters and fluid quantities.
% We refer to Appendix~\ref{app:action} for more explanations.
%The rescaled mass $\hat{m}_f$ has the same sign as $q_f$.
% The definitions~(\ref{eq:fluid-quantities}) for the fluid quantities are valid only for
% $\hat{\mu}_l$ of the same sign as $q_f$ with $|\hat{m}_f|<|\hat{\mu}_l|$.

Finally, the local chemical equilibrium and Thomas-Fermi approximation, necessary for this description
to be valid,  require that
\be\label{valid}
e^2 \sim \frac{\kappa}{L} \ll 1 \, .
\ee

\subsection{Taxonomy of non-interacting solutions} \label{taxonomy}

The situation where bosonic and fermionic components do not have a direct interaction was analyzed in \cite{Nitti:2013xaa}. In this case, the expression of the local chemical potential is (\ref{eq:local-mu-0}), and  the possible homogeneous, charged, zero-temperature solutions of this system are listed below (the reader is referred to \cite{Nitti:2013xaa} for more details and to Appendix~\ref{app:action} for conventions and definitions of the rescaled variables).

\paragraph{Extremal RN-AdS Black hole (ERN).}

 In this case all the matter fields are trivial, the solution to the action~(\ref{eq:action}) with metric
and gauge field of the form~(\ref{eq:ansatz-full}) is the extremal Reissner-Nordstr\"om black hole. The infrared  is an extremal horizon with geometry ${AdS_2\times \mathbb{R}^2}$. 
%One can add bosonic or fermionic matter to the model.

\paragraph{Holographic Superconductor (HSC).}
 These are solution with non-vanishing scalar field (with vev-like UV asymptotics) but zero fluid density in the bulk \cite{Hartnoll:2008vx,Hartnoll:2008kx}. At zero temperature, the asymptotic solution in the IR ($r\to\infty$) is~\cite{Horowitz:2009ij}
\begin{equation}
\begin{aligned}
\label{eq:sc-IR-sol}
f(r) &\sim \frac{1}{r^2}  \, ,  &  g(r) \sim -\frac{3}{2\hat{m}_s^2} \frac{1}{r^2 \log r} \, ,
 \\
h(r) &\sim  h_0 r^{\delta}\left(\log r\right)^{1/2} \, ,  & \psi(r) \sim 2\left(\log r
\right)^{1/2} \, , ~~~
\end{aligned}
\end{equation}
where ``hatted'' quantities are defined in Appendix~\ref{app:action}, and $\delta<-1$.  
%
% \begin{align}
% \delta = \frac{1}{2} - \frac{1}{2}\left(1-\frac{24\hat{q}^2}{\hat{m}_s^2}\right)^{1/2} \qquad \delta<-1
% \end{align}
%
These solutions require the condition
\begin{align}
\label{eq:condition-q-scalar}
 \hat{q}^2 > -\frac{\hat{m}_s^2}{3}
\end{align}
which will be always assumed in this paper.

\paragraph{Electron star (ES).} These solutions, first constructed in ~\cite{Hartnoll:2010gu},  have a trivial scalar field, $\psi(r)=0$ but a non-vanishing fluid density  in the bulk region where the local chemical potential $\hat{\mu}_l(r)$ exceeds the fermion mass $\hat{m}_f$. This happens in  an unbounded region  ${r_s < r<+\infty}$, $r_s$ representing  the star boundary where $\hat{\rho}(r_s)=\hat{\sigma}(r_s) = 0$   . Outside this region (i.e. for $0<r< r_s$) the solution coincides with the RN-AdS black hole described above. The region occupied by the fluid is unbounded, and in the far IR ($r\to\infty$), the fluid energy and charge density are constant and  the geometry is asymptotically Lifshitz:
 \begin{align}
\label{eq:sol-ES-Lif}
 f(r) \sim \frac{1}{r^{2z}} \, , \qquad g(r) \sim \frac{g_\infty}{r^2} \, , \qquad h(r) \sim
\frac{h_\infty}{r^z} \, ,
\end{align}
where $g_\infty$ and $h_\infty$ are constants depending on $\hat{m}_f$ and $\hat{\beta}$, and 
the dynamical  exponent $z$ is determined by
\begin{align}
 h_\infty = \sqrt{\frac{z-1}{z}} \, .
\end{align}

\paragraph{Compact electron stars (eCS).}

Solutions with both a non-trivial scalar profile {\it and} a non-zero fluid density were found in  \cite{Nitti:2013xaa}. The  fluid density is confined in a shell $r_1<r<r_2$, whose boundaries are determined by the equation
\be\label{cs}
\hat{\mu}_l(r_1) = \hat{\mu}_l(r_2)=\hat{m}_f \, .
\ee 
The non-trivial scalar field profile is similar to the one of the holographic superconductor,
and causes the local chemical potential to be non-monotonic: this allows equation   (\ref{cs})
to admit two solutions $r_{1,2}$, which represent the outer and inner star boundaries. Since
the star is confined to a shell in the holographic direction, these solutions were called {\it
compact electron stars} (eCS). 
In the UV ($0<r<r_1$) and in the IR  ($r_2<r< +\infty$) the fluid density is identically vanishes, and the solution is given by the holographic
superconductor described above, with IR asymptotics~(\ref{eq:sc-IR-sol}).
% In the UV, the fluid quantities are also vanishing and the metric, gauge field and scalar
% field asymptote to~(\ref{eq:f-g-h-uv}) and~(\ref{eq:psi-uv}), where the IR parameter is, as
% ftheor the holographic superconductor solution, chosen to set the non-normalizable mode of the
% scalar field to zero.
%  These solutions have been found to be thermodynamically favored when they exist.
% Below, we quickly review these bosonic, fermionic and mixed solutions. A detail presentation
% can be found in~\cite{Nitti:2013xaa}, and

Different types of solutions with both non-trivial scalar and fluid density  were found in \cite{Liu:2013yaa}, with a different choice of the scalar field potential (which included a quartic term). In this paper we limit ourselves  to a  quadratic scalar potential, and these solutions will not be considered. 

\section{Interacting Fermion-Boson Mixtures in AdS}
\label{sec:mixture}
%%%%%%%%%%%%%%%%%%%%%%%%%%%%%%%%%%%%%%%%%%%%%%%%%%%%%%%%%%%%%%%%%%%%%%%%%%%%%%%%%%%%%%%%%%%%%%%
%%%%%%%%%%%%%%%%%%%%%%%%%%%%%%%%%%%%%%%%%%%%%%%%%%%%%%%%%%%%%%%%%%%%%%%%%%%%%%%%%%%%%%%%%%%%%%%

We now  generalize the model of~\cite{Nitti:2013xaa} by coupling directly the fermion fluid to the scalar field.
%We consider now Einstein-Maxwell theory coupled to both the scalar field with mass squared
%$m_s^2$ and $U(1)$ charge $q$ of Section~\ref{sec:hsc}, and the perfect fluid made out of charged
%fermions of mass $m_f$ and charge $q_f$ of Section~\ref{sec:es}.

In order to be able to continue working in the fluid approximation for the fermions, we consider a direct coupling between the scalar field and the fluid through
their respective electromagnetic currents:
% $J^a_\textnormal{scalar}$ and
%$J^a_\textnormal{fluid}$: 
%
\begin{align}
\label{eq:L-int}
 \mathcal{L}_\textnormal{int} = \eta J_a^\textnormal{fluid}J^a_\textnormal{scalar} \, ,
\end{align}
where $\eta$ parametrizes the intensity of the coupling and can have either sign.

The fermionic current is given in~(\ref{eq:stress-tensor-fluid}).  The general expression for the scalar field current is  given in Eq.~(\ref{eq:current-psi}).  For a static and homogeneous radial profile $\psi(r)$, it is given by
\be\label{Jscalar}
J^a_\textnormal{scalar} =- q^2g^{ab} A_a |\psi|^2 \, .
\ee

In Appendix~\ref{app:action}, we derive the field equations by considering an action principle
for the fluid~\cite{Hartnoll:2010gu}, including the interaction Lagrangian~(\ref{eq:L-int})
in the fluid action.
The local chemical potential now depends on the scalar field through its electromagnetic
currrent,
\begin{align}
 \hat{\mu}_l = \hat{u}^a (\hat{A}_a+\hat{\eta}J_a^\textnormal{scalar}) \, , \qquad \hat{\eta}\equiv{\eta \over e^2 L^2}.
\end{align}

The interaction also contributes to the field equation of the scalar field $\psi$ and Einstein
equations.
Using the ansatz~(\ref{eq:ansatz-full}),  the
field equations for the rescaled fields and parameters, derived in Appendix~\ref{app:action}, are
\begin{subequations}
\label{eq:field-equations}
\begin{align}
\hat{\psi}'' + \left(\frac{f'}{2f}-\frac{g'}{2g}-\frac{2}{r}\right)\hat{\psi}' +
g\left(\frac{\hat{q}^2h^2}{f}-\hat{m}_s^2 - 2\hat{q}^2\hat{\eta}
\frac{h}{\sqrt{f}}\hat{\sigma}\right)\hat{\psi} &= 0 \, , \\
h'' - \frac{1}{2}\left(\frac{f'}{f}+\frac{g'}{g}+\frac{4}{r}\right)h' -
g\left(\sqrt{f}\hat{\sigma}+\hat{q}^2h\hat{\psi}^2 -
\hat{q}^2\hat{\eta}\sqrt{f}\hat{\sigma}\hat{\psi}^2 \right) &= 0 \, , \\
g' + \left( \frac{5}{r}+\frac{r h'^2}{2f}+\frac{r}{2}\hat{\psi}'^2 \right)g
+ \left[ \frac{r}{2}\left(\frac{\hat{q}^2h^2}{f}+\hat{m}_s^2\right)\hat{\psi}^2
+r(\hat{\rho}-3) \right] g^2 &= 0 \, , \\
f' + \left[ rg(\hat{p}+3) - \frac{1}{r}  + \frac{1}{2}r\hat{\psi}'^2 +
\frac{r}{2}g\left(\frac{\hat{q}^2h^2}{f}-\hat{m}_s^2\right)\hat{\psi}^2 \right]f -
\frac{1}{2}rh'^2 &= 0 \, ,
\end{align}
\end{subequations}
where primes denote derivatives with respect to the radial coordinate.

The fluid quantities are given by~(\ref{eq:fluid-quantities}) where the local chemical
potential is now
\begin{align}
\label{eq:local-mu}
 \hat{\mu}_l = \frac{h}{\sqrt{f}} \left(1 - \hat{\eta} \hat{q}^2\hat{\psi}^2 \right) \, .
\end{align}
The fluid parameters $m_f$, $q_f$ and $\beta$ appear in the field
equations~(\ref{eq:field-equations}) only through the rescaled quantities~(\ref{mf}).
When working with these rescaled quantities,  the fermionic charge $q_f$ drops out of the equations, and  its sign is encoded in the sign of $\hat{\mu}_l$ as discussed in Section \ref{sec:matter in AdS}.

The bulk fluid is made up of 4d Dirac fermions, thus we can have physical states with  either sign of the charge. 
In our conventions, Dirac particles (which we call electrons) have $q_f>0$,  antiparticles (or holes, or  positrons) have $q_f<0$.
Thus, a positive chemical potential will fill particle-like states, and a negative one hole-like states. 

As noted in Section \ref{sec:matter in AdS}, in the absence of current-current interactions, i.e. for $\hat\eta=0$, the sign of the local chemical potential in~(\ref{eq:local-mu}) is dictated by the sign of the electric potential. In both the electron  star  and compact star solutions this is the same throughout the bulk (for example, it is non-negative  if the  boundary value of the electric potential is positive). The same happens when $\hat\eta<0$, as one can see from Eq.~(\ref{eq:local-mu}). 

On the other hand, if we turn on a boson-fermion coupling  $\hat\eta >0$, the sign of the chemical potential is not determined, and there can be cases in which   $\hat\mu_l(r)$ has different signs in different bulk regions.
This is indeed what happens in the solutions we describe in subsection \ref{epsol}.

From equations  (\ref{eq:fluid-quantities}), we see that the fluid density is non-zero for
$|\hat{\mu}_l|>\hat{m}_f$. %  or for  $\hat{\mu}_l< \hat{m}_f <0$, with $\hat{m}_f$ and
% $\hat{\mu}_l$ having the same sign.
% Since, from Eq.~(\ref{mf}), the sign of $\hat{m}_f$ is the same as that of $q_f$, the
% first case
The case $\hat{\mu}_l(r)>m_f$ corresponds to a fluid made out of positively charged particles (electrons), whereas a negative $\hat{\mu}_l(r)<-\hat{m}_f$ 
leads to  a fluid of negatively charged particles (positrons). 
Notice that, in equations (\ref{eq:fluid-quantities}), the energy density $\hat{\rho}(r)$ and the pressure $\hat{p}(r)$ are positive in both
cases,  whereas  the charge density $\hat{\sigma}$ is positive or negative for the  electrons and positrons fluids, respectively. 

We will see in the next sections that, depending on the parameters, bulk solutions with various arrangements of differently charged fluids are possible (electrons, positrons, or both).

The relevant parameters of the model are thus:
\begin{align}
\textnormal{scalar:} \: \, (\hat{q},  \hat{m}_s) \, , \qquad \textnormal{fluid:} \: \,
(\hat{m}_f,
\hat{\beta}) \, , \qquad \textnormal{interaction:} \, \: \hat{\eta} \, ,
\end{align}
where the scalar field mass satisfies the $AdS_4$ BF bound $-9/4<\hat{m}_s^2<0$ (i.e. the
operator dual to $\psi$ is relevant).
We restrict the analysis to the case where the scalar parameters
satisfy~(\ref{eq:condition-q-scalar}) and the fermion mass satisfies $0<\hat{m}_f<1$.
Then, the holographic superconductor,  with IR
asymptotics~(\ref{eq:sc-IR-sol}), and the electron star are
solutions of the system when there is no fluid and the scalar field is trivial, respectively.

We assume that in the UV ($r\to0$) the solutions are asymptotically $AdS_4$; the metric, gauge
field and scalar field in this region are then given by~(\ref{eq:f-g-h-uv})
and~(\ref{eq:psi-uv}) where we have imposed that the non-normalizable mode of the scalar
field vanishes.
If the normalizable mode $\psi_+$ is non-trivial, this corresponds to a spontaneous breaking of
the boundary global $U(1)$ symmetry.

Of the solutions described in  Section \ref{sec:matter in AdS}, the ERN, HSC and ES  are still solutions for any value of $\hat{\eta}$, since in the absence of  either the scalar condensate, or the fluid density, or both, the current-current interaction terms drop out of the field equations. 
 The CS configurations are solutions  to the system~(\ref{eq:field-equations}) when the direct interaction~(\ref{eq:L-int})
between the scalar field and the fluid is turned off.
Below, we present new solutions which exhibit a non-trivial profile of
the scalar field and a compact  electron star, a compact positron star or both at the same time.

%%%%%%%%%%%%%%%%%%%%%%%%%%%%%%%%%%%%%%%%%%%%%%%%%%%%%%
\subsection{The electron-positron-scalar solution} \label{epsol}
%%%%%%%%%%%%%%%%%%%%%%%%%%%%%%%%%%%%%%%%%%%%%%%%%%%%%%

\begin{figure}[h!]
\centering
\subfloat[ ]{
\includegraphics[width=0.45\textwidth]{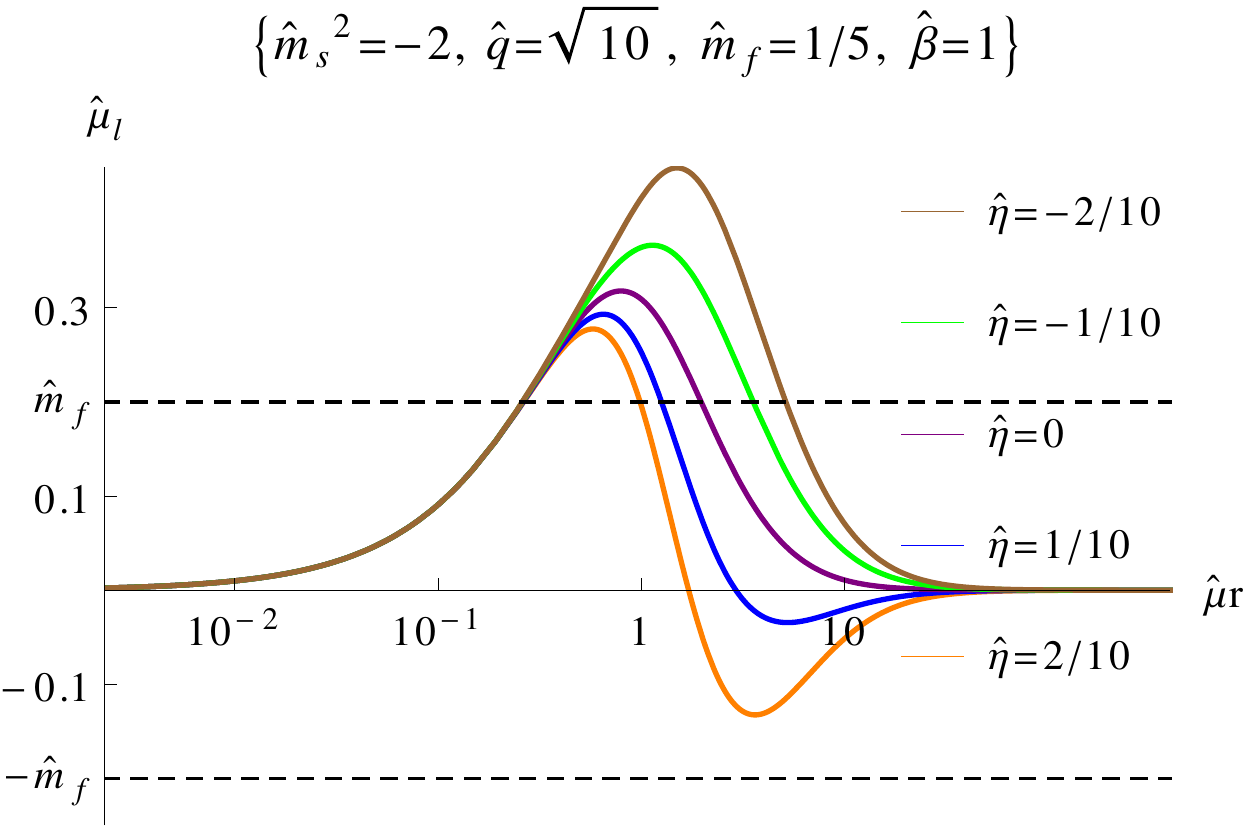}
\label{fig:profiles-e-localmu}
}
\subfloat[ ]{
\includegraphics[width=0.45\textwidth]{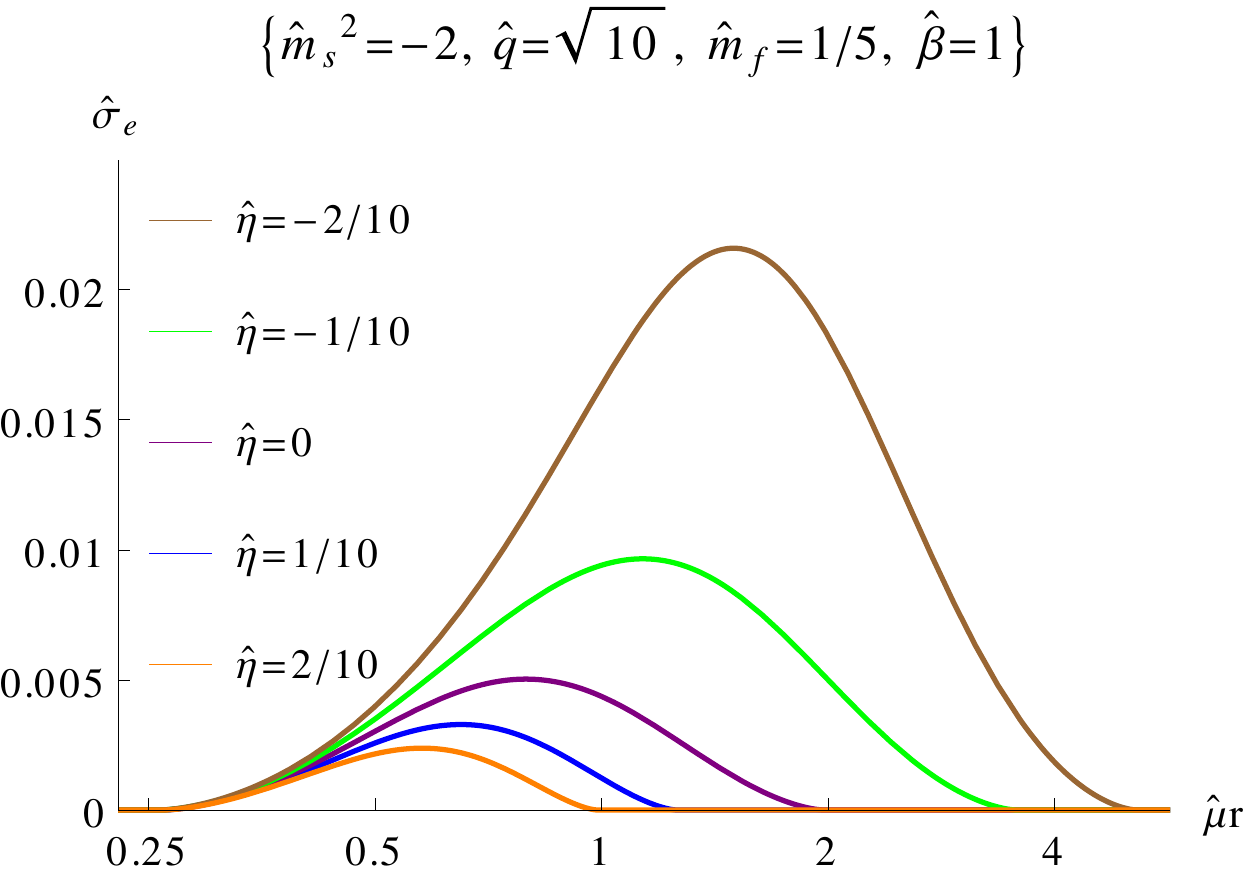}
\label{fig:profiles-e-chargee}
}\caption{Profiles of the local chemical potential (a) and the fluid charge densities (b) for  eCS solutions.}
\end{figure}

To see how the new solutions arise, we notice that in the HSC solution, the local chemical
potential~(\ref{eq:local-mu}) vanishes asymptotically both in the UV and in the IR and admits
at least one extremum value in the bulk.
In the IR,
\begin{align}
\label{eq:local-mu-IR}
\hat{\mu}_l \sim - 4 \, h_0 \, \hat{\eta} \, \hat{q}^2 r^{\delta+1} \left(\log r\right)^{3/2}
\, , \qquad r\to\infty \, .
\end{align}
For $\hat{\eta}>0$ the local chemical potential $\hat{\mu}_l$ is negative in the
IR for $h_0>0$\footnote{The system~(\ref{eq:field-equations}) is invariant under charge
conjugation which acts on the gauge field and the charges as
$(h,\hat{q})\to-(h,\hat{q})$.   Without loss of generality, we can then
restrict our analysis to the case where the gauge field $A_t$ is positive, which means that
$h_0>0$.}.
Moreover, since the asymptotics $\hat{\mu}_l\sim\hat{\mu}r$ in the UV ($r\to0$) are not
modified by $\hat{\eta}\neq0$, $\hat{\mu}_l$ has both a minimum and a maximum value at
finite radii in the bulk for $\hat{\eta}>0$ (see Figure \ref{fig:profiles-e-localmu}).
For $\hat{\eta}\leq0$, the behavior of $\hat{\mu}_l$ is similar to what happens in the
compact star solutions of~\cite{Nitti:2013xaa}, where it is always positive and admits
a maximum.

\begin{figure}[h!]
\centering
\subfloat[ ]{
\includegraphics[width=0.45\textwidth]{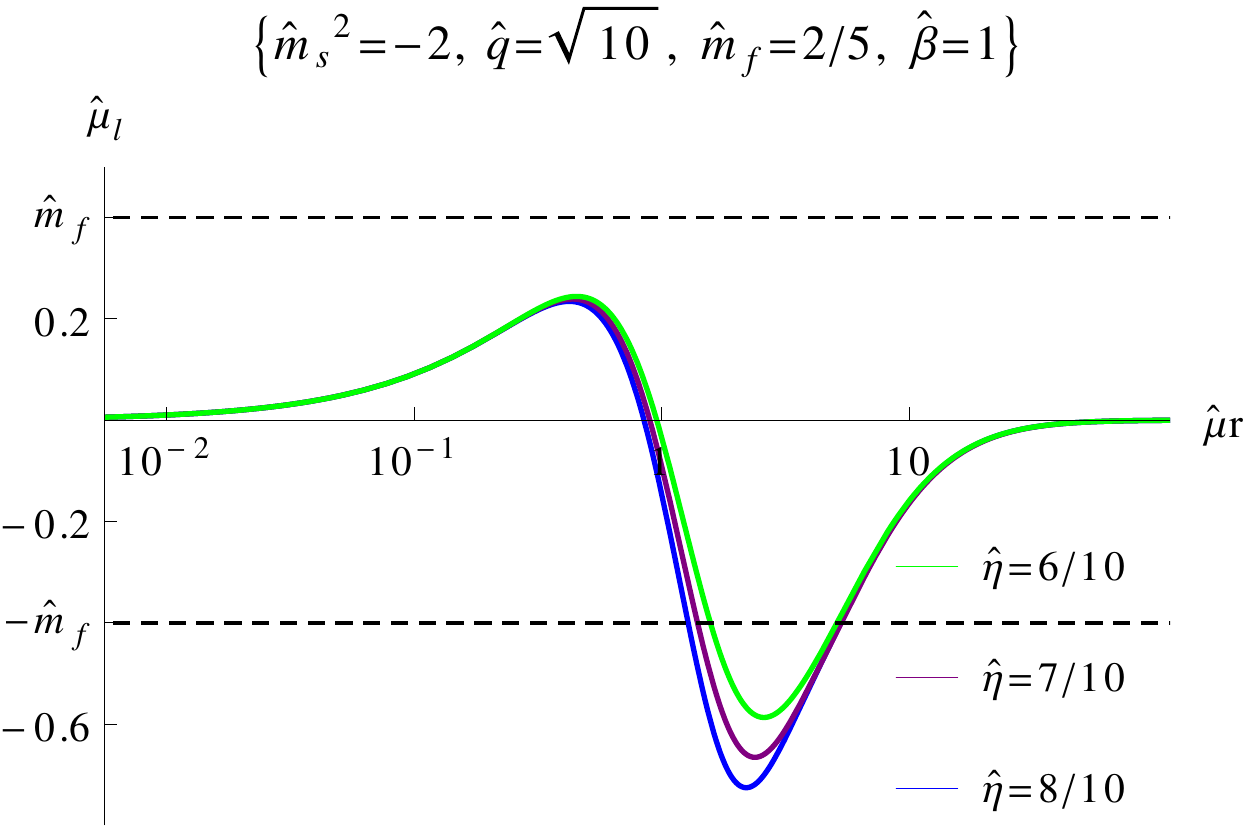}
\label{fig:profiles-p-localmu}
}
\subfloat[ ]{
\includegraphics[width=0.45\textwidth]{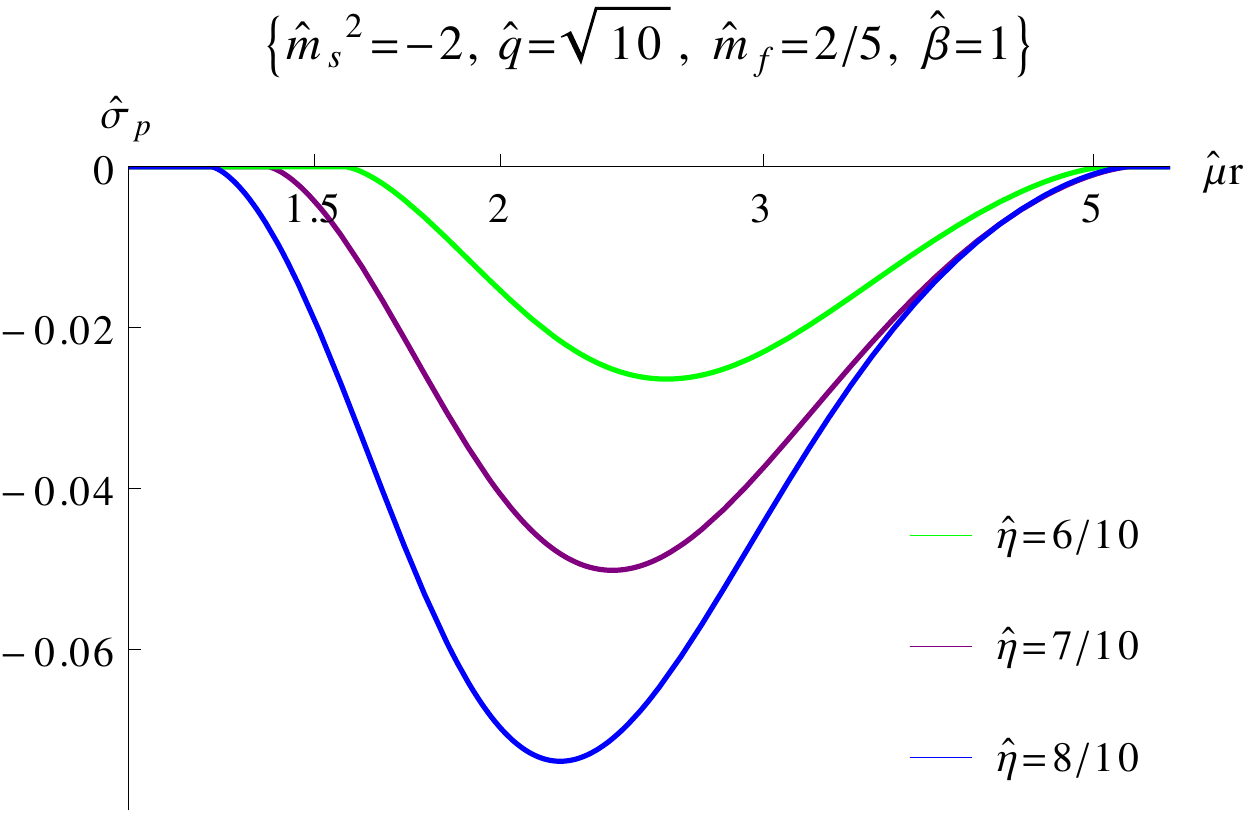}
\label{fig:profiles-p-chargep}
}\caption{Profiles of the local chemical potential (a) and the fluid charge densities (b) for pCS solutions.}
\end{figure}

\begin{figure}[h!]
\centering
\subfloat[ ]{
\includegraphics[width=0.45\textwidth]{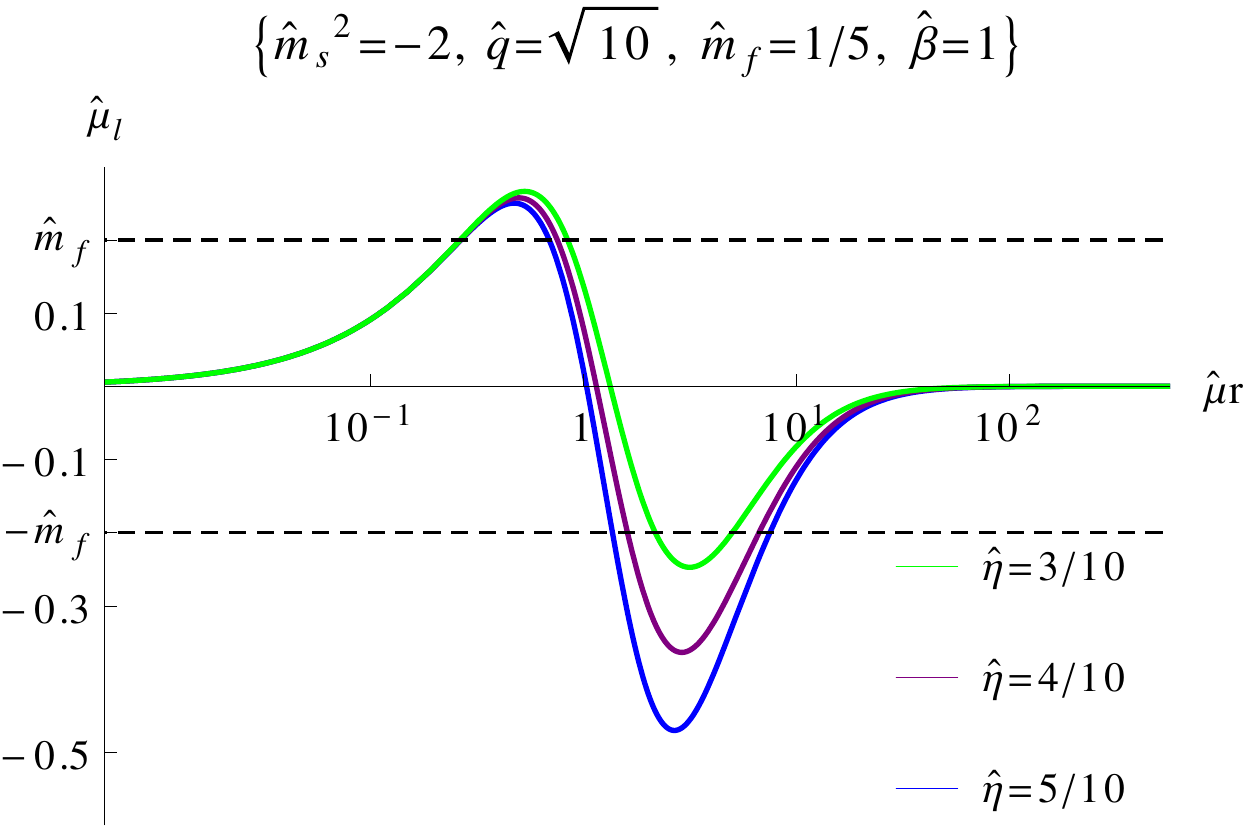}
\label{fig:profiles-pe-localmu}
}\\
\subfloat[ ]{
\includegraphics[width=0.45\textwidth]{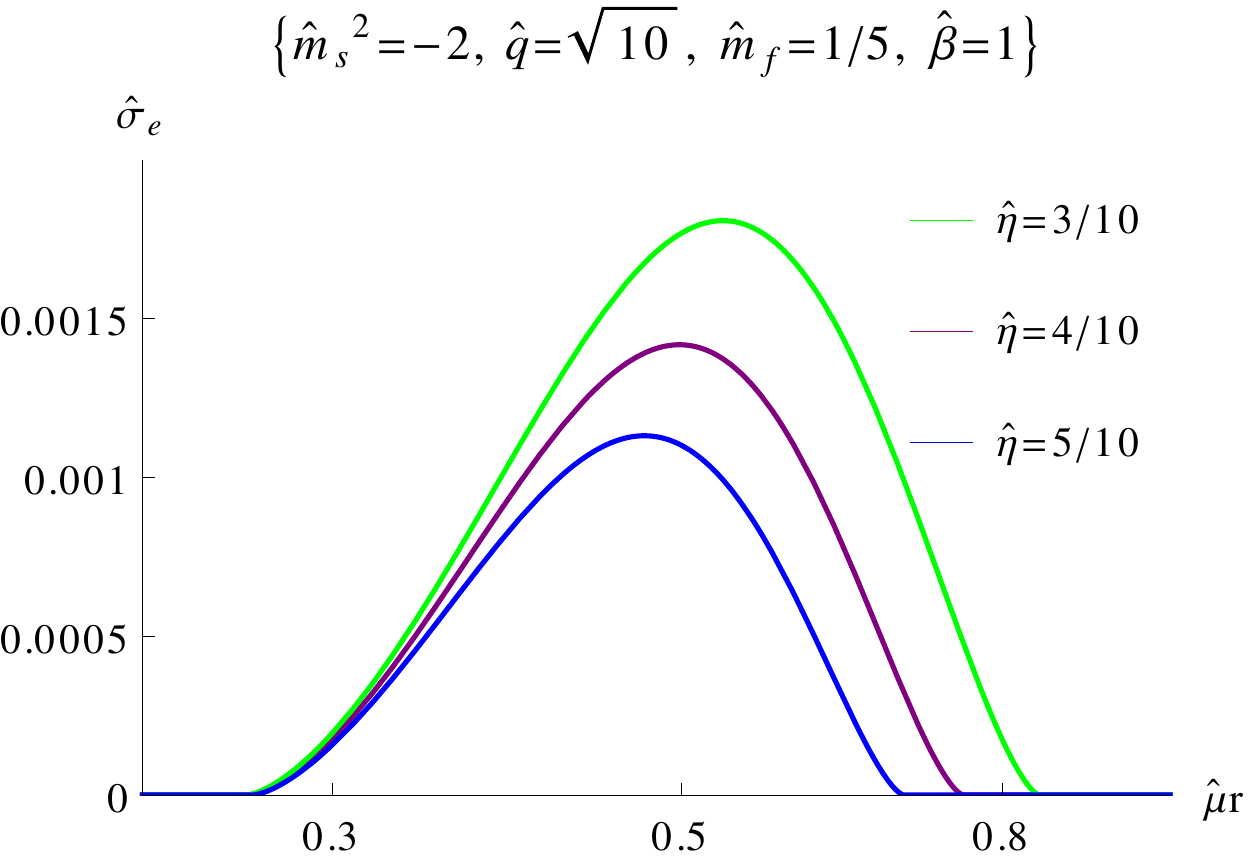}
\label{fig:profiles-pe-chargee}
}
\subfloat[ ]{
\includegraphics[width=0.45\textwidth]{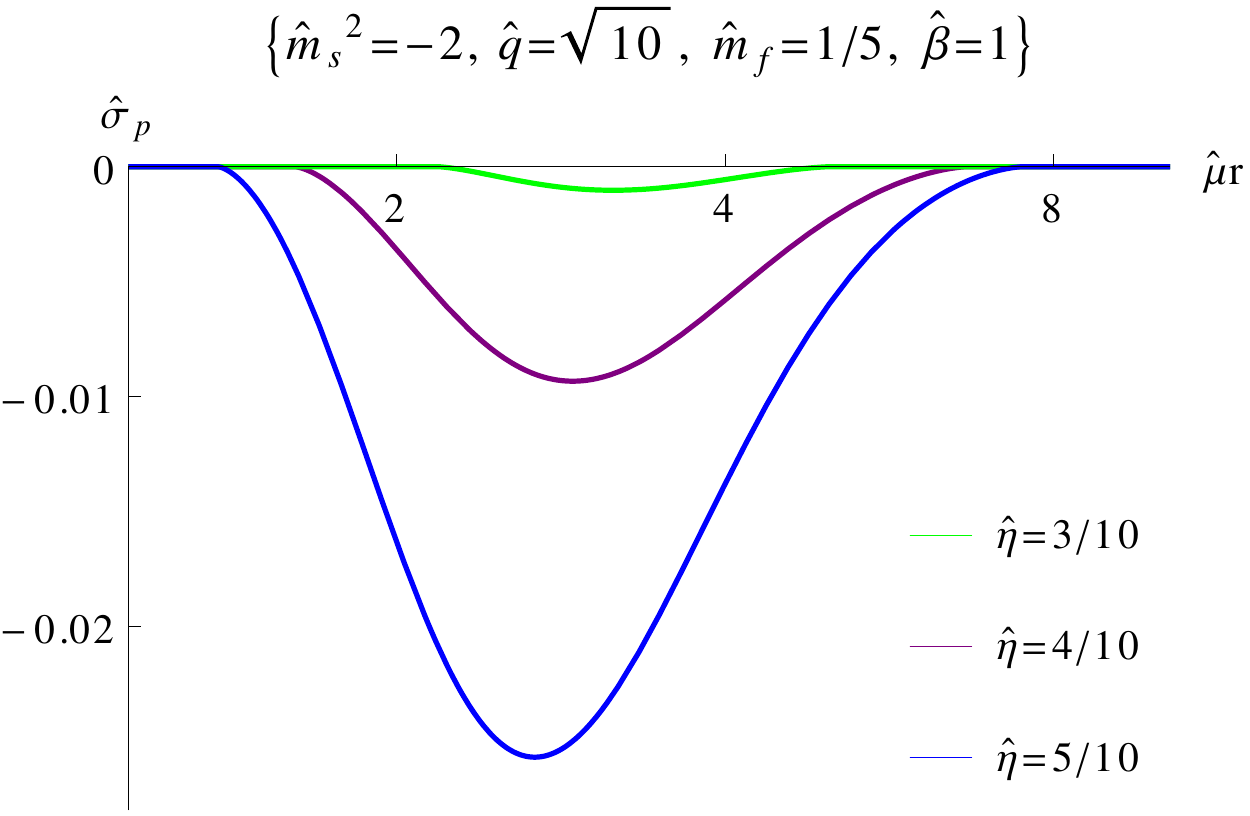}
\label{fig:profiles-pe-chargep}
}
\caption{Profiles of the local chemical potential (a) and the fluid charge densities (b), (c) for peCS solutions.}
%\label{fig:profiles-CS}
\end{figure}

Similarly to the non-interacting case~\cite{Nitti:2013xaa}, solutions with both a non-trivial
scalar and a non-trivial fluid exist when at least one local extremum of the local chemical
potential in the HSC solution is larger than the mass of the fermions.
Three kinds of compact star(s) solutions arise~:
\begin{itemize}
 \item \textbf{Compact electron star (eCS):} a fluid density is confined in a shell whose
boundaries $r_1$ and $r_2$ satisfy $\hat{\mu}_l(r_i)=\hat{m}_f$, $i=1,2$. The charge density
of the fluid is positive, then the fluid is made of electrons. This situation is displayed
in Figures~\ref{fig:profiles-e-localmu} and~\ref{fig:profiles-e-chargee}.

 \item \textbf{Compact positron star (pCS):} a fluid density is confined in a shell whose
boundaries $r_1$ and $r_2$ satisfy $\hat{\mu}_l(r_i)=-\hat{m}_f$, $i=1,2$. The charge density
of the fluid is negative, then the fluid is made of positrons. This situation is displayed in
Figures~\ref{fig:profiles-p-localmu} and~\ref{fig:profiles-p-chargep}.

 \item \textbf{Compact positron/electron stars (peCS):}  In this case the solution exhibits charge polarization  in the bulk: two fluid shells  of opposite charges are confined in
distinct regions of spacetime,  bounded respectively by $(r^{e}_1, r^{e}_2)$ and  $(r^{p}_1, r^{p}_2)$ determined by the equations:
\be\label{peCSbound}
\hat{\mu}_l(r^{e}_1) = \hat{\mu}_l(r^{e}_2) =\hat{m}_f\, , \qquad \hat{\mu}_l(r^{p}_1) = \hat{\mu}_l(r^{p}_2) =-\hat{m}_f. 
\ee
The  fluid in one region is made of electrons,  the one in the other region of positrons. Clearly, for this solutions to exist, the chemical potential must change sign in the bulk.  
 Due to fixed UV asymptotics of the local chemical potential, the
fluid of electrons is situated closer to the UV boundary than the fluid of positrons. 

This
situation is displayed in Figures~\ref{fig:profiles-pe-localmu}, \ref{fig:profiles-pe-chargee}
and \ref{fig:profiles-pe-chargep}.
\end{itemize}
The kind of compact star(s) solutions that may exist depends on the maximum and minimum value  of the local chemical potential,  
\begin{align}
\hat{\mu}_\textnormal{max}(\hat{m}_s^2,\hat{q},\hat{m}_f,\hat{\beta},\hat{\eta}) \equiv
\max_{0<r<\infty} \hat{\mu}_l(r) \, , \qquad \hat{\mu}_\textnormal{min}(\hat{m}_s^2,\hat{q},\hat{m}_f,\hat{\beta},\hat{\eta}) \equiv
\min_{0<r<\infty} \hat{\mu}_l(r) \, .
\end{align}
% %
% and the minimum value
% %
% \begin{align}
% \hat{\mu}_\textnormal{min}(\hat{m}_s^2,\hat{q},\hat{m}_f,\hat{\beta},\hat{\eta}) \equiv
% \min_{0<r<\infty} \hat{\mu}_l(r)
% \end{align}
%
The possible outcomes are summarized  in Table~\ref{tab:existence}.
We denote the case where no compact star(s) exists by noCS.
Notice that $\hat{\mu}_\textnormal{max}>0$ for all $\hat{\eta}$; $\hat{\mu}_\textnormal{min}$
is negative when $\hat{\eta}>0$ and vanishes for $\hat{\eta}\leq 0$.
\begin{table}[ht!]
\centering
\begin{tabular}{ c | c | c | }
\cline{2-3}
  & $\hat{\mu}_\textnormal{min}<-\hat{m}_f$ &
$\hat{\mu}_\textnormal{min}>-\hat{m}_f$ \\
\hline
 \multicolumn{1}{|c|}{$\hat{\mu}_\textnormal{max}>\hat{m}_f$} & peCS, pCS, eCS & eCS \\
\hline
 \multicolumn{1}{|c|}{$\hat{\mu}_\textnormal{max}<\hat{m}_f$} & pCS & noCS \\
\hline
 \end{tabular}
\caption{Conditions on the minimal and maximal values of the local chemical potential
for the existence of the compact star(s) solutions. For given parameters, if the peCS solution exists, the pCS and eCS are also solutions to the system.  \lq\lq noCS'' means that no compact star(s) solution exists, because 
$\left|\hat{\mu}_l(r)\right|<\hat{m}_f$ everywhere in the bulk.}
\label{tab:existence}
\end{table}
%

% {\bf \large In the following paragraph we need to explain how the various charges are defined. It is not enough to point to the appendix}\\

%
%

%%%%%%%%%%%%%%%%%%%%%%%%
\subsection{Phase diagrams of charged  solutions}
\label{sec:free-energy}
%%%%%%%%%%%%%%%%%%%%%%%%

In the previous section we have seen that different arrangements  of  fermionic fluids are possible at zero temperature. Depending on the parameter values, we can go from the pure condensate with no fermions (holographic superconductor, or HSC), purely positive or purely negative confined  fluid shells (electron or positron compact stars, or eCS and pCS respectively), and polarized shells of positive/negative charged fluid regions  (compact positron-electron compact stars, or peCS). In all these configurations, the fermionic charges are surrounded by the scalar condensate, which dominates the UV and IR geometry and confines the fluid in finite regions of the bulk. 

Here we address the question about which, for a given choice of parameter, is the solution that has the lowest free energy and dominates the  grand-canonical ensemble. We work at zero temperature and fixed (boundary) chemical potential $\hat\mu$. Thus, different solutions will in general have different charge. 

As was noted in \cite{Nitti:2013xaa}, comparing the free energy of different solutions is relatively simple,  due to the existence of a scaling symmetry of the field equations that allows to change the value of $\hat\mu$ within a given class of solutions. Thanks to this solution-generating symmetry, one can show that, within each class of solution, the free energy $F(\hat\mu)$ has the simple expression
\be\label{free}
F_i(\hat\mu) =  -\frac{1}{3} c_i \hat\mu^3 \, ,
\ee 
where $c_i$ are $\hat\mu$-independent constants which depend only on the class of solutions. The index $i$ runs over all  solutions which exist at a given point in parameter space (HCS, eCS, pCS and/or peCS). 

Equation (\ref{free}) shows that there can be no non-trivial phase transitions between the solutions as $\hat\mu$ is varied. On the other hand, the constants $c_i$ depend non-trivially  on the parameters of the model, and there can be phase transitions between different solutions  as these parameters are varied. 

To observe the transition, it is sufficient to compute $ F(\hat\mu)$ in units of $\hat\mu$ for one representative of each  class of solution, and for a given point in parameter space the solution with larger $c_i$ will be the preferred one, as $F_i - F_j \propto -(c_i-c_j)\hat\mu^3$. 

We performed this analysis numerically on the solutions described in the previous section\footnote{Extremal RN black hole solutions were left out of this analysis, since they are always dominated by the HCS solutions in the region of parameter space where the latter exist, and to which we are restricting.}. The results of this analysis  reveals a rich phase diagram, with the interplay of various phase transitions. 
% as displayed in figures \ref{fig:free-energy} and \ref{fig:phase-diagram}.
  
\begin{figure}[h]
\centering
\subfloat[ ]{
\includegraphics[width=0.48\textwidth]{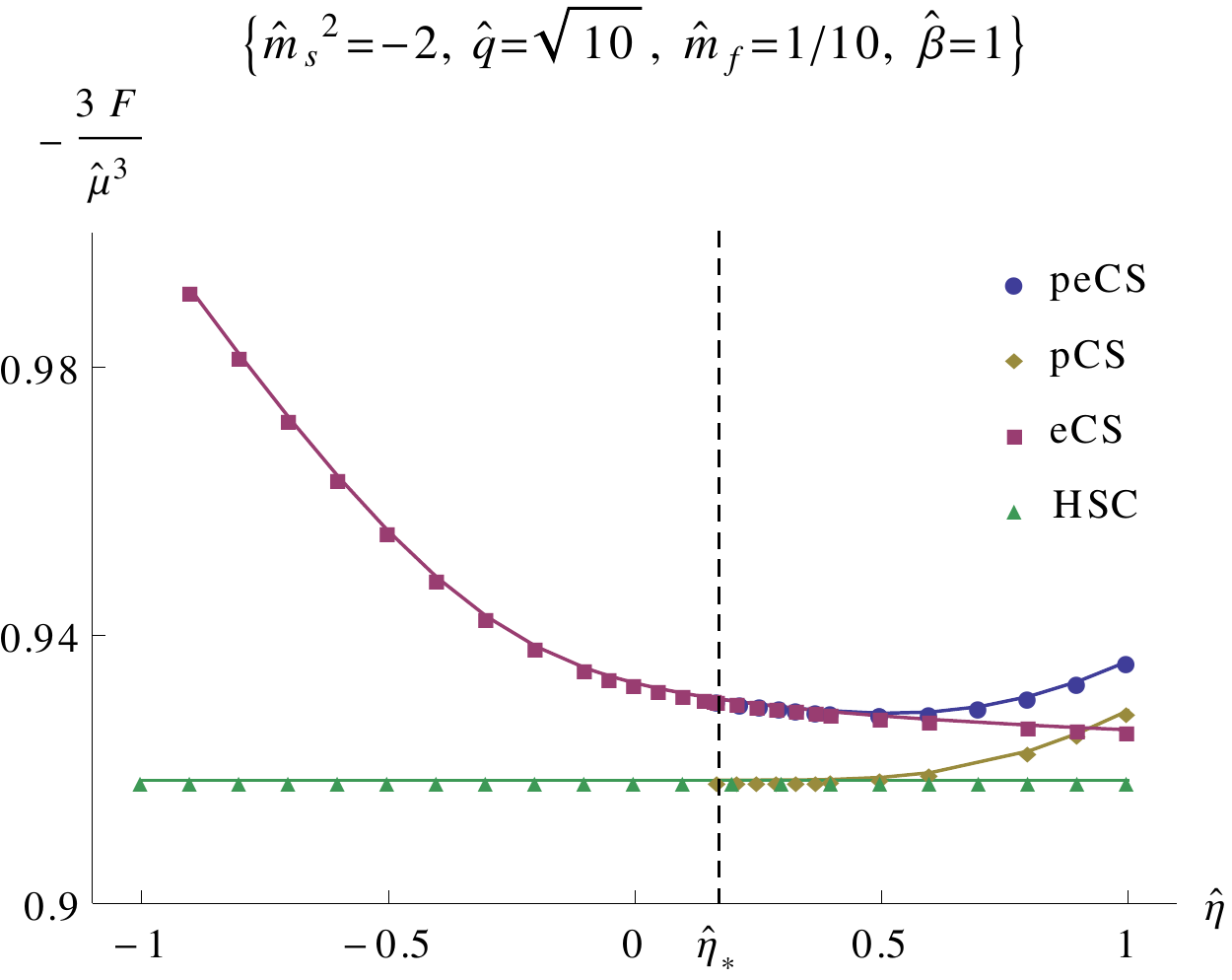}
\label{fig:free-energy-1}
}
\subfloat[ ]{
\includegraphics[width=0.48\textwidth]{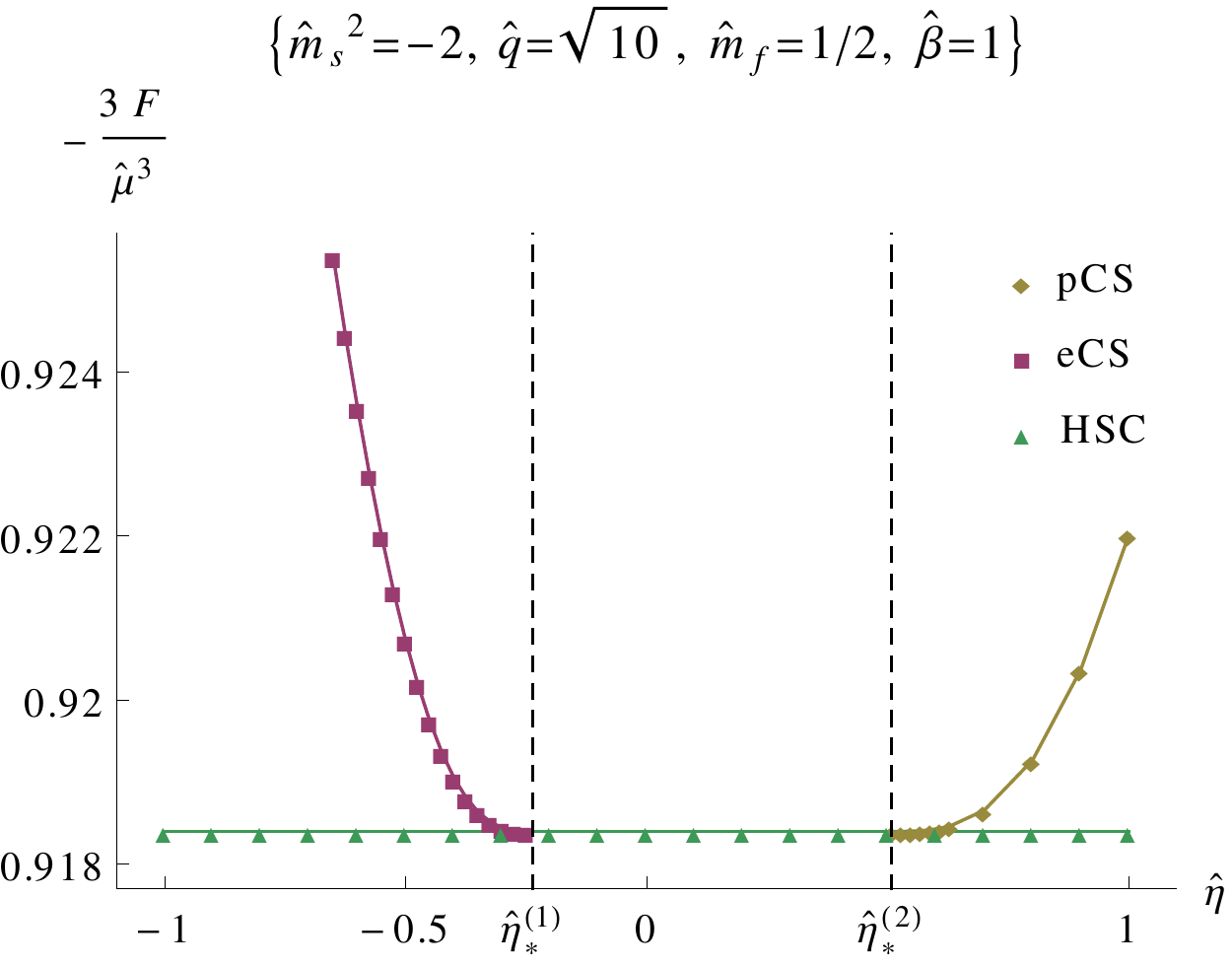}
\label{fig:free-energy-2}
}
\\
\subfloat[ ]{
\includegraphics[width=0.48\textwidth]{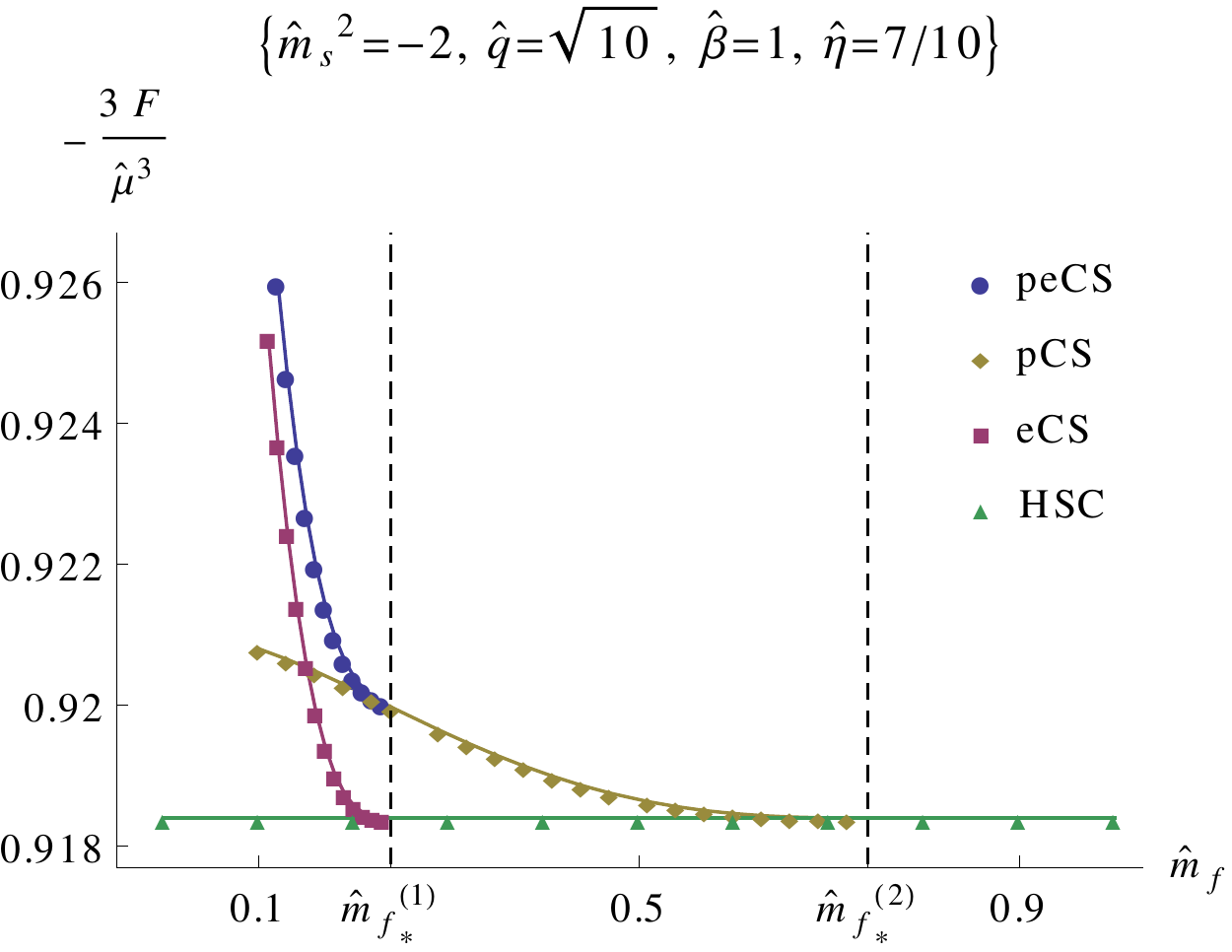}
\label{fig:free-energy-3}
}
\caption{Free energy (normalized to the chemical potential).
(a) displays the free energy as a function of $\hat{\eta}$ in the transition between eCS and
peCS solutions, where $\hat{\eta}_*\simeq0.17$ is the critical coupling constant.
(b) displays the free energy as a function of $\hat{\eta}$ in the transition between eCS, HSC
and pCS solutions; $\hat{\eta}_*^{(1)}\simeq-0.24$ and $\hat{\eta}_*^{(2)}\simeq0.51$ are the
critical coupling constants between the eCS and HSC solutions, and the HSC and pCS solutions,
respectively.
(c) displays the free energy as a function of $\hat{m}_f$ in the transition between peCS,
pCS and HSC solutions; $\hat{m}_{f*}^{(1)}\simeq0.24$ and
$\hat{m}_{f*}^{(2)}\simeq0.74$ are the critical coupling constants between the peCS and pCS
solutions, and the pCS and HSC solutions, respectively. To avoid clutter, 
in (a), (b) and (c) we did not display the (normalized) free energy of the ES and the ERN
solutions. They are much smaller than the free energy of the CS and HSC
solutions : in (a) and (b), the free energy of the ES solution is equal to 0.49 and 0.42,
respectively. In (c), it is between 0.41 and 0.49 depending on $\hat{m}_f$. In all cases, the
free energy of the ERN solution is $1/\sqrt{6}\simeq0.41$.}
\label{fig:free-energy}
\end{figure}

We will first analyze what happens if we vary $\hat\eta$ while keeping other parameters fixed
(Figures~\ref{fig:free-energy-1} and~\ref{fig:free-energy-2}). As we know from
\cite{Nitti:2013xaa}, for $\hat{\eta}=0$,  whenever compact electron star solutions exist, they dominate the ensemble. Otherwise, the preferred solution is the HSC. 

Let us first  choose the parameters so that, for $\hat\eta=0$, the compact electron star is the
preferred  solution (Figure~\ref{fig:free-energy-1}): if we dial up a positive interaction 
term $\hat\eta$, eventually the effect of the condensate polarizes the star and, at a critical value $\hat{\eta}_*>0$, we find a
continuous transition to the peCS solution, which now is the one dominating the ensemble.
The eCS keeps existing beyond the critical point $\hat{\eta}_*$, where we also see the emergence of a new pCS solution which starts
dominating over the HSC solution but not over the peCS
solution.

On the other hand, if we start from a point where, for $\hat\eta=0$, there is no eCS solution
(Figure~\ref{fig:free-energy-2}), we see that dialing up $\hat\eta$ either way one will get to
a critical point where either a positive or a  negative charged star will be formed, and
dominate the ensemble henceforth. 
 
For a given (positive) value of $\hat\eta$, the type of solution depends on the fermion mass
$\hat{m}_f$, as shown in Figure \ref{fig:free-energy-3}.
At small mass, the polarized peCS solution dominates over the eCS and pCS solutions.
As the mass is increased, one first encounters a continuous transition from the peCS to the pCS
solution at the critical value $\hat{m}_{f_*}^{(1)}$.
At this point, the (subdominant) eCS solution merges into the (subdominant) HSC solution.
Then, at the second critical point $\hat{m}_{f_*}^{(2)}$ the charged fluid 
disappears and the solution merges into the HCS solution.

\begin{figure}[h!]
\centering
\subfloat[ ]{
\includegraphics[width=0.48\textwidth]{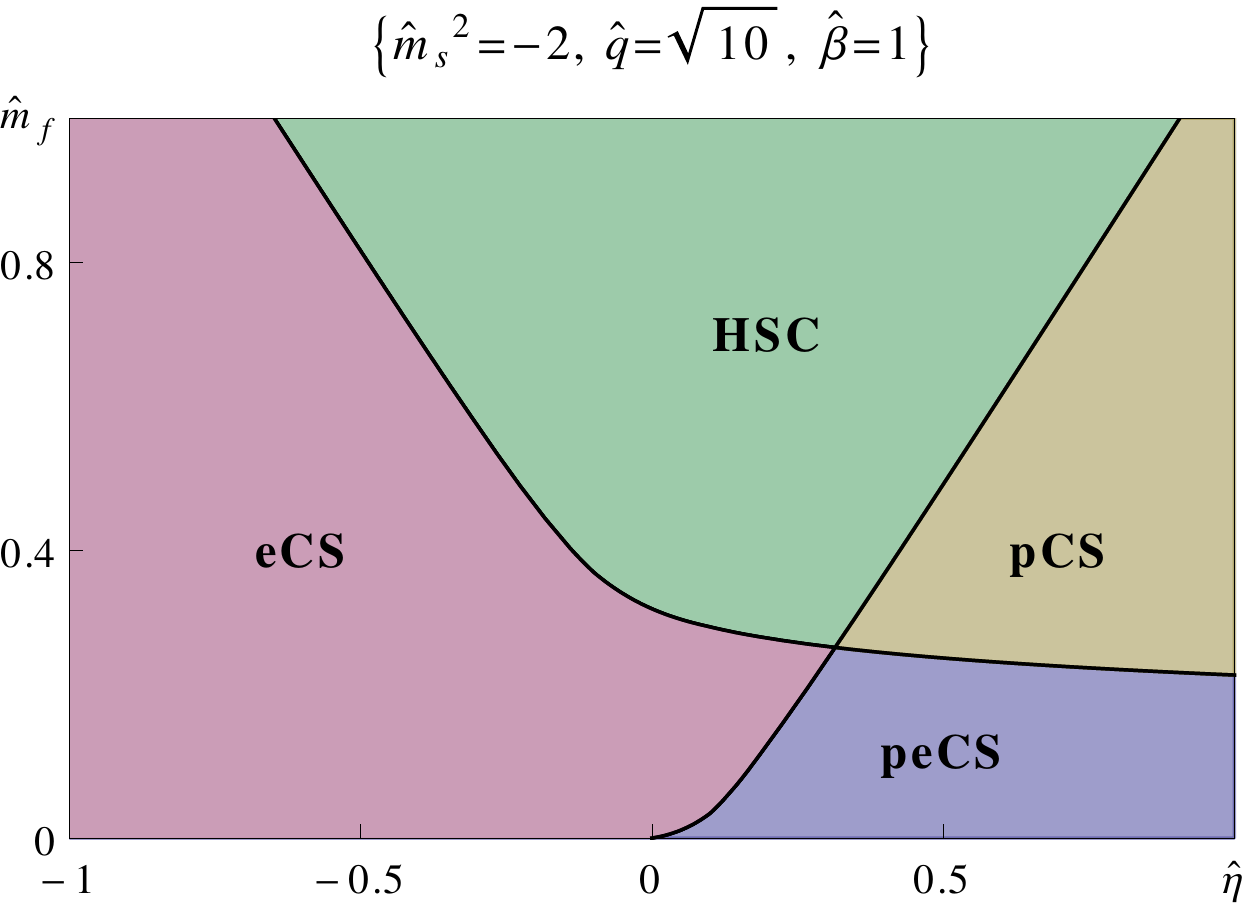}
\label{fig:phase-diagram-1}
}
\subfloat[ ]{
\includegraphics[width=0.48\textwidth]{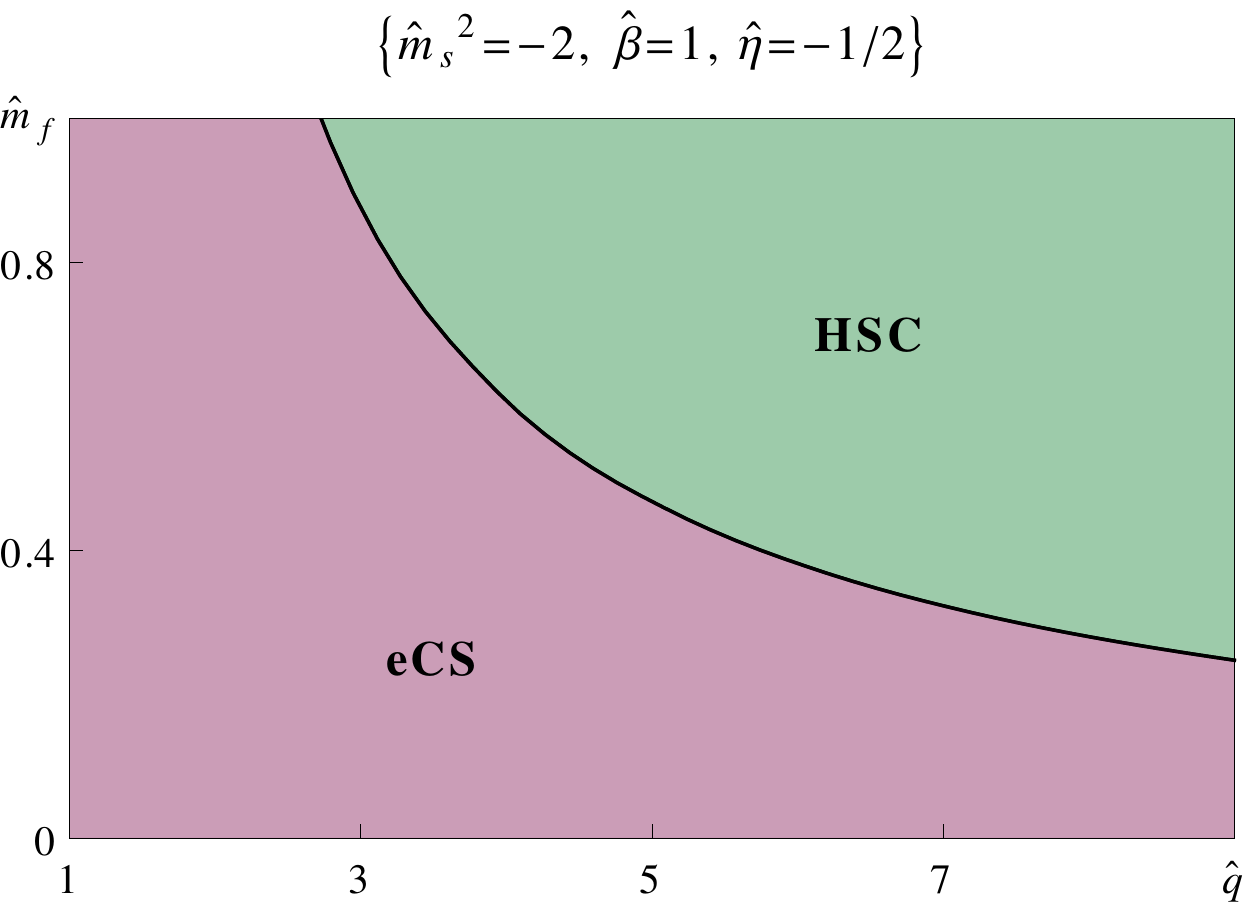}
\label{fig:phase-diagram-2}

}\\
\subfloat[ ]{
\includegraphics[width=0.48\textwidth]{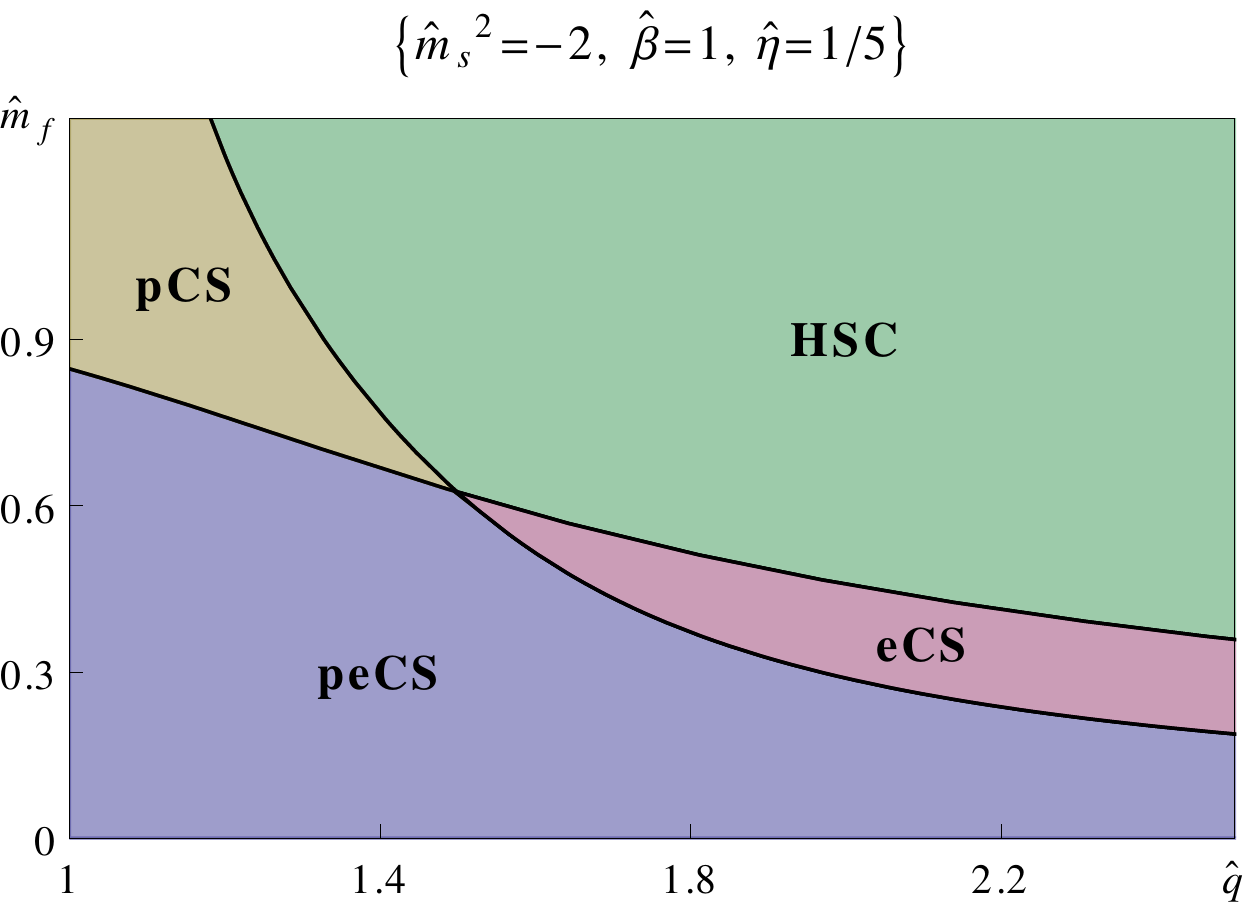}
\label{fig:phase-diagram-3}

}
\subfloat[ ]{
\includegraphics[width=0.48\textwidth]{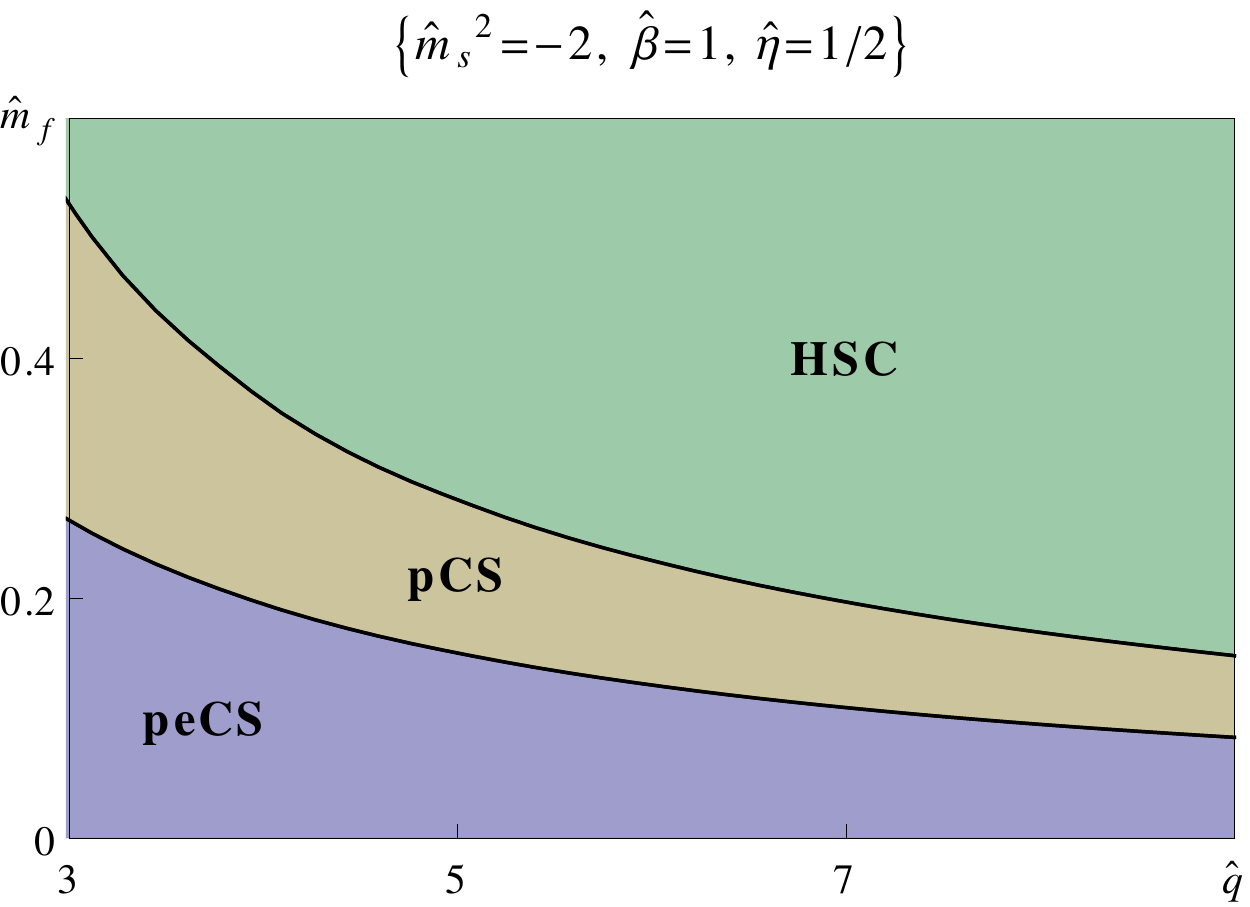}
\label{fig:phase-diagram-4}
}
\caption{Phase diagrams of the CS/HSC transitions. The critical lines (solid black
lines) correspond to the minimum and maximum values of the local chemical potential in
the HSC solution, which can be seen from the plots the free energy of the
different solutions (see Figure~\protect\ref{fig:free-energy}).}
\label{fig:phase-diagram}
\end{figure}

We have also analyzed the phase diagram as a function of the scalar charge $\hat{q}$. Phase
diagrams of the system are displayed in Figure \ref{fig:phase-diagram}, in the plane
$(\hat{m}_f, \hat{\eta})$ at fixed scalar charge $\hat{q}$ (Figure~\ref{fig:phase-diagram-1})
and in the plane  $(\hat{m}_f, \hat{q})$ at fixed $\hat{\eta}$
(Figures~\ref{fig:phase-diagram-2}-\ref{fig:phase-diagram-4}). The critical lines separating
the various phases correspond to the points where the maxima and minima of the local chemical
potential are equal in absolute value to $\hat{m}_f$. The different colors correspond to the
dominant phase in each region. Thus, all these transitions are continuous and take place at the
points where it is possible to fill the charged fermion states: whenever a fluid solution is
possible by the condition $|\hat{\mu}_l| > \hat{m}_f$, that solution will form. Furthermore,
solutions in which the charge is distributed between more fluid components are preferred.


\subsection{Charge distribution and screening}\label{scr}

Let us briefly discuss how the total charge of the system is divided among the various bulk
components. From the boundary field theory point of view, the only invariant definition of the
charge is the {\it total} charge of the solution, that one can read off from the asymptotic
behavior of the gauge field. However, it is still instructive to define, in the bulk, the
separate   charge of each component by the expression it would have if that were the only
component present.%  Thus, we define the charge associated to the electron fluid, positron
%fluid, and scalar field

With the ansatz~(\ref{eq:ansatz-full}), the electric charge carried by the scalar field in
the bulk is
\begin{align}
\label{eq:charge-scalar}
 \hat{Q}_\textnormal{scalar} = \hat{q}^2 \int_0^\infty \D r\,
\frac{\sqrt{g}}{r^2\sqrt{f}}\,h\,|\hat{\psi}|^2
\end{align}
and the electric charges of the electron and positron fluids are respectively
\begin{equation}
\label{eq:charge-fluids}
\begin{aligned}
 \hat{Q}_\textnormal{e} &= \int_{r_1^e}^{r_2^e} \D r\,
\frac{\sqrt{g}}{r^2}\,\hat{\sigma}_\textnormal{e} \, , \qquad 
\hat{Q}_\textnormal{p} &= \int_{r_1^p}^{r_2^p} \D r\,
\frac{\sqrt{g}}{r^2}\,\hat{\sigma}_\textnormal{p} \, ,
\end{aligned}
\end{equation}
where $r_1^e$ and $r_2^e$ are the boundaries of the electron star, and similarly for the
positron star.
The charge densities of the electron fluid $\hat{\sigma}_\textnormal{e}$ and the positron fluid
$\hat{\sigma}_\textnormal{p}$ are respectively positive and negative, and they are given in Eq.~(\ref{eq:fluid-quantities}) with $\hat{\mu}_l$ respectively positive or negative.   

Additionally, due to the current-current interaction term (\ref{eq:L-int}), and the fact that by Eq.~(\ref{Jscalar})  the scalar current is linear in the gauge field,  there are  \textit{screening electric charges}, given by
\begin{equation}
\label{eq:charge-int}
\begin{aligned}
 \hat{Q}_\textnormal{int,e} &= -\hat{\eta}\int_{r_1^e}^{r_2^e} \D r\,
\frac{\sqrt{g}}{r^2}\,\hat{q}^2\hat{\psi}^2 \,\hat{\sigma}_\textnormal{e} \, , \qquad 
\hat{Q}_\textnormal{int,p} &= -\hat{\eta}\int_{r_1^p}^{r_2^p} \D r\,
\frac{\sqrt{g}}{r^2}\,\hat{q}^2\hat{\psi}^2 \,\hat{\sigma}_\textnormal{p} \, ,
\end{aligned}
\end{equation}
which reflect the interaction between the scalar and the fluid made of electrons and positrons, respectively, and it gets contributions from the regions where the fluid density is non-vanishing.

The total electric charge of the system\footnote{In all our solutions except the extremal
Reissner-Nordstr\"om black hole, there is no charged event horizon and the electric charge is
shared between the fluid(s) and the scalar field.}
\begin{align}\label{Qtotal}
 \hat{Q} = \hat{Q}_\textnormal{scalar} + \left(\hat{Q}_\textnormal{e} + \hat{Q}_\textnormal{int,e} \right) + 
\left(\hat{Q}_\textnormal{p} +  \hat{Q}_\textnormal{int,p}\right).
\end{align}
matches the UV asymptotic behavior of the gauge field (\ref{eq:f-g-h-uv}),   
\begin{align}
\label{eq:charge-from-h}
 \hat{Q} = - %  \frac{1}{c}
 h'(0) \, .
\end{align}
This  has been verified numerically in all solutions we have constructed.

%
% where $c>0$ satisfies
% %
% \begin{align}
%  c^2 = \lim_{r\to0} r^2 f(r) \, .
% \end{align}
% %
% The constant $c$ in~(\ref{eq:charge-from-h}) corresponds to the rescaling of
% the time coordinate to get the asymptotic $AdS_4$ metric in Poincar\'e coordinates.

In Eq.~(\ref{Qtotal}), the second and third terms represent the total contributions from each charged fluid. Despite the possible presence of local negative charge components,  we will show below that all three terms  in Eq.~(\ref{Qtotal}) are positive for all solutions under considerations. This is consistent with  our choice  $\hat{\mu}>0$ for the boundary chemical potential  in the UV asymptotics (\ref{eq:f-g-h-uv}), which implies that   the boundary charge, i.e.  total charge of the system $\hat{Q}$ must be positive in all solutions. 

% The expressions for the charges are given in Eq.~(\ref{eq:charge-scalar}-\ref{eq:charge-int}).
Let us first consider the scalar field condensate contribution to the total charge.   As a consequence of the choice $\hat{\mu}>0$ in the UV,  the electric potential $h(r)$ is positive throughout the bulk\footnote{There can be, in principle, solutions with HSC IR and UV asymptotics in which $h(r)$ changes sign, but these are expected to have larger free energy\cite{Horowitz:2009ij}.}.  Thus, from Eq.~(\ref{eq:charge-scalar}), one can  see that the electric charge of the scalar field $\hat{Q}_\textnormal{scalar}$ is positive.

Due to the signs of the local charge densities in (\ref{eq:charge-fluids}),  $\hat{Q}_\textnormal{e}$ is positive, and $\hat{Q}_\textnormal{p}$ is negative. However, the latter is
over-screened by the scalar field through the charge of interaction $\hat{Q}_\textnormal{int,p}$,
so that $\left(\hat{Q}_\textnormal{p}+\hat{Q}_\textnormal{int,p}\right)>0$, as can be seen from
Eq.~(\ref{eq:charge-fluids}-\ref{eq:charge-int}) and the fact that, inside the positron fluid, $(1-\hat{\eta}\hat{q}^2) <0$, since by Eq.~(\ref{eq:local-mu}) this quantity  determines the sign of the chemical potential. 
 
By a similar reasoning, $\hat{Q}_\textnormal{int,e}$ is positive (negative) for $\hat{\eta}<0$
($\hat{\eta}>0$), but $\left(\hat{Q}_\textnormal{e}+\hat{Q}_\textnormal{int,e}\right)$ is always positive. Thus, for electrons, there may be charge screening or anti-screening, but never over-screening. 

Given the previous discussion, we can have a qualitative understanding of why the polarized electron-positron compact stars are stable configurations:   although the positron and electron parts of the fluid are made up of positive and negative charge fermionic constituents respectively, the screening of the negative electric charge by the  scalar condensate renders the total charge positive in both fluids. In particular, for peCS solutions, the two charged shells  experience electromagnetic repulsion, rather than attraction. Gravitational and electromagnetic forces are competing, and this makes the solution stable.

\begin{figure}[t!]
\centering
\subfloat[ ]{
\includegraphics[width=0.48\textwidth]{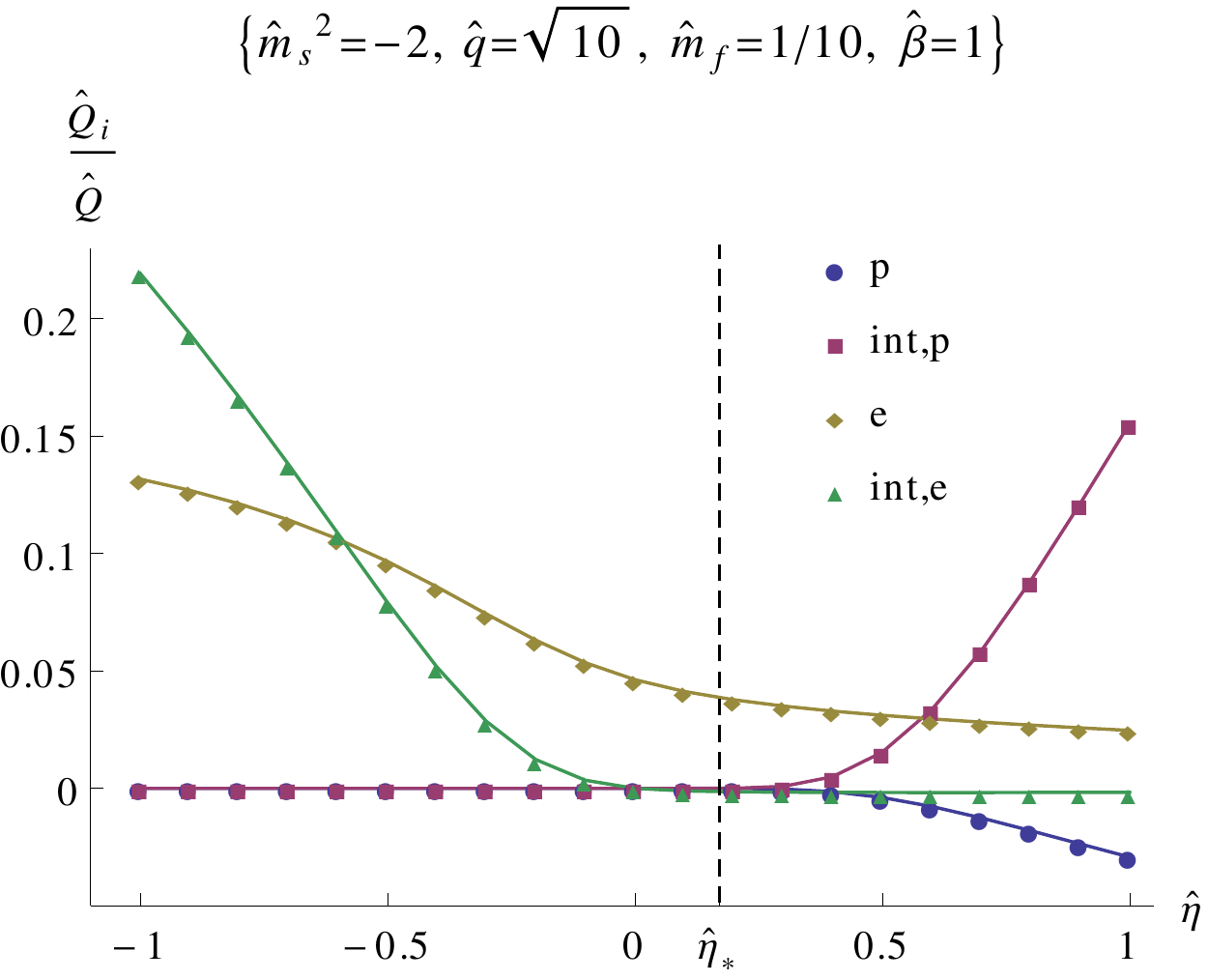}
}
\subfloat[ ]{
\includegraphics[width=0.48\textwidth]{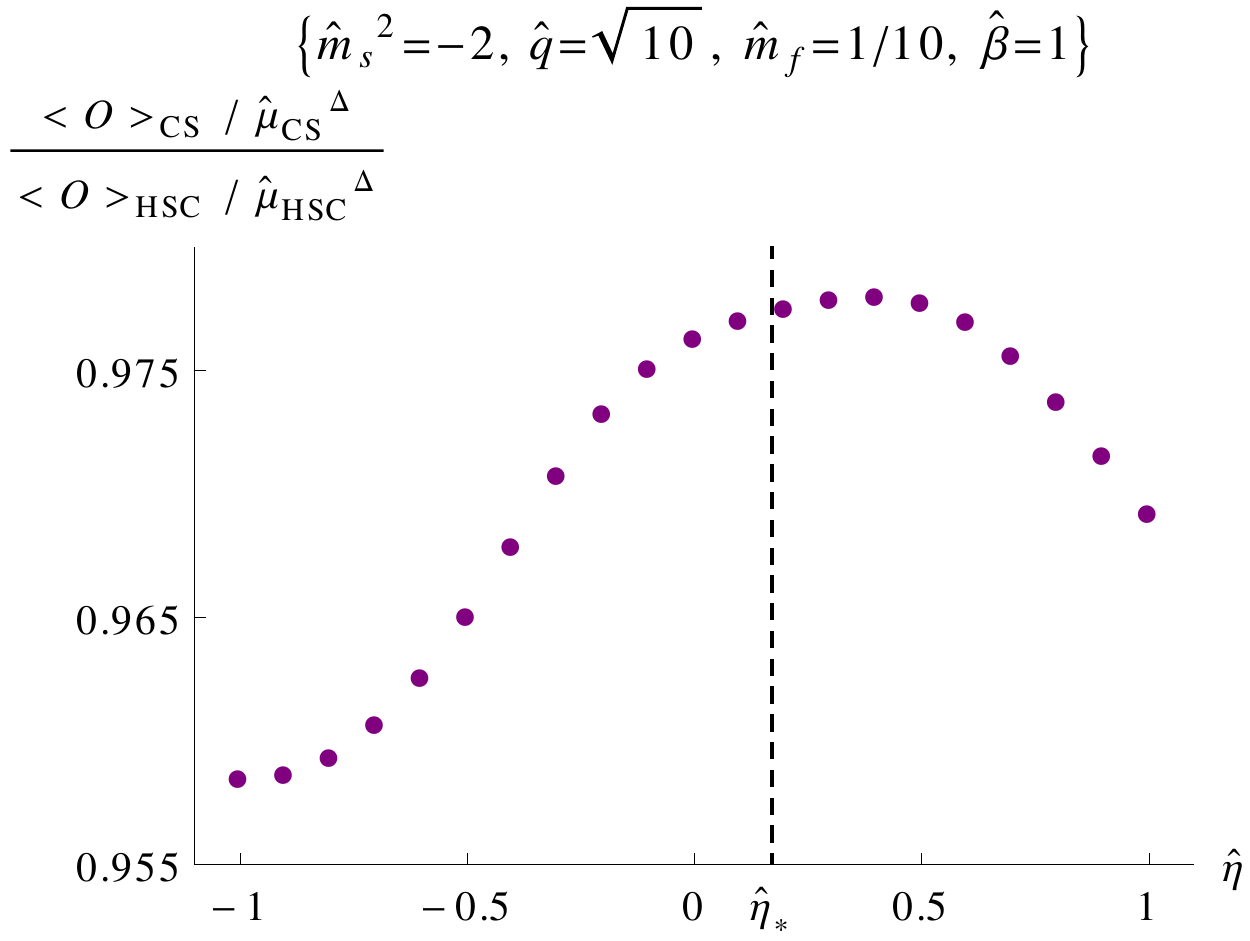}
}%\\
\caption{\protect Distribution of the electric charge components (a) and value of the
condensate (b) in the compact star(s) solutions corresponding to the choice of parameters in
Figure \protect\ref{fig:free-energy-1}.  
% in Figure 
To avoid cluttering of the figures,  the electric charge of the scalar field $\hat{Q}_\textnormal{scalar}$ is not displayed. . The dashed line at the critical  value 
 $\hat{\eta}_*\simeq0.17$  marks the transition between the eCS and peCS visible in
Figure~\protect\ref{fig:free-energy-1}.}
\label{fig:ratio-Q-and-vev-1}
\end{figure}
\begin{figure}[t!]
\subfloat[ ]{
\includegraphics[width=0.48\textwidth]{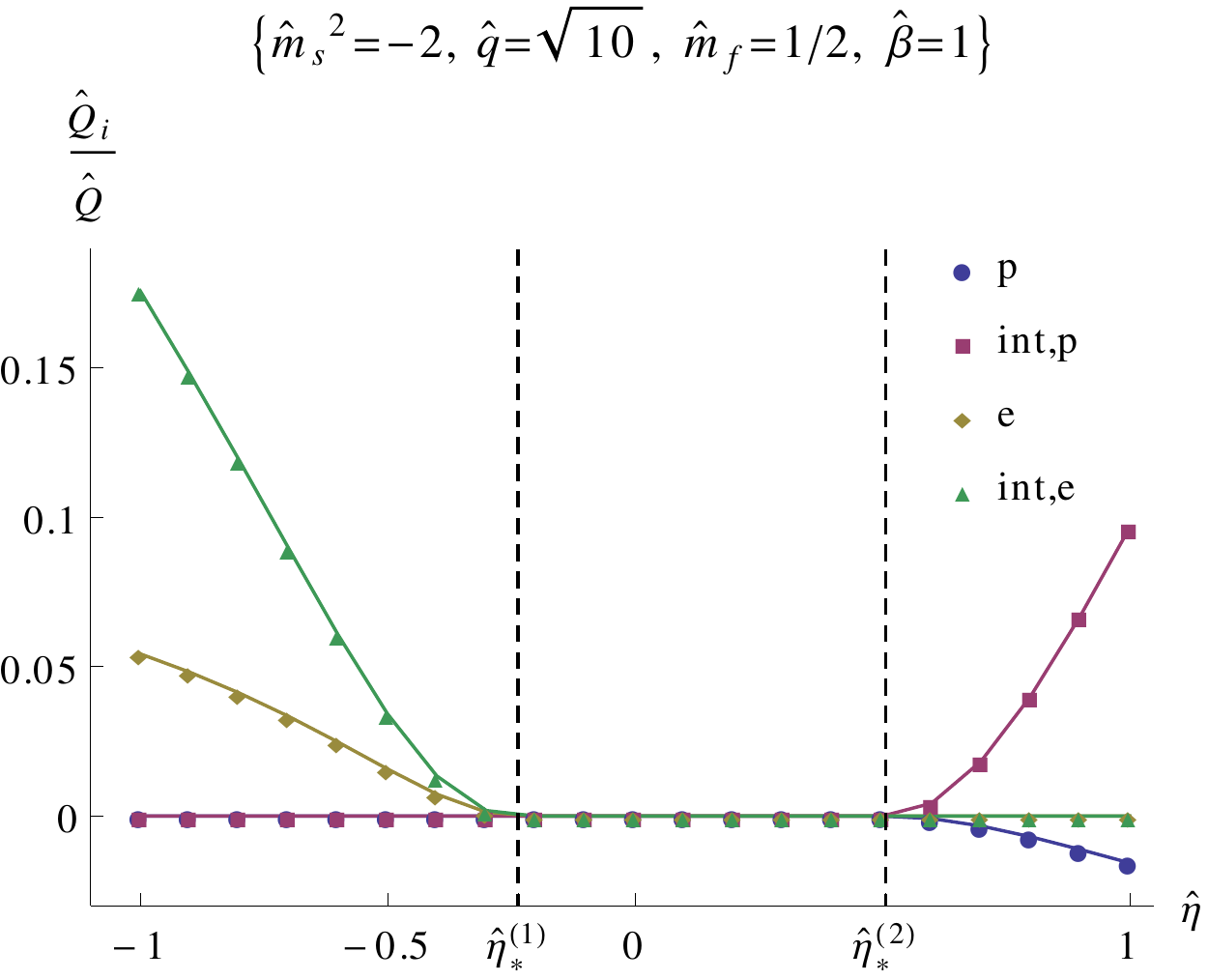}
}
\subfloat[ ]{
\includegraphics[width=0.48\textwidth]{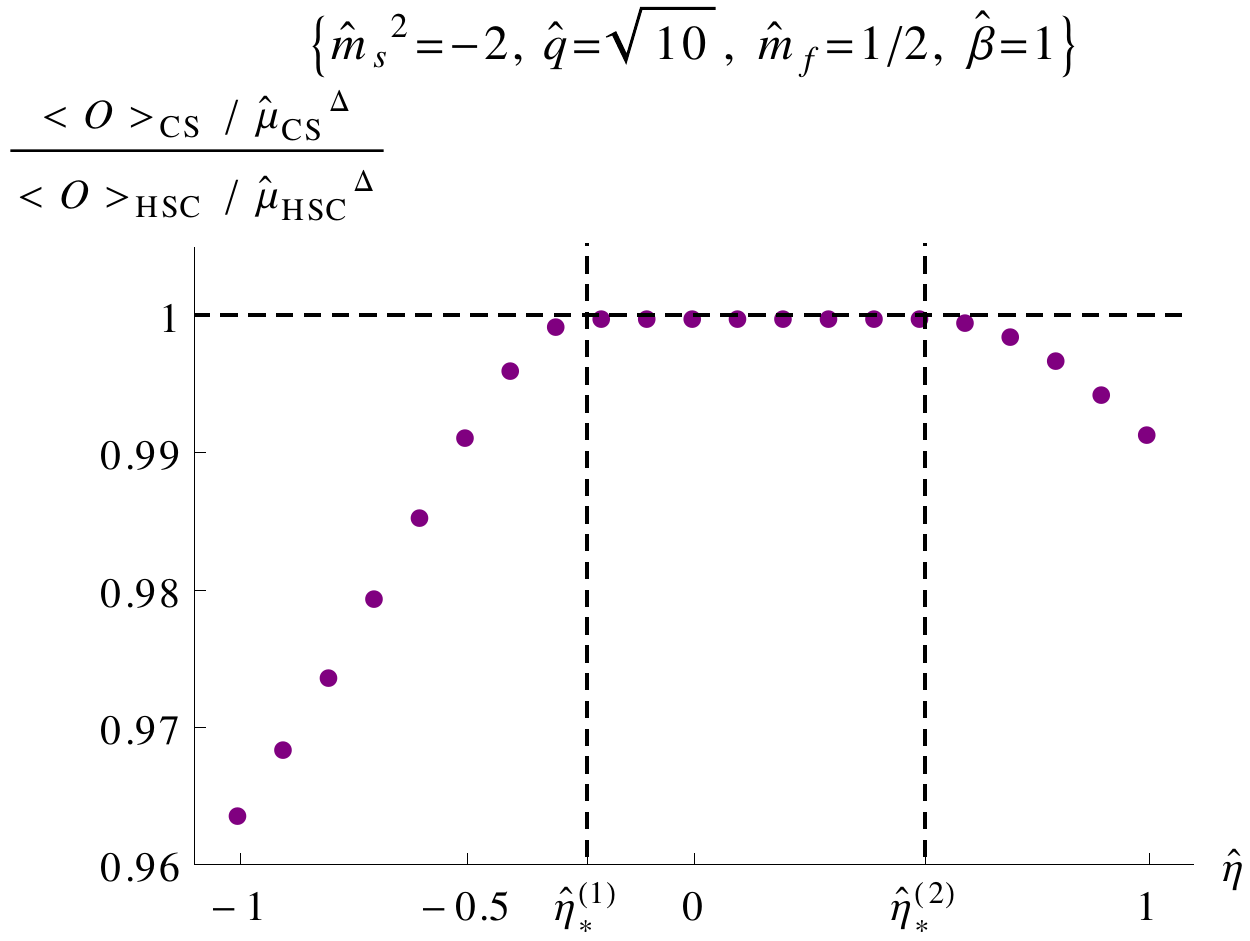}
}\caption{Distribution of the electric charge (a) and value of the condensate (b)  in the
compact star(s)
solutions corresponding to Figure \protect\ref{fig:free-energy-2}. 
% We do not display the electric charge of the scalar field $\hat{Q}_\textnormal{scalar}$ for
% visual aid.
% (a) and (b) display the transition between the eCS and peCS solutions, where
% $\hat{\eta}_*\simeq0.17$ is the critical coupling constant.
% Notice that for $\hat{\eta}>0$, $\hat{Q}_{\textnormal{int,e}}<0$ even if it not visible on
the
% plot
 vertical dashed lines indicate phase  the phase transitions visible in
Figure~\protect\ref{fig:free-energy-2}: the values  
$\hat{\eta}_*^{(1)}\simeq-0.24$ and $\hat{\eta}_*^{(2)}\simeq0.51$ are the critical coupling
constants marking the transition between the eCS and HSC solutions, and the HSC and pCS
solutions, respectively.}
\label{fig:ratio-Q-and-vev-2}
\end{figure}
\begin{figure}[t!]
\subfloat[ ]{
\includegraphics[width=0.48\textwidth]{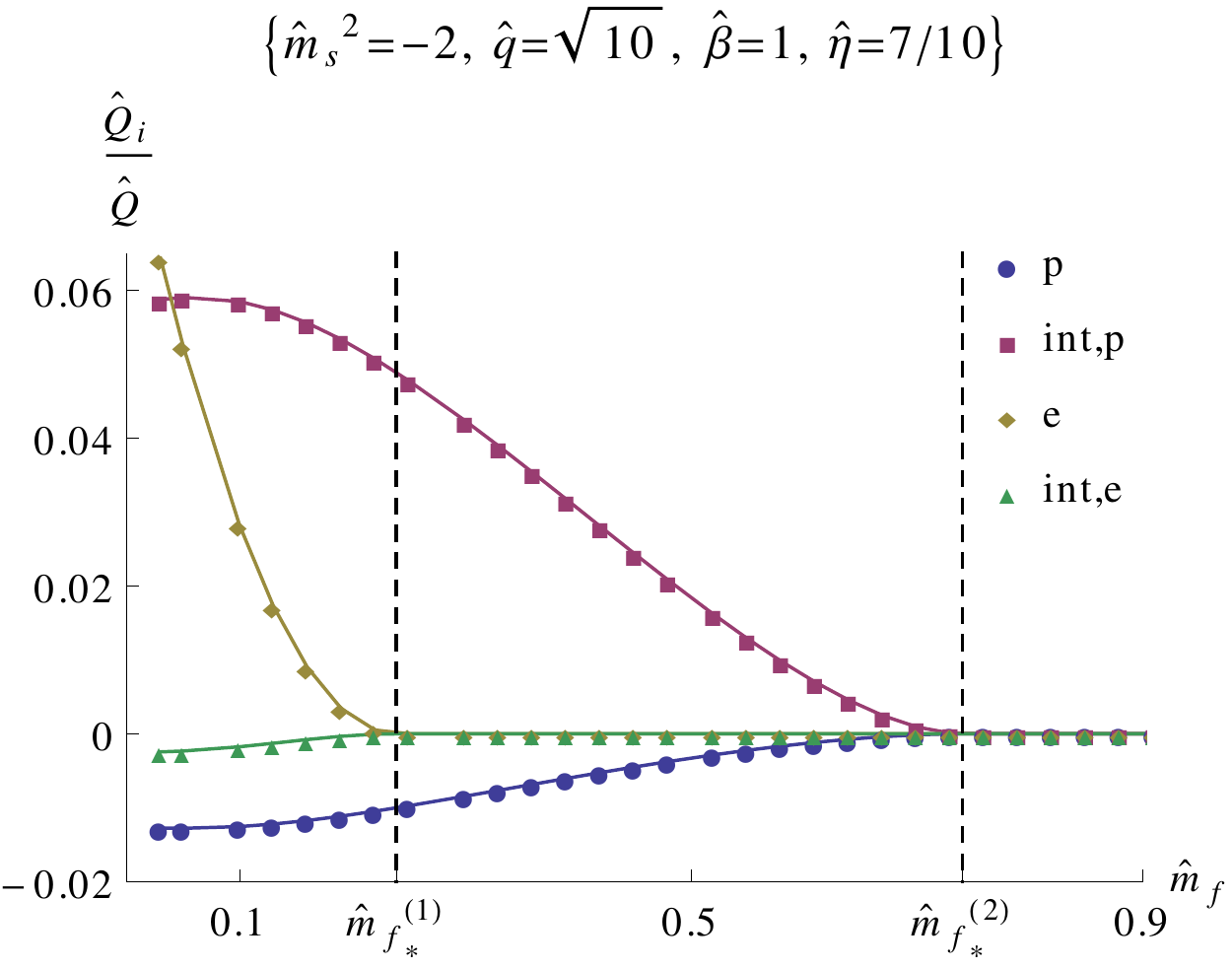}
}
\subfloat[ ]{
\includegraphics[width=0.48\textwidth]{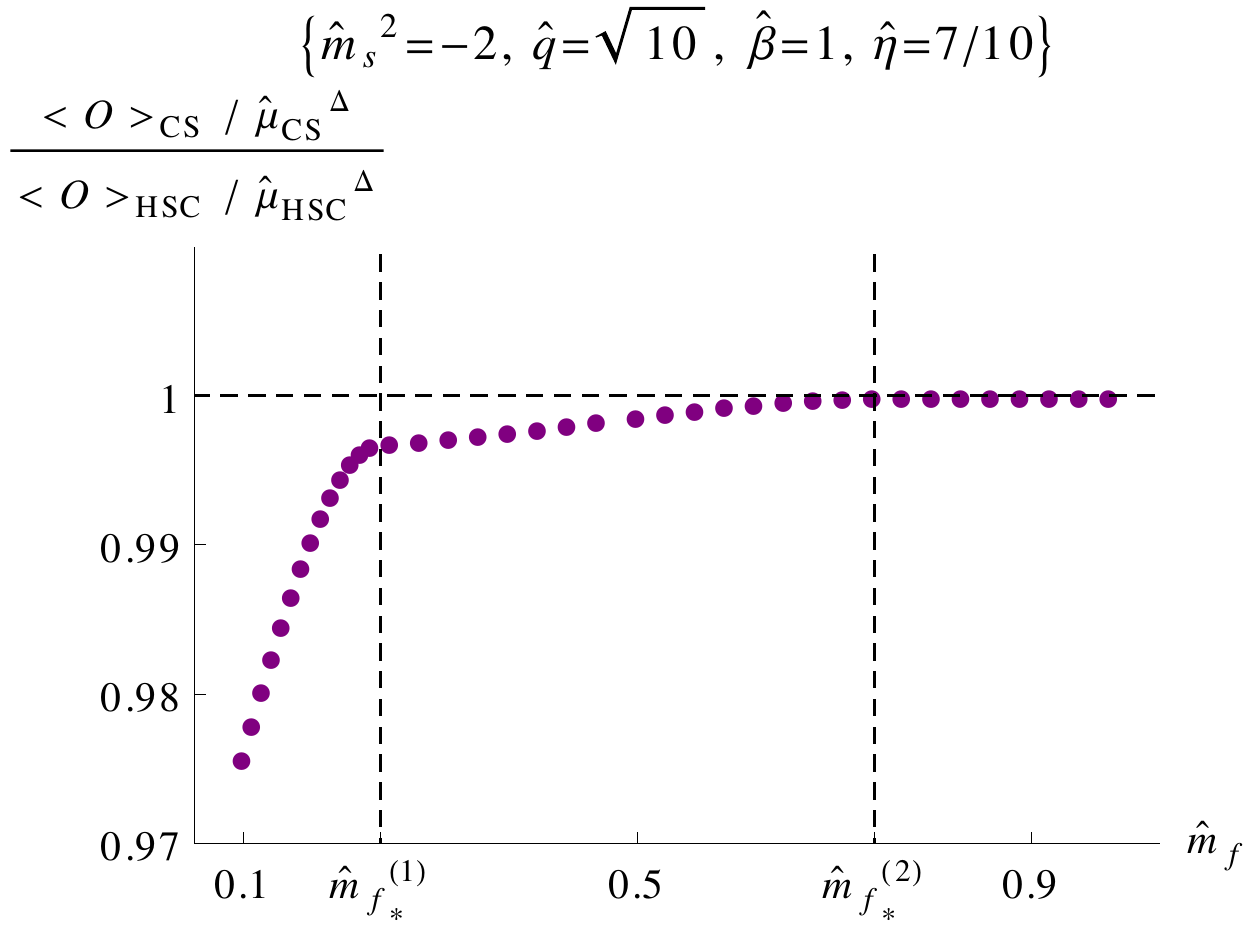}
}
\caption{Distribution of the electric charge (a) and value of the condensate (b)  in the
compact star(s)
solutions corresponding to Figure~\protect\ref{fig:free-energy-3}. 
% We do not display the electric charge of the scalar field $\hat{Q}_\textnormal{scalar}$ for
% visual aid.
% (a) and (b) display the transition between the eCS and peCS solutions, where
% $\hat{\eta}_*\simeq0.17$ is the critical coupling constant.
% Notice that for $\hat{\eta}>0$, $\hat{Q}_{\textnormal{int,e}}<0$ even if it not visible on
the
% plot.
The dashed lines at the critical values 
$\hat{m}_{f*}^{(1)}\simeq0.24$ and $\hat{m}_{f*}^{(2)}\simeq0.74$ mark the transitions between
the peCS and pCS solutions  and the pCS and HSC solutions, respectively. The same value
$\hat{m}_{f*}^{(1)}$ corresponds to the (subdominant) transition between the eCS and HSC
solutions}
\label{fig:ratio-Q-and-vev-3}
\end{figure}

In Figures~\ref{fig:ratio-Q-and-vev-1}-\ref{fig:ratio-Q-and-vev-3} (left)  we present the
distribution of the total electric charge of
the system between the scalar field, the fluids of electrons and positrons and the charges of
interaction for different  values of the parameters (the same that were used in
Figures~\ref{fig:free-energy-1}-\ref{fig:free-energy-3}). The boundary condensate
$\langle\mathcal{O}\rangle$ is also shown in those figures (right). 
It is interesting to note that it is lower in the CS solutions than in the HSC solution.
In Section~\ref{sec:spectrum} we will show that the presence of fermions in the bulk maps to the
formation of Fermi surfaces in the dual field theory.
If one interprets the scalar operator $\mathcal{O}$ as being a composite operator of the
fermionic operator,  decreasing of the condensate in the CS solutions can be thought of as
coming from the breaking of part of the scalar operator excitations. 

% {\bf can we break up into smaller pieces Figures 1 and 2 ? otherwise there will always  be blank spaces}

%
%

%%%%%%%%%%%%%%%%%%%%%%%%%%%%%%%%%%%%%%%%%%%%%%%%%%%%%%%%%%%%%%%%%%%%%%%%%%%%%%%%%%%%%%%%%%%%%%%
%%%%%%%%%%%%%%%%%%%%%%%%%%%%%%%%%%%%%%%%%%%%%%%%%%%%%%%%%%%%%%%%%%%%%%%%%%%%%%%%%%%%%%%%%%%%%%%
\section{Fermionic low energy spectrum}
\label{sec:spectrum}
%%%%%%%%%%%%%%%%%%%%%%%%%%%%%%%%%%%%%%%%%%%%%%%%%%%%%%%%%%%%%%%%%%%%%%%%%%%%%%%%%%%%%%%%%%%%%%%
%%%%%%%%%%%%%%%%%%%%%%%%%%%%%%%%%%%%%%%%%%%%%%%%%%%%%%%%%%%%%%%%%%%%%%%%%%%%%%%%%%%%%%%%%%%%%%%

In this section, we compute the Fermi surfaces and the fermionic low energy excitations of
our model.
This can be done by solving the equation of motion of a probe fermion in the WKB approximation. We assume the probe fermion is a constituent  of positive charge~$|q_f|$. 

%%%%%%%%%%%%%%%%%%%%%%%%%%%%%%%%%%%%%%%%%%%%%%%%%%%%%%%%%%%%%%%%%%%%%%%%%%%%%%%%%%%%%%%%%%%%%%%
%%%%%%%%%%%%%%%%%%%%%%%%%%%%%%%%%%%%%%%%%%%%%%%%%%%%%%%%%%%%%%%%%%%%%%%%%%%%%%%%%%%%%%%%%%%%%%%
\subsection{Probe fermion and the Dirac equation}
\label{sec:dirac-eq}
%%%%%%%%%%%%%%%%%%%%%%%%%%%%%%%%%%%%%%%%%%%%%%%%%%%%%%%%%%%%%%%%%%%%%%%%%%%%%%%%%%%%%%%%%%%%%%%
%%%%%%%%%%%%%%%%%%%%%%%%%%%%%%%%%%%%%%%%%%%%%%%%%%%%%%%%%%%%%%%%%%%%%%%%%%%%%%%%%%%%%%%%%%%%%%%

The electromagnetic current of bulk elementary fermions $\chi$ with charge $|q_f|$ is given
by
\begin{align}
\label{eq:current-chi}
 J_\textnormal{ferm}^a = - |q_f|\langle \bar{\chi}\Gamma^a\chi\rangle \, .
\end{align}
To take into account the current-current interaction between the fermions and the bosons, it
is natural to add to the action for free probe fermions $\chi$ the interaction
\begin{align}
 S_{\textnormal{int}} = \eta \int d^4x\, \sqrt{-g} J_a^\textnormal{ferm}J^a_\textnormal{scal}
\end{align}
where $J^a_\textnormal{ferm}$ and $J^a_\textnormal{scal}$ are given
by~(\ref{eq:currents-probe}).

\begin{figure}[h!]
\centering
\subfloat[ ]{
\includegraphics[width=0.48\textwidth]{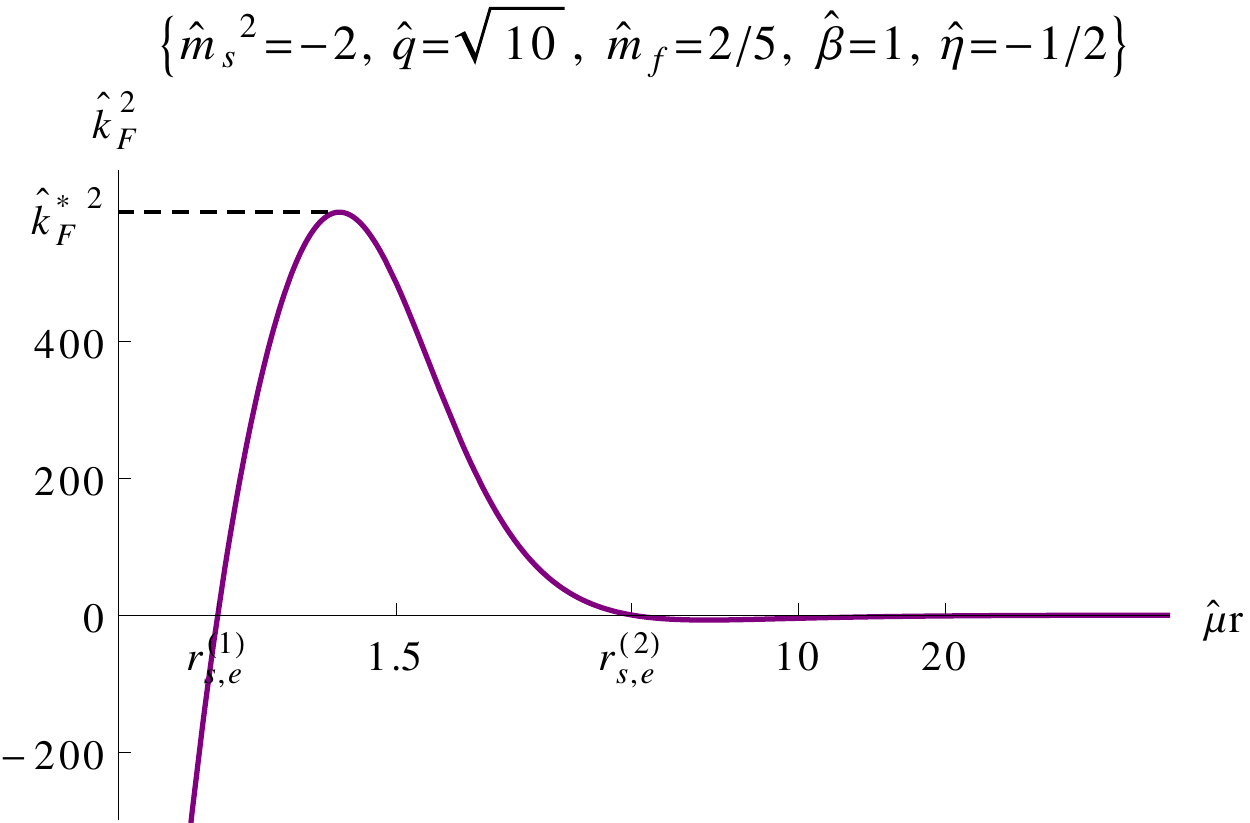}
}
\subfloat[ ]{
\includegraphics[width=0.48\textwidth]{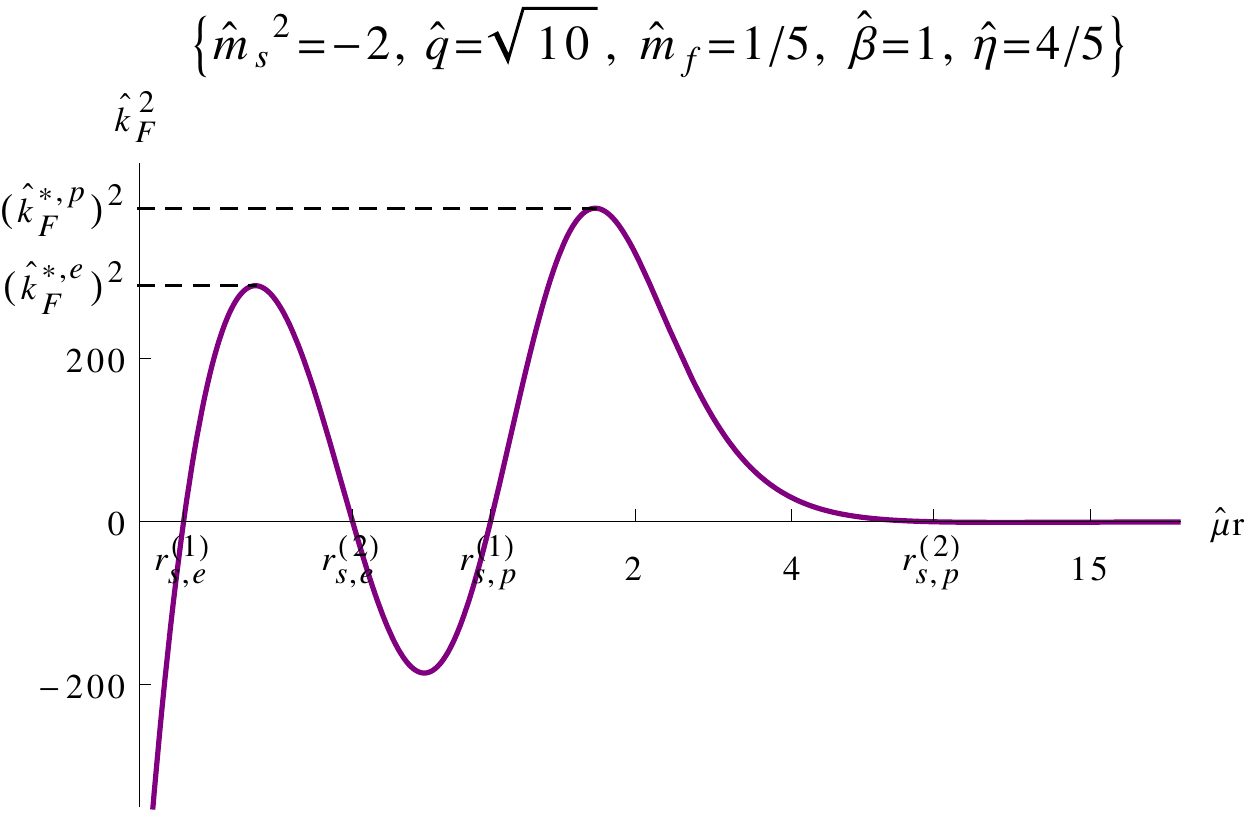}
}
\caption{Local Fermi momentum squared $\hat{k}_F^2(r)$ for (a) an eCS solution and (b) a peCS
solution. It is positive inside the star(s) and negative outside. The points $r_{s,e}^{(1)}$,
$r_{s,e}^{(2)}$, $r_{s,p}^{(1)}$ and $r_{s,p}^{(2)}$ are the boundaries of the electron and the
positron stars.}
\label{fig:kF2}
\end{figure}

In Appendix~\ref{app:dirac-eq}, we obtain in details the Schr\"odinger-like equation,
but we give here the key steps.
By choosing correctly the basis of Gamma matrices, the Dirac equation for a probe spinor
field $\chi$ on top of the background solution can be written as an equation for the
two-component spinor $\Phi=r^{-1}f^{-1/4}\chi_1$,
\begin{align}
\label{eq:Dirac-eq}
\left(\partial_r + \gamma\hat{m}_f g^{1/2}\sigma^3\right)\Phi =
g^{1/2}\left\{i\gamma\sigma^2 \left[\hat{\omega}f^{-1/2}+\hat{\mu}_l
\right] - \gamma\hat{k}r\sigma^1\right\} \Phi
\end{align}
where we have rescaled the momentum and frequency,
\begin{equation}
\hat{\omega} = \frac{\omega}{\gamma} \, , \qquad \hat{k} = \frac{k}{\gamma}
\end{equation}
by the parameter
\begin{equation}
\label{eq:def-gamma}
\gamma \equiv \frac{|q_f| eL}{\kappa} \gg 1
\end{equation}
which is large in the Thomas-Fermi approximation applied to the bulk fermions.
In this limit, the Dirac equation~(\ref{eq:Dirac-eq}) is equivalent to the Schr\"odinger-like
equation
\begin{align}
\label{eq:wkb-eq}
\Phi_2'' &= \gamma^2 V(r) \Phi_2
\end{align}
together with the expression
\begin{align}
\Phi_1 &= \frac{1}{\frac{\hat{\omega}}{\sqrt{f}}+\hat{\mu}_l+\hat{k}r}
\left(\hat{m}_f\Phi_2-\frac{1}{\gamma}\frac{1}{\sqrt{g}}\Phi_2'\right)
\end{align}
for the components $\Phi_1$ and $\Phi_2$ of the spinor
\begin{align}
 \Phi =
\left( \begin{array}{c}
\Phi_1 \\
\Phi_2
\end{array} \right) \, .
\end{align}
The potential $V(r)$ can be expressed as
\begin{align}
\label{eq:potential}
 V(r) = g(r)\left\{r^2\left[\hat{k}^2-\hat{k}_F^2(r)\right] -
\frac{\hat{\omega}}{f(r)}\left[\hat{\omega}+2\sqrt{f(r)}\hat{\mu}_l(r)\right]\right\}
+\mathcal{O}(\gamma^{-1})
\end{align}
where the local Fermi momentum~$\hat{k}_F$ is defined as~\cite{Hartnoll:2011dm}
\begin{align}
\label{eq:local-Fermi-mom}
 \hat{k}_F^2(r) \equiv \frac{1}{r^2}\left(\hat{\mu}_l^2-\hat{m}_f^2\right) \, .
\end{align}
Notice that $\hat{k}_F^2>0$ inside the stars only; these are the regions where $\hat{k}_F$ is
relevant for our considerations.
The local Fermi momentum is displayed in Figure~\ref{fig:kF2} for an eCS solution and a peCS
solution.
The momentum $\hat{k}$ appears only through $\hat{k}^2$ in the potential~(\ref{eq:potential}),
so we can restrict the analysis to $\hat{k}>0$ without loss of generality.

%%%%%%%%%%%%%%%%%%%%%%%%%%%%%%%%%%%%%%%%%%%%%%%%%%%%%%%%%%%%%%%%%%%%%%%%%%%%%%%%%%%%%%%%%%%%%%%
\subsubsection{The Schr\"odinger equation in a standard form}
\label{sec:Schro-true-form}
%%%%%%%%%%%%%%%%%%%%%%%%%%%%%%%%%%%%%%%%%%%%%%%%%%%%%%%%%%%%%%%%%%%%%%%%%%%%%%%%%%%%%%%%%%%%%%%

The potential~(\ref{eq:potential}) depends on the momentum $\hat{k}$.
In order to see the physical interpretation of Eq.~(\ref{eq:wkb-eq}), we put it in a
Schr\"odinger form where $\hat{k}$ plays the role of the energy by introducing the new
coordinate $y$, defined by
\begin{align}
 \frac{\D y}{\D r} = r\sqrt{g(r)} \, .
\end{align}
We then obtain the equation
\begin{align}
\label{eq:true-schro-eq}
 -\partial_y^2{\varphi} + \gamma^2 \tilde{V}(y) \varphi = -\gamma^2 \hat{k}^2 \varphi
\end{align}
for the rescaled field $\varphi \equiv r^{1/2}g^{1/4}\Phi_2$, where
\begin{align}
\label{eq:true-potential}
\tilde{V}(y) = - \hat{k}_F^2(y) -
\frac{\hat{\omega}}{r(y)^2f(y)}\left(\hat{\omega}+2\sqrt{f(y)}\hat{\mu}_l(y)\right) +
\mathcal{O}\left(\gamma^{-2}\right)
\end{align}
in the large-$\gamma$ limit.
Herein and in the following, $r(y)$ is the inverse map of
\begin{align}
 y(r) = \int^r\D r' r'\sqrt{g(r')} \, .
\end{align}
Notice that $y\to\infty$ when $r\to\infty$.
The equation~(\ref{eq:true-schro-eq}) has to be seen as a Schr\"odinger equation with
negative eigenvalue of
the energy $E=-\gamma^2\hat{k}^2$.
The potential~(\ref{eq:true-potential}) is now independent of the momentum $\hat{k}$.

\begin{figure}[ht!]
\centering
\subfloat[ ]{
\label{fig:true-potential-omega-zero}
\includegraphics[width=0.48\textwidth]{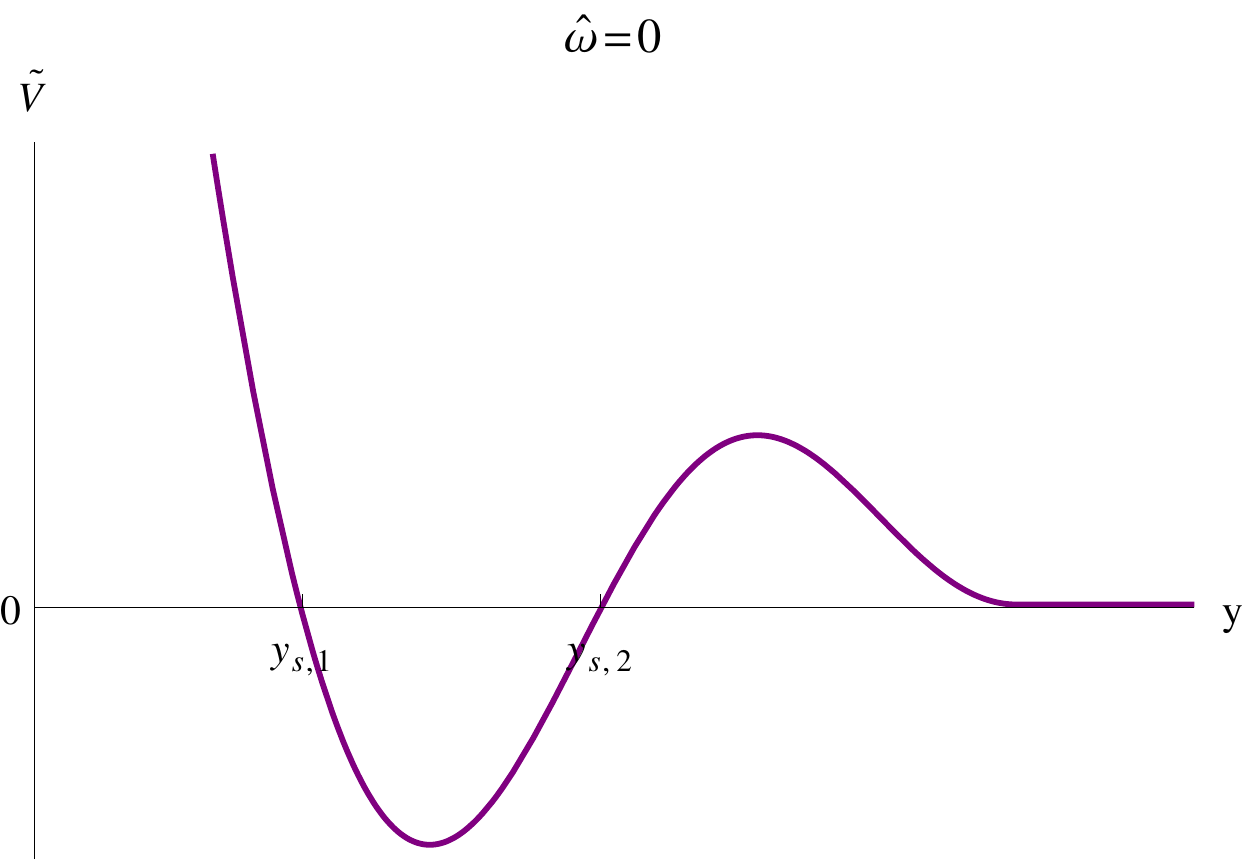}
}
\subfloat[ ]{
\label{fig:true-potential-omega-small}
\includegraphics[width=0.48\textwidth]{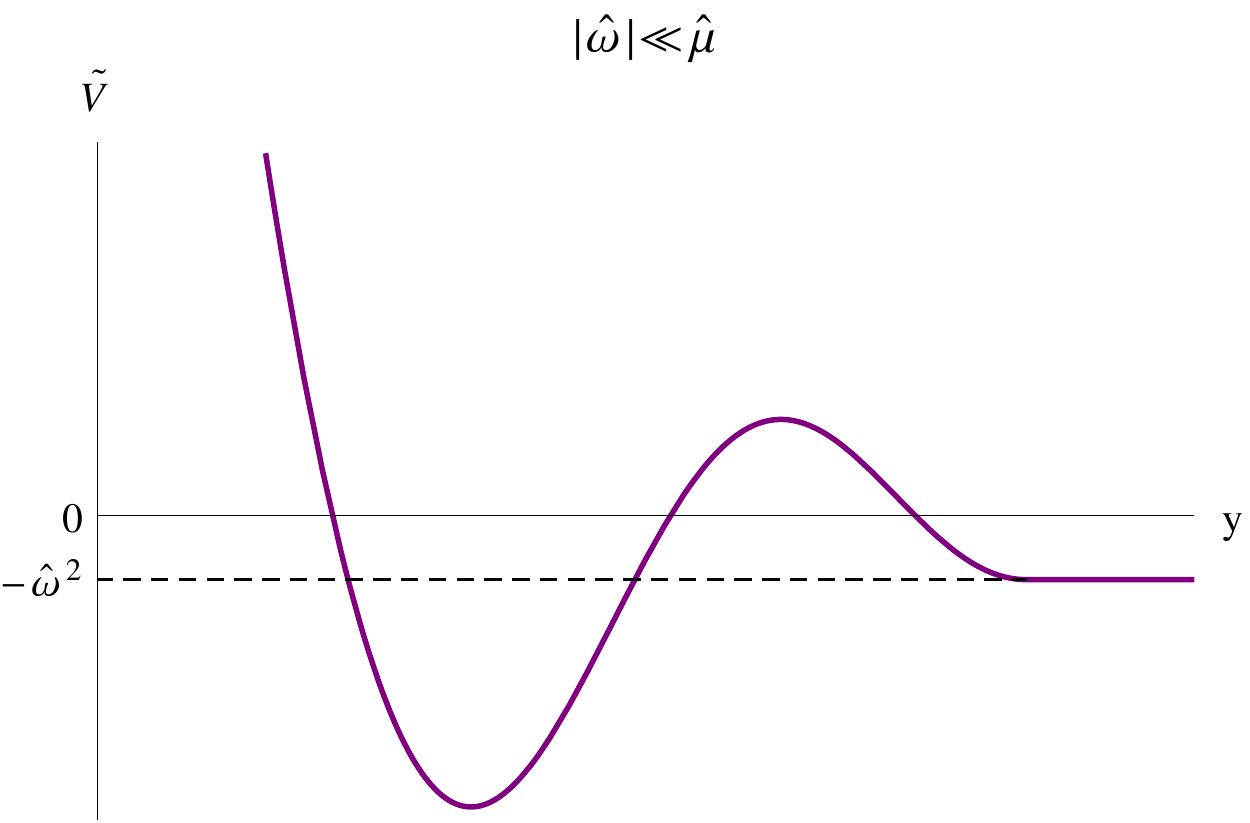}}
\caption{Profile of the potential of the Schr\"odinger equation for an eCS or a pCS
solution at (a) zero frequency, (b) small frequency $0<|\hat{\omega}|\ll\hat{\mu}$.
The points $y_{s,1}$ and $y_{s,2}$ are the star boundaries.}
\label{fig:true-potential}
\end{figure}

At zero frequency, the potential $\tilde{V}$ is, up to a minus sign, given by the local
Fermi momentum squared~(\ref{eq:local-Fermi-mom}).
It is negative inside the star and positive outside, and the zero-energy turning points are the
star boundaries where $|\hat{\mu}_l(y)|=\hat{m}_f$.
Since the local chemical potential $\hat{\mu}_l$ vanishes in the IR as
in~(\ref{eq:local-mu-IR}), the local Fermi momentum behaves in this region as
$\hat{k}_F^2(y)\sim-\hat{m}_f^2/r(y)^2$ where in the IR, the inverse map of $r(y)$ is
\begin{align}
y(r) \sim \sqrt{\frac{3}{2|\hat{m}_s^2|}}\frac{r}{\sqrt{\log r}} \, , \qquad r\to \infty \, ,
\end{align}
and the
potential $\tilde{V}(y) \to 0^+$  at infinity.
The potential $\tilde{V}$ for $\hat{\omega}=0$ is displayed schematically in
Figure~\ref{fig:true-potential-omega-zero} for a compact star solution involving one star (eCS
or pCS).

A non-zero frequency $0<|\hat{\omega}|\ll\hat{\mu}$ affects the behavior of the
potential~(\ref{eq:true-potential}) in the near-horizon region.
Indeed,
\begin{align}
\tilde{V}(y)\sim-\hat{\omega}^2+\frac{\hat{m}_f^2}{r(y)^{2}} \, , \qquad r\to\infty
\, ,
\end{align}
so the zero-frequency limit and the near-horizon limit do
not commute.
This is also observed in the electron star phase~\cite{Hartnoll:2011dm} but with a different
asymptotic behavior with respect to the potential~(\ref{eq:true-potential}), leading to a
different
conclusion as we shall see below.
Outside the near-horizon region, the effects of small frequency are small and do not affect
much the shape of the potential.
In particular, the zero-energy turning points of the potential $\tilde{V}$ are close to the
star
boundaries.
In the UV, the potential~(\ref{eq:true-potential}) behaves as
\begin{align}
\label{eq:true-potential-IR}
\tilde{V}(y)\sim\frac{\hat{m}_f^2}{y^2} \, , \qquad y\to0 \, ,
\end{align}
and effects of $\hat{\omega}$ are subleading.
The potential for $0<|\hat{\omega}|\ll\hat{\mu}$ is displayed in
Figure~\ref{fig:true-potential-omega-small} for a compact star solution involving one star (eCS
or pCS).

Our aim is to solve the Schr\"odinger equation~(\ref{eq:true-schro-eq}) for the
field $\varphi$ to compute the poles of the retarded two-point Green function of the
gauge-invariant field operator dual to the bulk fermionic particles.
To do so, we shall impose the Dirichlet condition on $\varphi$ at the UV boundary and the
in-falling condition in the near-horizon region.
This second condition can be applied when the solution to Eq.~(\ref{eq:true-schro-eq}) is
oscillating in the IR, that is when $\tilde{V}(y=\infty)=-\hat{\omega}^2<-\hat{k}^2$, i.e. for
$|\hat{\omega}|>\hat{k}$.
The dispersion relation of the Green function corresponds in this case to quasinormal modes of
the wave equation~(\ref{eq:true-schro-eq}).
For $|\hat{\omega}|<\hat{k}$, the solution is exponential in the near-horizon region and one
shall impose the regularity condition, leading to normal modes of the
equation~(\ref{eq:true-schro-eq}).

The discussion of the previous paragraph allows us to distinguish three different regimes for
the spectrum of the Schr\"odinger equation~(\ref{eq:true-schro-eq}). Let us define the extremal local Fermi momentum $\hat{k}^{\star}_F$ by
\begin{align}
\hat{k}^{\star}_F \equiv \max_{y_{s,1}<y<y_{s,2}} \hat{k}_F(y)
\end{align}
For $\hat{k}>\hat{k}^{\star}_F$, the ``energy'' $-k^2$ lies everywhere below the Schr\"odinger potential, and   there are no eigenstates.
In the intermediate region $|\hat{\omega}|<\hat{k}<\hat{k}_F^\star$, we expect to have a
discrete spectrum of bound states, which are normal modes of the Schr\"odinger
equation~(\ref{eq:true-schro-eq}).
Since the region of spacetime where $\tilde{V}<E$ is compact for any frequency, the number of
bound states is finite at fixed frequency and is almost independent of the frequency since
$\hat{\omega}\ll\hat{\mu}$ only affects the near-horizon region.
For $\hat{k}<|\hat{\omega}|\ll\hat{\mu}$, the spectrum of the Schr\"odinger
equation~(\ref{eq:true-schro-eq}) is continuous but there is a discrete set of quasinormal
modes, which dissipate in the IR region by quantum tunnelling.
The number of quasinormal modes is finite at fixed frequency for the same reasons as for the
intermediate region.
By setting $\hat{\omega}=0$, one can count the number of Fermi surfaces.
From the qualitative behavior of the Schr\"odinger potential in  Figure~\ref{fig:true-potential-omega-zero},  the number of boundary
Fermi momenta is finite.

% In fact, there exists a critical extremal local Fermi momentum $\hat{k}_F^c$ such that if
% $\hat{k}_F^\star<\hat{k}_F^c$, the number of boundary Fermi momenta is zero.
In certain parameter regions,  although the fluid density is non-zero,  there may be no negative energy bound states (thus no Fermi surfaces): 
this happens for \lq small stars', for which the potential $\tilde{V}$ is not deep enough inside the star to allow any bound state.

We then expect to obtain a finite number of boundary Fermi momenta $\hat{k}_n$
($n=0,\dots,N-1$) satisfying $\hat{k}_n<\hat{k}_F^\star$ in the large-$\gamma$ limit.
Each boundary Fermi surface admits particle excitations which are stable at small energy
$|\hat{\omega}|<\hat{k}$.
At larger frequency $|\hat{\omega}|>\hat{k}$ with $|\hat{\omega}|\ll\hat{\mu}$, the excitations
are resonances, they can dissipate.
This dissipation maps in the bulk to the possible quantum tunnelling of the modes into the
near-horizon region.
The modes dissipate only at sufficiently large frequency due to the fact that the compact star
does not occupy the inner region of the bulk spacetime.
From the field theory point of view, the bosonic modes, represented in the bulk by the scalar
field, the metric and the gauge field, do not interact with the fermions at sufficiently low
energy and the excitations around the $n$-th Fermi surface are stable up to
$|\hat{\omega}|\sim\hat{k}_n$.

\begin{figure}[t!]
\centering
\subfloat[ ]{
\label{fig:true-potential-omega-zero-ES}
\includegraphics[width=0.48\textwidth]{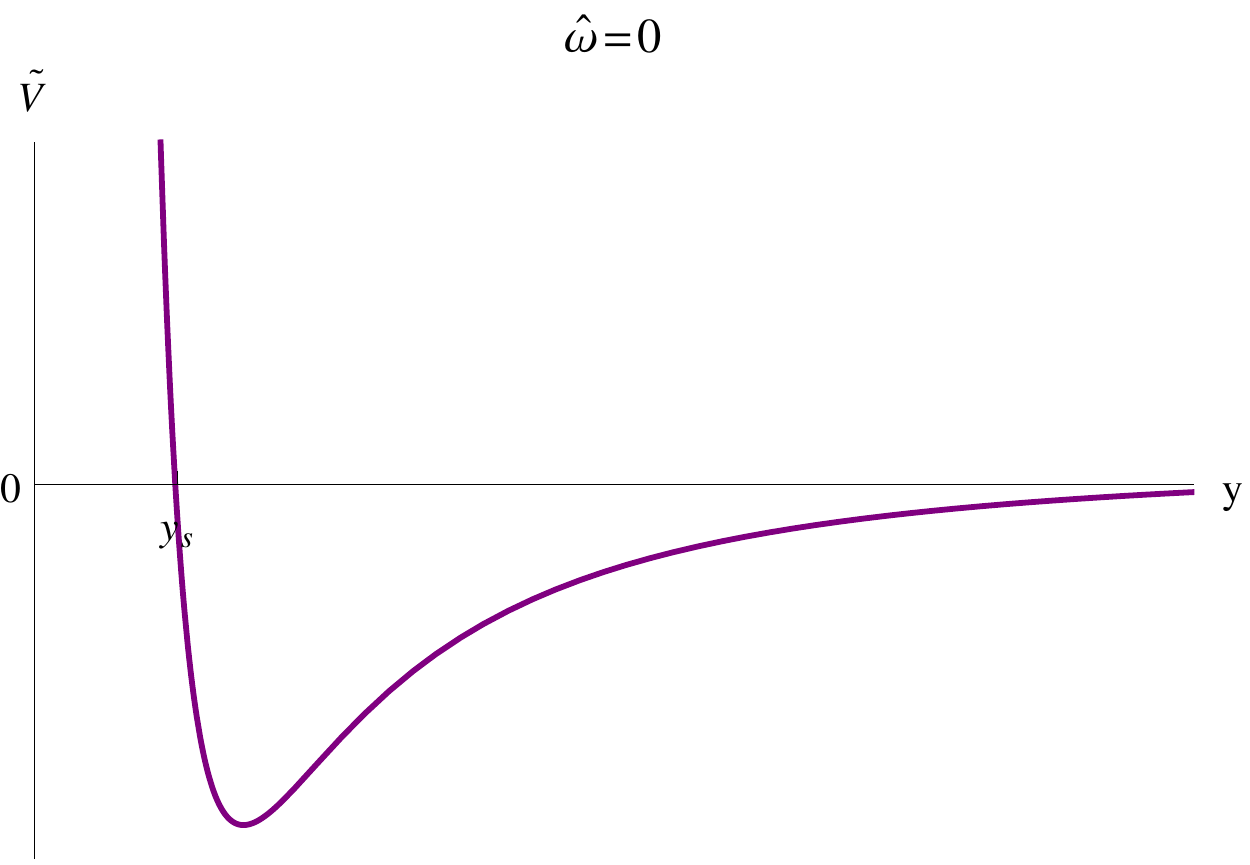}
}
\\
\subfloat[ ]{
\label{fig:true-potential-omega-negative-ES}
\includegraphics[width=0.48\textwidth]{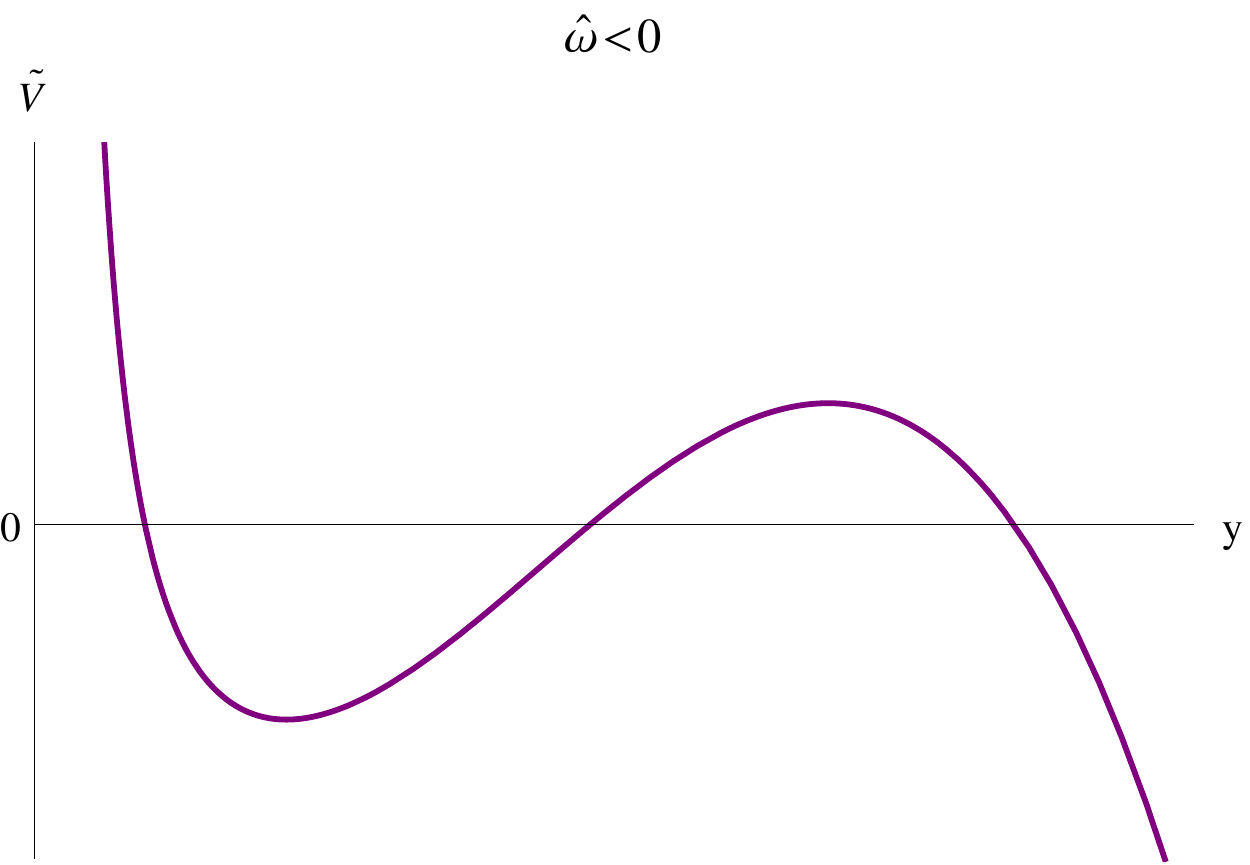}}
\subfloat[ ]{
\label{fig:true-potential-omega-positive-ES}
\includegraphics[width=0.48\textwidth]{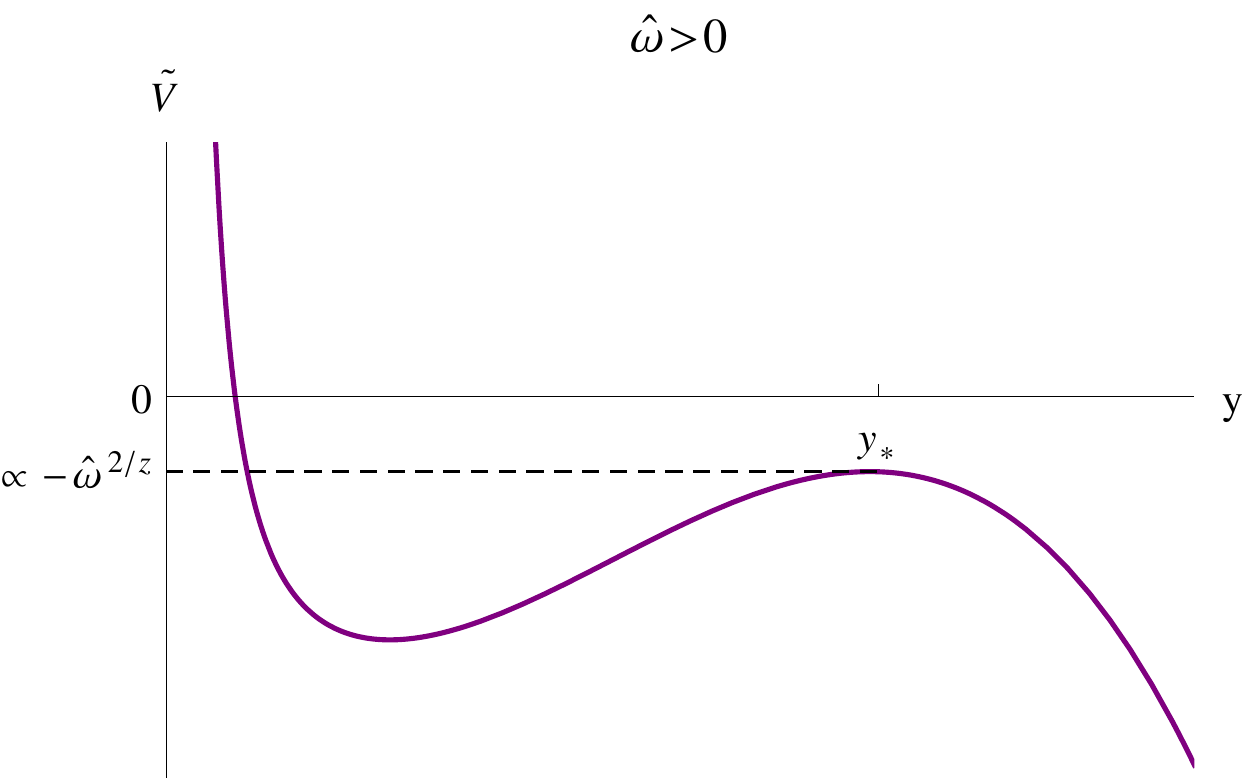}}
\caption{Profile of the potential of the Schr\"odinger equation in the electron star
phase at, (a) zero frequency, (b) negative frequency and (c) positive frequency.
The point $y_s$ is the star boundary and $y_\star$ is the point where $\tilde{V}$ admits a
local maximum in the Lifshitz region for $\hat{\omega}<0$.}
\label{fig:true-potential-ES}
\end{figure}

Let us compare the situation to the electron star phase, studied in detail
in~\cite{Hartnoll:2011dm}.
At zero frequency the potential is negative for $y>y_s$, where $y_s$ is the star boundary, and
$\tilde{V}(y)\to0^{-}$ for $y\to\infty$.
For $\hat{\omega}\neq0$, the potential diverges to $-\infty$ in the IR  so dissipation occurs
for any non-zero frequency.
This is because the fluid does occupy the inner region of spacetime and the boundary fermions
interact with the bosonic modes at low energy.
For $\hat{\omega}>0$, the potential admits a local maximum at a point\footnote{The
corresponding point in the original $r$ variable has been computed by Hartnoll et al.
in~\cite{Hartnoll:2011dm}.} $y_\star\propto\hat{\omega}^{-1/z}$
%%%%%
\begin{comment}
\begin{align}
 y_\star = \left[\frac{g_\infty^{z/2}}{2(z-1)\hat{\omega}}\left(h_\infty(2-z) +
\sqrt{h_\infty^2 z^2-4\hat{m}_f^2 (z-1)}\right) \right]^{1/z}
\end{align}
\end{comment}
%%%%%
where $\tilde{V}(y_\star)\propto-\hat{\omega}^{2/z}<0$.
The constant of proportionality can be easily computed analytically.
From the Schr\"odinger equation~(\ref{eq:true-schro-eq}), it means that for positive
frequency,  the modes are unstable for $\hat{\omega}>\hat{k}^z$ because, as explained
in~\cite{Hartnoll:2011dm}, the fermionic excitations strongly interact with the critical modes
of the Lifshitz sector.
On the other hand, for $\hat{\omega}<0$, the potential admits two turning points in the
Lifshitz region, so there is a region where it is positive.
The conclusion is that there is no strong dissipation and one finds a set of quasinormal modes
for all $\hat{k}\lesssim\hat{k}_F^\star$~\cite{Hartnoll:2011dm}.
These situations are displayed in Figure~\ref{fig:true-potential-ES}.

We should notice that for the compact star solutions, $\tilde{V}$ also admits a local maximum,
as shown in Figure~\ref{fig:true-potential}.
However, this local maximum is positive at zero frequency and not modified at small frequency
$|\hat{\omega}|\ll\hat{\mu}$ because it does not belong to the asymptotic IR region.
At larger frequencies, the effects of $\hat{\omega}$ are also relevant outside the
near-horizon region.
It may happen that for sufficiently large frequencies, this local maximum becomes negative.
In this case, the potential would be negative for all $y$ larger than the first turning point,
leading to unstable states at small momentum.

We have argued above that the number of boundary Fermi momenta dual to the compact star
solutions is finite.
In the electron star phase on the other hand, at zero frequency we see
from~(\ref{eq:sol-ES-Lif}) that the potential asymptotes to zero in the IR as
\begin{align}
\label{eq:true-potential-ES-IR}
\tilde{V}(y) \sim -\frac{g_\infty\left(h_\infty^2-\hat{m}_f^2\right)}{y^2} \, , \qquad
y\to\infty \, .
\end{align}
It means that there is accumulation of levels at small momentum and, as we will explicitly
show in Section\ref{sec:FS-PT}, the number of Fermi surfaces turns out to be
infinite~\cite{Landau1977quantum}.

Here, we focus on the regime $|\hat{\omega}|<\hat{k}$ with
$|\hat{\omega}|\ll\hat{\mu}$.
In this case, all the modes are bound states.
We leave the detailed study of the case $|\hat{\omega}|>\hat{k}$ for future
considerations~\cite{Nitti:wip}.

%%%%%%%%%%%%%%%%%%%%%%%%%%%%%%%%%%%%%%%%%%%%%%%%%%%%%%%%%%%%%%%%%%%%%%%%%%%%%%%%%%%%%%%%%%%%%%%
\subsubsection{Bound states for $|\hat{\omega}|\ll\hat{k}$}
%%%%%%%%%%%%%%%%%%%%%%%%%%%%%%%%%%%%%%%%%%%%%%%%%%%%%%%%%%%%%%%%%%%%%%%%%%%%%%%%%%%%%%%%%%%%%%%

In the electron star phase, the dispersion relation of the fermionic excitations around the
Fermi surfaces was found to be linear, up to a (imaginary) dissipative
term~\cite{Hartnoll:2011dm}.
We will see later that this is also what we obtain for the compact star phases.
In the rest of this paper, we will be interested in the behavior of the particle excitations
close to the Fermi surfaces, that is for
$|\hat{\omega}|\sim|\hat{k}_n-\hat{k}|\ll\hat{k}_n$.
In this case, $|\hat{\omega}|\ll\hat{k}$ and the dependence on $\hat{\omega}$ of
the potential~(\ref{eq:true-potential}) can be neglected everywhere, in particular in the IR
region, and one can consistently set $\hat{\omega}=0$.
The potential is then simply given by
\begin{align}
 \tilde{V}(y) = -\hat{k}_F^2(y) \, .
\end{align}
It is easy to generalize the above discussion to peCS solutions.
In Figure~\ref{fig:kF2}, we display the local Fermi momentum for an eCS solution and a peCS
solution.
For convenience, we plot it in the original variable $r$.
Doing so does not affect the analysis as it only changes the shape of the local Fermi momentum
in the radial direction and does not modify the extremal values of it.
It is clear from~(\ref{eq:true-schro-eq}) that for $\hat{k}>\hat{k}_F(r)$, the solution is
exponential while it is oscillating for $\hat{k}<\hat{k}_F(r)$.
There can exist zero, one or two regions where $\hat{k}<\hat{k}_F(r)$ depending on the
background solution and the value of $\hat{k}$ compared to the local maxima of $\hat{k}_F$.
For eCS and pCS solutions, the extremal local Fermi momentum is
\begin{align}
\label{eq:extremal-momenta-eCS-pCS}
\hat{k}^{\star}_F \equiv \max_{r_s^{(1)}<r<r_s^{(2)}} \hat{k}_F(r)
\end{align}
and for peCS solutions, there exist two local extrema
\begin{align}
\label{eq:extremal-momenta-peCS}
\hat{k}_F^{\star,p} \equiv \max_{r_{s,p}^{(1)}<r<r_{s,p}^{(2)}}\hat{k}_F(r) \qquad
\textnormal{and}
\qquad \hat{k}_F^{\star,e} \equiv \max_{r_{s,e}^{(1)}<r<r_{s,e}^{(2)}}\hat{k}_F(r) \, .
\end{align}
The bounds in the maxima are the star boundaries of the star(s) of the different solutions.
The conditions for oscillations are detailed in Table~\ref{tab:conditions-oscillations}.
\renewcommand{\arraystretch}{1.08}% Wider
\begin{table}[ht!]
\centering
\begin{tabular}{ c | c | c | }
\cline{2-3}
& eCS/pCS & peCS \\
\hline
\multicolumn{1}{|c|}{$\hat{k}<\hat{k}_F^\star$} & oscillations & $\times$ \\
\hline
\multicolumn{1}{|c|}{$\hat{k}_F^\star<\hat{k}$} & no oscillations & $\times$ \\
\hline
\multicolumn{1}{|c|}{$\hat{k}<\hat{k}_F^{\star,e}$ and $\hat{k}<\hat{k}_F^{\star,p}$} &
$\times$ &
oscillations in electron and positron stars \\
\hline
\multicolumn{1}{|c|}{$\hat{k}_F^{\star,e}<\hat{k}<\hat{k}_F^{\star,p}$} & $\times$ &
oscillations in positron star \\
\hline
\multicolumn{1}{|c|}{$\hat{k}_F^{\star,p}<\hat{k}<\hat{k}_F^{\star,e}$} & $\times$ &
oscillations in electron star \\
\hline
\multicolumn{1}{|c|}{$\hat{k}_F^{\star,p}<\hat{k}$ and $\hat{k}_F^{\star,e}<\hat{k}$} &
$\times$ & no oscillations \\
\hline
 \end{tabular}
\caption{Conditions on the momentum for oscillations. When possible, oscillations occur in
the region(s), located in the star(s), where $\hat{k}<\hat{k}_F(r)$.}
\label{tab:conditions-oscillations}
\end{table}
When oscillations can occur, as discussed above we expect in the large-$\gamma$ limit that we
are considering to obtain a finite number of  eigenvalues $k_n$, bounded above by the
extremal local Fermi momenta.
They correspond to oscillations which are localized in the star for pCS and eCS solutions.
For peCS solutions, for $\min(\hat{k}_F^{\star,p},\hat{k}_F^{\star,e}) < \hat{k} <
\max(\hat{k}_F^{\star,p},\hat{k}_F^{ \star,e})$ these oscillations are localized in one of the
two stars while for smaller momentum there can be quantum tunnelling between the two stars.

%%%%%%%%%%%%%%%%%%%%%%%%%%%%%%%%%%%%%%%%%%%%%%%%%%%%%%%%%%%%%%%%%%%%%%%%%%%%%%%%%%%%%%%%%%%%%%%
%%%%%%%%%%%%%%%%%%%%%%%%%%%%%%%%%%%%%%%%%%%%%%%%%%%%%%%%%%%%%%%%%%%%%%%%%%%%%%%%%%%%%%%%%%%%%%%
\subsection{The WKB analysis for $|\hat{\omega}|\ll\hat{k}<\hat{k}_F^\star$}
\label{sec:wkb-analysis}
%%%%%%%%%%%%%%%%%%%%%%%%%%%%%%%%%%%%%%%%%%%%%%%%%%%%%%%%%%%%%%%%%%%%%%%%%%%%%%%%%%%%%%%%%%%%%%%
%%%%%%%%%%%%%%%%%%%%%%%%%%%%%%%%%%%%%%%%%%%%%%%%%%%%%%%%%%%%%%%%%%%%%%%%%%%%%%%%%%%%%%%%%%%%%%%

Even if the Schr\"odinger problem in the standard form allows to have a clear analysis of the
spectrum, we can equivalently obtain the Green function for\footnote{From
now on, we also denote by $\hat{k}_F^\star$ the maximum of $\hat{k}_F^{\star,e}$ and
$\hat{k}_F^{\star,p}$ for peCS solutions.} $\hat{k}<\hat{k}_F^\star$ from the original
Schr\"odinger-like equation~(\ref{eq:wkb-eq}) with potential~(\ref{eq:potential}) by looking
for zero-energy solutions.
As noticed above, the parameter $\gamma$ defined by~(\ref{eq:def-gamma}) is large,
\begin{align}
 \gamma \gg 1 \, .
\end{align}
This allows us to solve Eq.~(\ref{eq:wkb-eq}) for $\hat{k}<\hat{k}_F^\star$ by applying the WKB
approximation if the conditions~\cite{Hartnoll:2011dm}
\begin{align}
\label{eq:conditions-wkb-approx}
 |V'(r)| \ll \gamma|V(r)|^{3/2} \qquad \textnormal{and} \qquad |V''(r)| \ll \gamma^2|V(r)|^2
\end{align}
are satisfied.
This is the case if the momentum $\hat{k}$ is not too close to the local extremal Fermi
momenta~(\ref{eq:extremal-momenta-eCS-pCS}) and~(\ref{eq:extremal-momenta-peCS}).
This means in particular that the potential vanishes linearly at the turning
points.

\begin{figure}[h!]
\centering
\subfloat[ ]{
\includegraphics[width=0.48\textwidth]{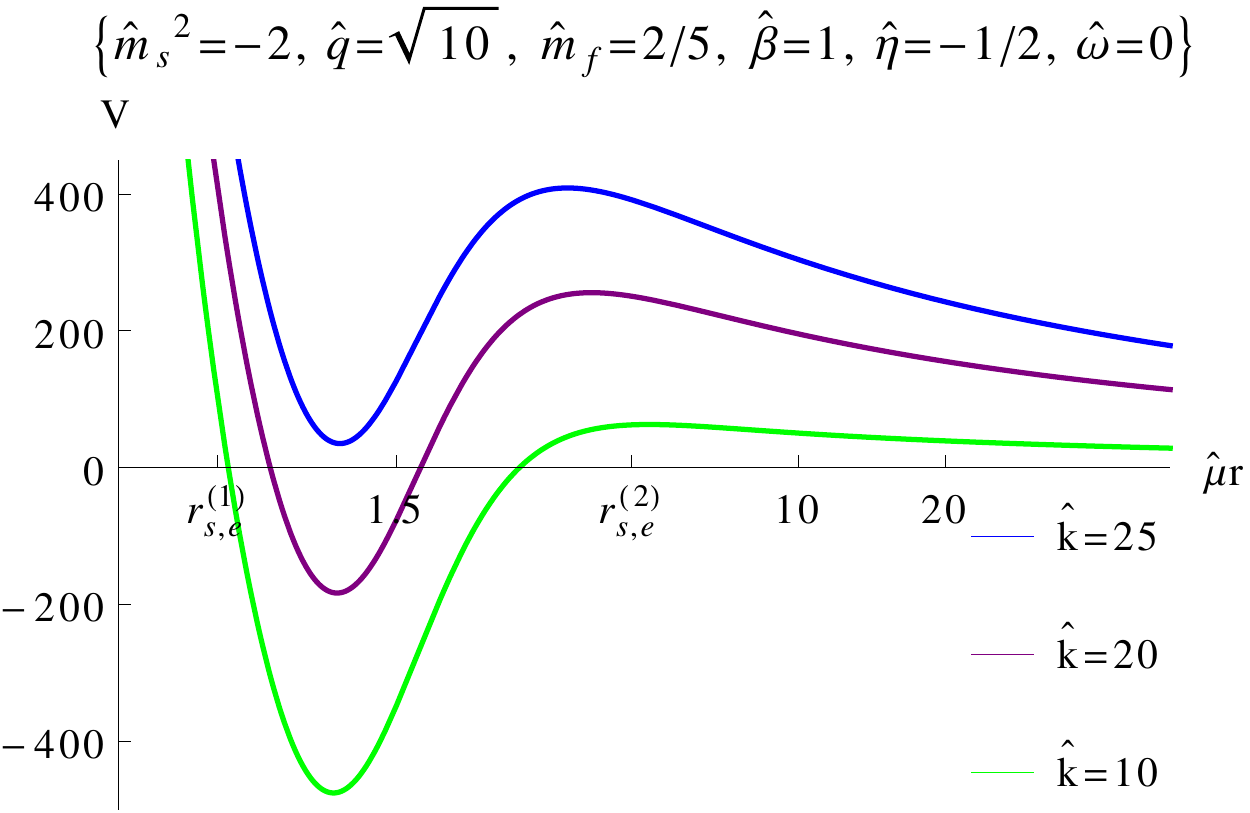}
}
\subfloat[ ]{
\includegraphics[width=0.48\textwidth]{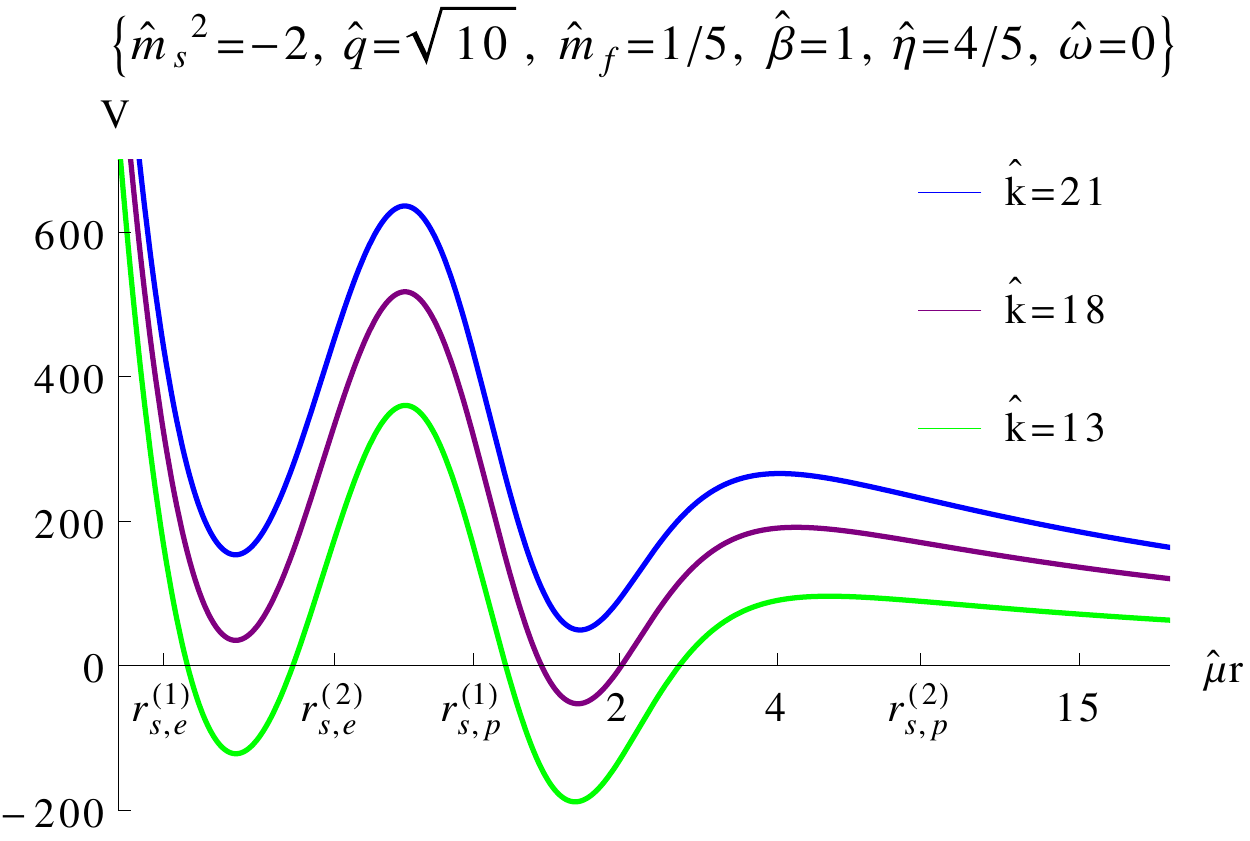}
}
\caption{Profiles of the potential~(\protect\ref{eq:potential}) at zero frequency.
(a) The potential for an eCS solutions. The extremal Fermi momentum is in this case
$\hat{k}_F^\star\simeq24.3$.
(b) The potential for a peCS solution.  The extremal Fermi momenta are
$\hat{k}_F^{\star,e}\simeq17.0$ and $\hat{k}_F^{\star,p}\simeq19.6$.
In (a) and (b), the points $r_{s,e}^{(1)}$, $r_{s,e}^{(2)}$, $r_{s,p}^{(1)}$ and
$r_{s,p}^{(2)}$ are the boundaries of the electron and the positron stars.}
\label{fig:wkb-pot}
\end{figure}

The potential~(\ref{eq:potential}) is positive both in the UV and the IR, where it behaves as
\begin{equation}
\label{eq:asympt-V-UV}
V \sim \frac{\hat{m}_f^2}{r^2} \, , \qquad r\sim0 \, ,
\end{equation}
and
\begin{equation}
\label{eq:asympt-V-IR}
V \sim \frac{3}{2|\hat{m}_s^2|}\frac{\hat{k}^2}{\log r} \, , \qquad r\sim\infty \, ,
\end{equation}
where we have used the fact that $|\hat{\omega}|\ll\hat{k}$.
As explained in the previous section, for $\hat{k}<\hat{k}_F^\star$ there can be
one or two regions where $V<0$; in these regions, the solution is oscillating and we expect to
observe bound states.
In Figure~\ref{fig:wkb-pot}, we display these possible situations by plotting the
potential~(\ref{eq:potential}) for the compact star(s) solutions at zero frequency and several
values of the momentum.

To get the Green function of the gauge-invariant fermionic operators dual to the probe bulk
fermion $\chi$, we shall relate the UV asymptotics of $\Phi$ with the normalizable and
non-normalizable modes of $\chi$.
In the UV ($r\to0$), the asymptotic solution to the bulk Dirac
equation~(\ref{eq:Dirac-eq}) is~\cite{Faulkner:2009wj}
\begin{align}
\label{eq:spinor-UV-sol}
 \Phi \simeq A(\hat{k},\hat{\omega})\,r^{-\gamma\hat{m}_f}
\left[
\left( \begin{array}{c}
1 \\
0
\end{array} \right)
+ \dots
\right]
+ B(\hat{k},\hat{\omega})\,r^{\gamma\hat{m}_f}
\left[
\left( \begin{array}{c}
0 \\
1 \end{array} \right)
+ \dots
\right]
\end{align}
where $A$ and $B$ are independent of the radial coordinate.
The Green function of the fermionic operator dual to the bulk spinor $\Phi$ is then given by
\begin{align}
 G^R(\hat{k},\hat{\omega}) = \frac{B(\hat{k},\hat{\omega})}{A(\hat{k},\hat{\omega})} \, .
\end{align}
As shown in Appendix~\ref{app:wkb}, $\Phi_2$ behaves in the UV as
\begin{align}
\label{eq:asympt-spinor1}
\Phi_2(r) \simeq a_+^\epsilon(\hat{k},\hat{\omega}) \left(\frac{r}{\epsilon}\right)^{\gamma
\hat{m}_f}
+ a_-^\epsilon(\hat{k},\hat{\omega})\left(\frac{r}{\epsilon}\right)^{-\gamma \hat{m}_f+1} \,
,
\end{align}
where $\epsilon\ll1$ is a UV cut-off.
Matching this solution with~(\ref{eq:spinor-UV-sol}), the Green function is fully expressed in
terms of the UV asymptotics of the scalar field $\Phi_2$ as
\begin{align}
\label{eq:Green1}
 G^R(\hat{k},\hat{\omega}) =
\frac{\hat{\mu}+\hat{\omega}+\hat{k}}{2\hat{m}_f}
\lim_{\epsilon\to0}\frac{a_+^\epsilon(\hat{k},\hat{\omega})}{a_-^\epsilon(\hat{k},\hat{\omega})
}
\epsilon^{-2\gamma\hat{m}_f} \, .
\end{align}
Notice that the functions $a_+^\epsilon(\hat{k},\hat{\omega})$ and
$a_-^\epsilon(\hat{k},\hat{\omega})$ depend on
the UV cutoff $\epsilon$.

To obtain the ratio $a_+^\epsilon(\hat{k},\hat{\omega})/a_-^\epsilon(\hat{k},\hat{\omega})$ in
Eq.~(\ref{eq:Green1}), we must impose normalizability of the wave function in the IR and
integrate out Eq.~(\ref{eq:wkb-eq}) from IR to UV.
This can be done in the WKB approximation with large parameter $\gamma$.
The details of the computation are given in Appendix~\ref{app:wkb}.
The general form of the Green function is
\begin{align}
\label{eq:green-function}
 G_R(\hat{k},\hat{\omega}) = \frac{\hat{\mu}+\hat{\omega}+\hat{k}}{2\hat{m}_f} \,
\mathcal{G}(\hat{k},\hat{\omega}) \,
\lim_{r\to0}r^{-2\gamma\hat{m}_f}\exp\left[-2\gamma\int_r^{r_1}\D
s\sqrt{V(s)}\right] \, .
\end{align}
Here, $r_1$ is the turning point of the potential $V$ the closest to the UV boundary.
The exponential suppression is due to the fact that the potential is large close to the
boundary; the fermion has to tunnel into the spacetime.
The function $\mathcal{G}(\hat{k},\hat{\omega})$ depends on the behavior of the potential $V$
at momentum $\hat{k}$ and frequency $\hat{\omega}$.

%%%%%%%%%%%%%%%%%%%%%%%%%%%%%%%%%%%%%%%%%%%%%%%%%%%%%%%%%%%%%%%%%%%%%%%%%%%%%%%%%%%%%%%%%%%%%%%
%%%%%%%%%%%%%%%%%%%%%%%%%%%%%%%%%%%%%%%%%%%%%%%%%%%%%%%%%%%%%%%%%%%%%%%%%%%%%%%%%%%%%%%%%%%%%%%
\subsection{The Green function for eCS and pCS solutions}
\label{sec:green-function-1}
%%%%%%%%%%%%%%%%%%%%%%%%%%%%%%%%%%%%%%%%%%%%%%%%%%%%%%%%%%%%%%%%%%%%%%%%%%%%%%%%%%%%%%%%%%%%%%%
%%%%%%%%%%%%%%%%%%%%%%%%%%%%%%%%%%%%%%%%%%%%%%%%%%%%%%%%%%%%%%%%%%%%%%%%%%%%%%%%%%%%%%%%%%%%%%%

For eCS and pCS solutions, the Green function has poles when $V<0$ in one region.
In this case, we find in Appendix~\ref{app:wkb} that
\begin{align}
\label{eq:mathcalG1}
 \mathcal{G}(\hat{k},\hat{\omega}) = \frac{1}{2}\tan W(\hat{k},\hat{\omega})
\end{align}
where
\begin{align}
 W(\hat{k},\hat{\omega}) = \gamma\int_{r_1}^{r_2}\D r\sqrt{|V(r)|} \, .
\end{align}
The two points $r_1$ and $r_2$ ($r_1<r_2$) are the turning points of the potential $V$.

The poles of the Green function are situated at
\begin{align}
\label{eq:eq-poles1}
W(\hat{k},\hat{\omega}) = \frac{\pi}{2}+n\pi \, , \quad n\in\mathbb{N} \, .
\end{align}
This equation defines $N$ boundary Fermi momenta $\hat{k}=\hat{k}_n$,
where $n=0,\dots,N-1$, satisfying
\begin{align}
0 < \hat{k}_{N-1} < \dots < \hat{k}_0 < \hat{k}_F^\star \, .
\end{align}
Since $\gamma\gg1$, the number $N$ of boundary Fermi momenta is large and the
WKB analysis is reliable for large $n$.
By computing explicitly the normal modes of the equation~(\ref{eq:wkb-eq}), we have verified
that the poles of the Green function are well-approximated by the WKB analysis.
Expanding~(\ref{eq:mathcalG1}) around the poles~(\ref{eq:eq-poles1}), we obtain the Green
function
\begin{align}
\label{eq:green-function-1region}
 G^R(\hat{k},\hat{\omega}) = \frac{\hat{\mu}+\hat{\omega}+\hat{k}}{2\hat{m}_f}
\sum_{0<\hat{k}_n<\hat{k}_F^\star}\frac{\gamma^{-1}c_ne^{-2\gamma a_n}}{\hat{\omega} -
v_n(\hat{k}-\hat{k}_n)}
\end{align}
where
\begin{align}
v_n &= - \frac{\partial_{\hat{k}}\,W(\hat{k}_n,0)}{\partial_{\hat{\omega}}\,W(\hat{k}_n,0)} \, ,
\qquad c_n = -\frac{\gamma}{2}\left[\partial_{\hat{\omega}} W(\hat{k}_n,0)\right]^{-1} , \\
a_n &= \int_0^{r_1}\D r \sqrt{V(\hat{k}_n,0)} + \hat{m}_f \log r_1 \, .
\end{align}
Notice that $v_n>0$ for eCS and $v_n<0$ for pCS which means that, as expected, the electron
star and the positron star form respectively electronic and positronic Fermi surfaces in the
dual field theory\footnote{The electronic excitations have positive energy $\hat{\omega}$. For
them to be particle-like excitations ($\hat{k}>\hat{k}_n$), one must have $v_n>0$. Thus Fermi
momenta with $v_n>0$ ($v_n<0$) correspond to electron-like (hole-like) Fermi surfaces.}.

%%%%%%%%%%%%%%%%%%%%%%%%%%%%%%%%%%%%%%%%%%%%%%%%%%%%%%%%%%%%%%%%%%%%%%%%%%%%%%%%%%%%%%%%%%%%%%%
%%%%%%%%%%%%%%%%%%%%%%%%%%%%%%%%%%%%%%%%%%%%%%%%%%%%%%%%%%%%%%%%%%%%%%%%%%%%%%%%%%%%%%%%%%%%%%%
\subsection{The Green function for peCS solutions}
\label{sec:green-function-2}
%%%%%%%%%%%%%%%%%%%%%%%%%%%%%%%%%%%%%%%%%%%%%%%%%%%%%%%%%%%%%%%%%%%%%%%%%%%%%%%%%%%%%%%%%%%%%%%
%%%%%%%%%%%%%%%%%%%%%%%%%%%%%%%%%%%%%%%%%%%%%%%%%%%%%%%%%%%%%%%%%%%%%%%%%%%%%%%%%%%%%%%%%%%%%%%

For peCS solutions, the Green function has poles of the same kind as for the eCS and the pCS
solutions for intermediate momentum.
Indeed, these boundary Fermi momenta $\hat{k}_n$ are bounded as
\begin{align}
\hat{k}_F^{\star,e} < \hat{k}_{N-1} < \dots < \hat{k}_0 < \hat{k}_F^{\star,p}
\end{align}
where the oscillations happen in the positron star and vice-versa when they happen in the
electron star.
Moreover, for small momentum, $V<0$ in two regions and, as shown in Appendix~\ref{app:wkb}, the
Green function reads
\begin{align}
\label{eq:mathcalG2}
 \mathcal{G}(\hat{k},\hat{\omega}) =
\frac{4 e^{2X}\sin Y \cos Z + \cos Y \sin Z}{8e^{2X}\cos Y \cos Z - 2\sin Y\sin Z}
\end{align}
where
\begin{align}
X(\hat{k},\hat{\omega}) &= \gamma\int_{r_2}^{r_3} \D r \sqrt{V(r)} \, , \qquad
Y(\hat{k},\hat{\omega}) = \gamma\int_{r_1}^{r_2} \D r \sqrt{|V(r)|} \, , \\
Z(\hat{k},\hat{\omega}) &= \gamma\int_{r_3}^{r_4} \D r \sqrt{|V(r)|} \, .
\end{align}
The potential $V$ vanishes linearly at the turning points $r_1<r_2<r_3<r_4$.
From (\ref{eq:mathcalG2}) we see that the conditions for $\mathcal{G}$ to have a pole are 
\begin{align}
\label{eq:poles-2-1}
 (Y,Z) = \left(\frac{\pi}{2}+n\pi , m\pi\right) \quad \textnormal{and} \quad (Y,Z) =
\left(n\pi , \frac{\pi}{2}+m\pi\right) \, , \quad n,m\in\mathbb{N}
\end{align}
for all $X$, and
\begin{align}
\label{eq:poles-2-2}
 e^{2X} = \frac{1}{4}\tan Y \tan Z \, , \qquad \tan Y \tan Z \geq 0 \, .
\end{align}
However the conditions ~(\ref{eq:poles-2-1}) are generically not satisfied for any pair 
 $\{\hat{k},\hat{\omega}\}$ and so they do not give rise to poles. 

The boundary Fermi momenta
$\hat{k}_{\bar{n}}$ are then found by solving ~(\ref{eq:poles-2-2}) at zero frequency.
Around $\hat{k}=\hat{k}_{\bar{n}}$ and $\hat{\omega}=0$, (\ref{eq:mathcalG2}) becomes
\begin{align}
 \mathcal{G}(\hat{k},\hat{\omega}) \simeq
\frac{\gamma^{-1} c_{\bar{n}}}{\hat{\omega}-v_{\bar{n}}(\hat{k}-\hat{k}_{\bar{n}})}
\end{align}
where
\begin{align}
v_{\bar{n}} = -
\frac{8\,e^{2X_{\bar{n}}}\,\partial_{\hat{k}}X_{\bar{n}} - \frac{\tan Z_{\bar{n}}}{\cos^2
Y_{\bar{n}}}
\,\partial_{\hat{k}} Y_{\bar{n}} - \frac{\tan Y_{\bar{n}}}{\cos^2 Z_{\bar{n}}}
\,\partial_{\hat{k}} Z_{\bar{n}}}
{8\,e^{2X_{\bar{n}}}\,\partial_{\hat{\omega}}X_{\bar{n}} - \frac{\tan Z_{\bar{n}}}{\cos^2
Y_{\bar{n}}}
\,\partial_{\hat{\omega}} Y_{\bar{n}} - \frac{\tan Y_{\bar{n}}}{\cos^2 Z_{\bar{n}}}
\,\partial_{\hat{\omega}} Z_{\bar{n}}}
\end{align}
and
\begin{align}
c_{\bar{n}} = \gamma \tan Z_{\bar{n}}\left(1+\tan^2 Y_{\bar{n}}\right)
\left[16\,e^{2X_{\bar{n}}}\,\partial_{\hat{\omega}}X_{\bar{n}} - 2\frac{\tan
Z_{\bar{n}}}{\cos^2 Y_{\bar{n}}}  
\,\partial_{\hat{\omega}} Y_{\bar{n}} - 2\frac{\tan Y_{\bar{n}}}{\cos^2 Z_{\bar{n}}}
\,\partial_{\hat{\omega}}
Z_{\bar{n}} \right]^{-1}
\end{align}
with
\begin{align}
 X_{\bar{n}} \equiv X(\hat{k}_{\bar{n}},0)
\end{align}
and similarly for $Y$ and $Z$.

The Green function for peCS solutions with $\hat{k}_F^{\star,e}<\hat{k}_F^{\star,p}$ can
therefore be written as
\begin{align}
\label{eq:green-function-peCS-e<p}
 G^R(\hat{k},\hat{\omega}) = \frac{\hat{\mu}+\hat{\omega}+\hat{k}}{2\hat{m}_f}
&\left(
\sum_{\hat{k}_F^{\star,e}<\hat{k}_n^Z<\hat{k}_F^{\star,p}}\frac{\gamma^{-1}c_n^Ze^{
-2\gamma a_n^Z}}{\hat{\omega} - v_n^Z(\hat{k}-\hat{k}_n^Z)}
+
\sum_{0<\hat{k}_{\bar{n}}<\hat{k}_F^{\star,e}}\frac{\gamma^{-1}c_{\bar{n}}
e^{-2\gamma a_{\bar{n}}}}{\hat{\omega} - v_{\bar{n}}(\hat{k}-\hat{k}_{\bar{n}})}
 \right)
\end{align}
where
\begin{subequations}
\label{eq:def-vcaZ}
\begin{align}
v_n^Z &= -
\frac{\partial_{\hat{k}}\,Z(\hat{k}_n^Z,0)}{\partial_{\hat{\omega}}\,Z(\hat{k}_n^Z,0)}
\,
,
\qquad c_n^Z = -\frac{\gamma}{2}\left[\partial_{\hat{\omega}} Z(\hat{k}_n^Z,0)\right]^{-1} , \\
a_n^Z &= \int_0^{r_1}\D r \sqrt{V(\hat{k}_n^Z,0)} + \hat{m}_f \log r_1 \, .
\end{align}
\end{subequations}
We have denoted by $\hat{k}_n^Z$ the boundary Fermi momenta obtained when the potential $V$ is
negative in one region; in this case we have $Z(\hat{k}_n^Z,0)=\pi/2+n\pi$.
When $\hat{k}_F^{\star,p}<\hat{k}_F^{\star,e}$, the Green function for peCS solutions is
\begin{align}
\label{eq:green-function-peCS--p<e}
 G^R(\hat{k},\hat{\omega}) = \frac{\hat{\mu}+\hat{\omega}+\hat{k}}{2\hat{m}_f}
&\left(
\sum_{\hat{k}_F^{\star,p}<\hat{k}_n^Y<\hat{k}_F^{\star,e}}\frac{\gamma^{-1}c_n^Ye^{
-2\gamma a_n^Y}}{\hat{\omega} - v_n^Y(\hat{k}-\hat{k}_n^Y)}
+
\sum_{0<\hat{k}_{\bar{n}}<\hat{k}_F^{\star,p}}\frac{\gamma^{-1}c_{\bar{n}}
e^{-2\gamma a_{\bar{n}}}}{\hat{\omega} - v_{\bar{n}}(\hat{k}-\hat{k}_{\bar{n}})}
 \right)
\end{align}
where $v_n^Y$, $c_n^Y$ and $a_n^Y$ are defined by~(\ref{eq:def-vcaZ}) where one has to replace
$Z$ by $Y$.

\begin{figure}[t]
\centering
\subfloat[ ]{
\includegraphics[width=0.4\textwidth]{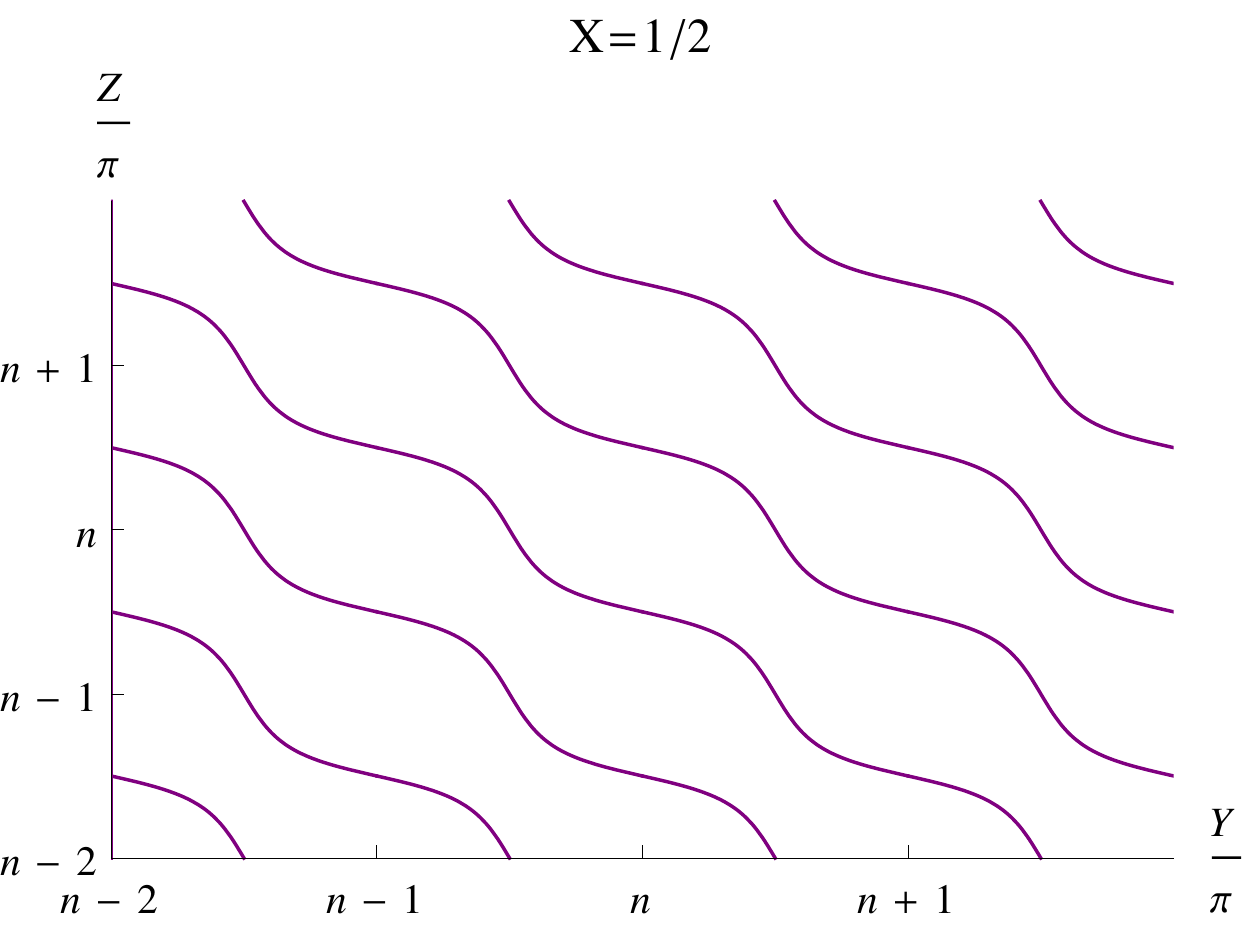}
\label{fig:poles-1}}
\subfloat[ ]{
\includegraphics[width=0.4\textwidth]{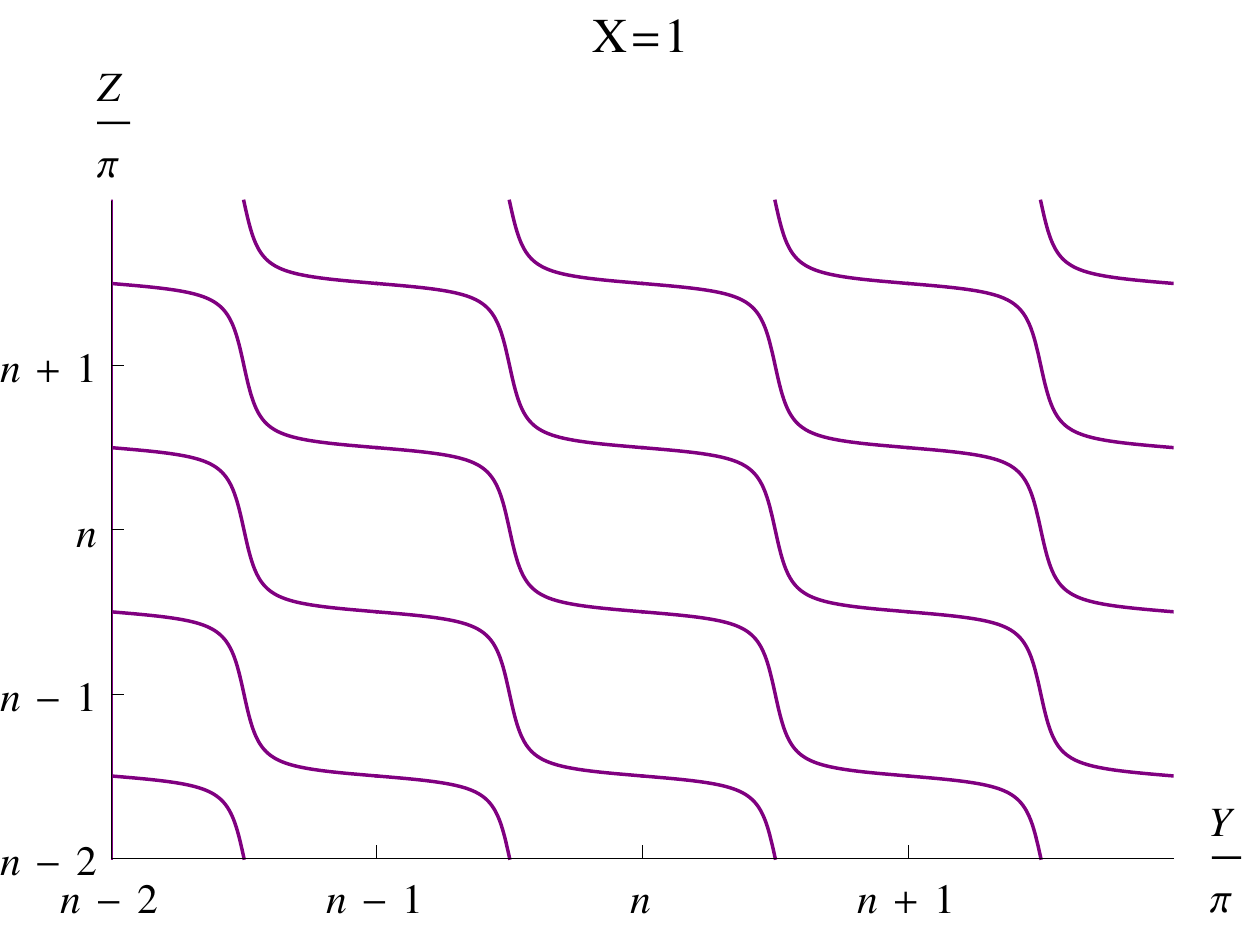}
\label{fig:poles-2}}\\
\subfloat[ ]{
\includegraphics[width=0.4\textwidth]{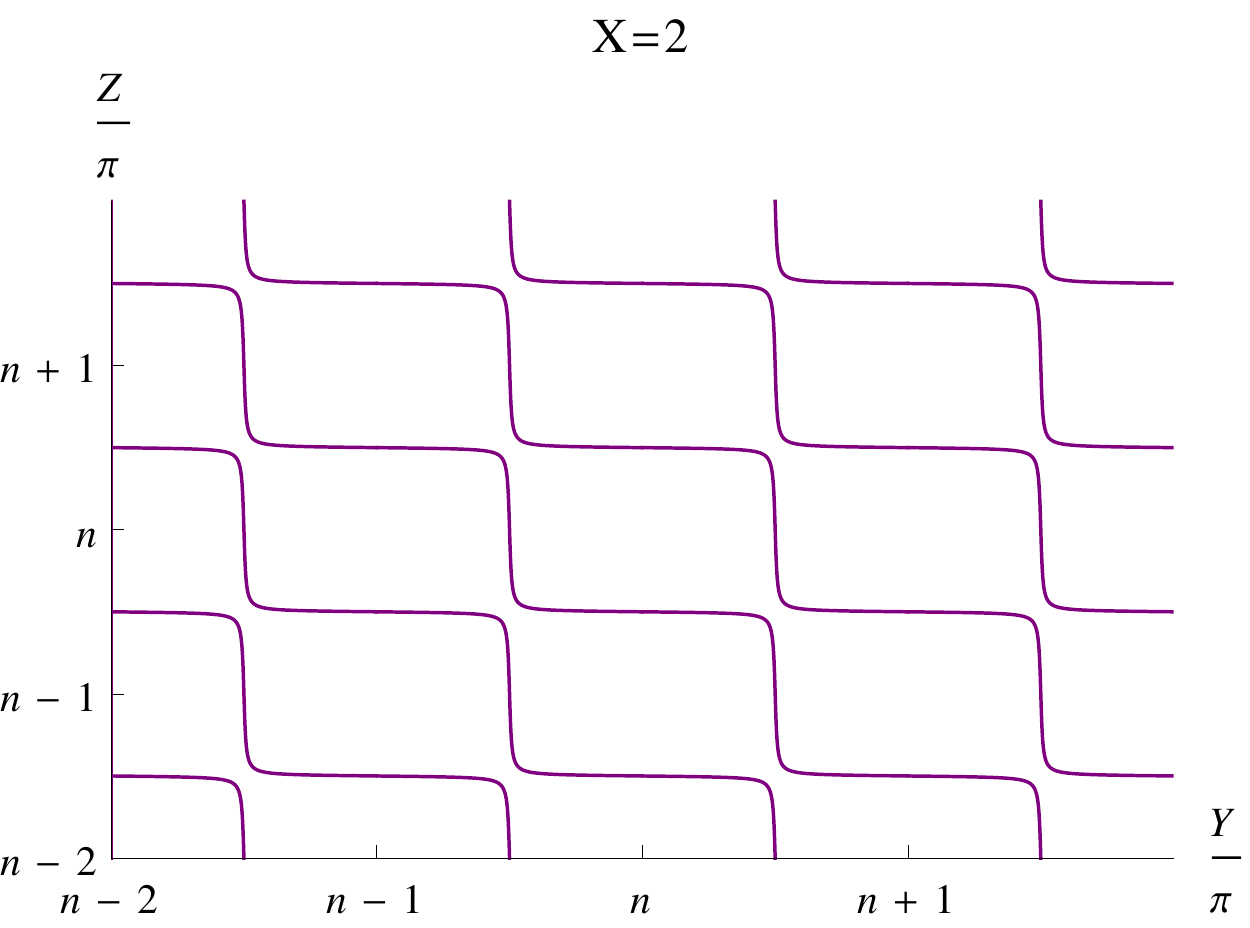}
\label{fig:poles-3}}
\subfloat[ ]{
\includegraphics[width=0.4\textwidth]{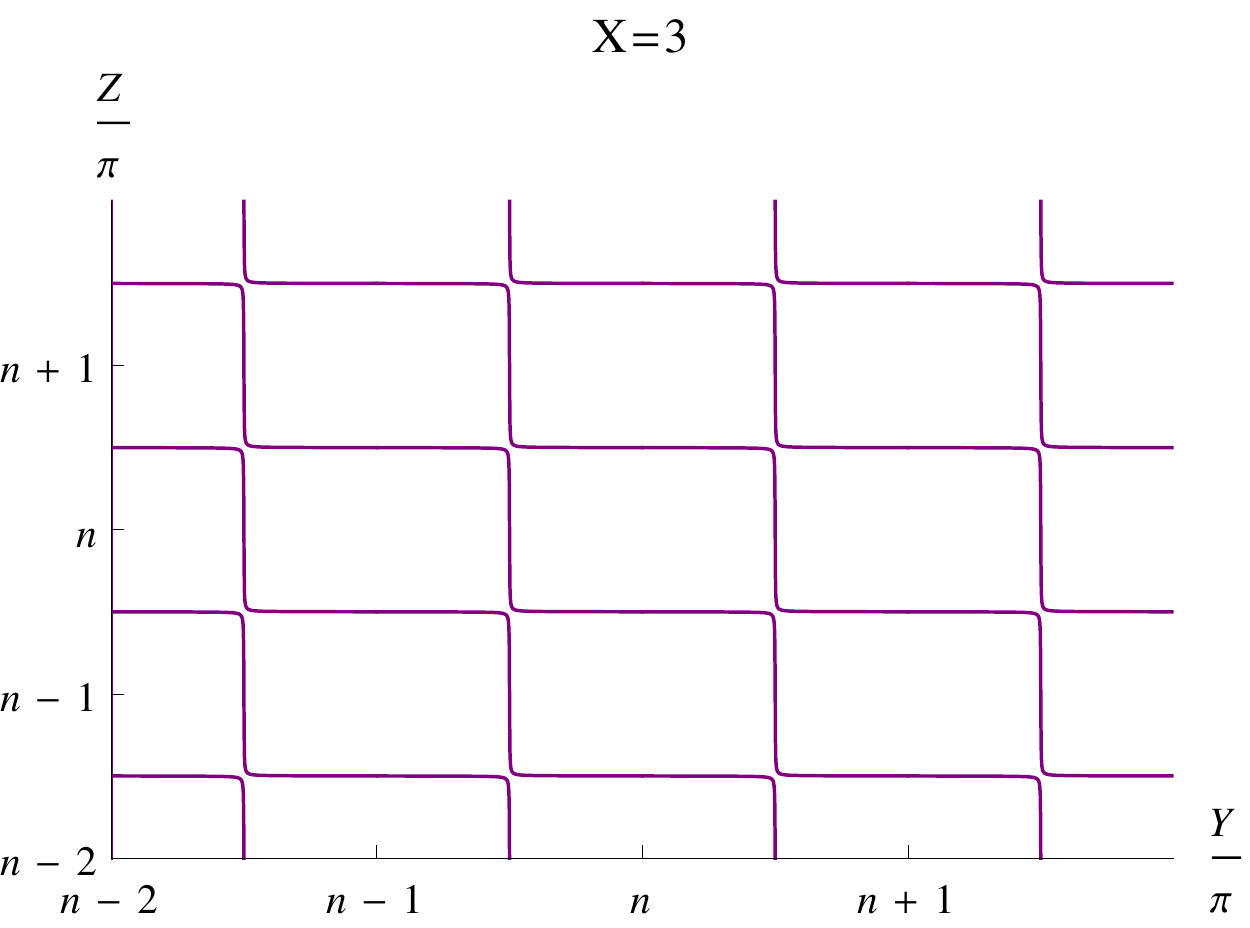}
\label{fig:poles-4}}
\caption{Poles of $\mathcal{G}$ for peCS solutions in the (Y-Z)-plan for different values of
$X$.}
\label{fig:poles}
\end{figure}
\begin{figure}[ht!]
\centering
\subfloat[ ]{
\includegraphics[width=0.44\textwidth]{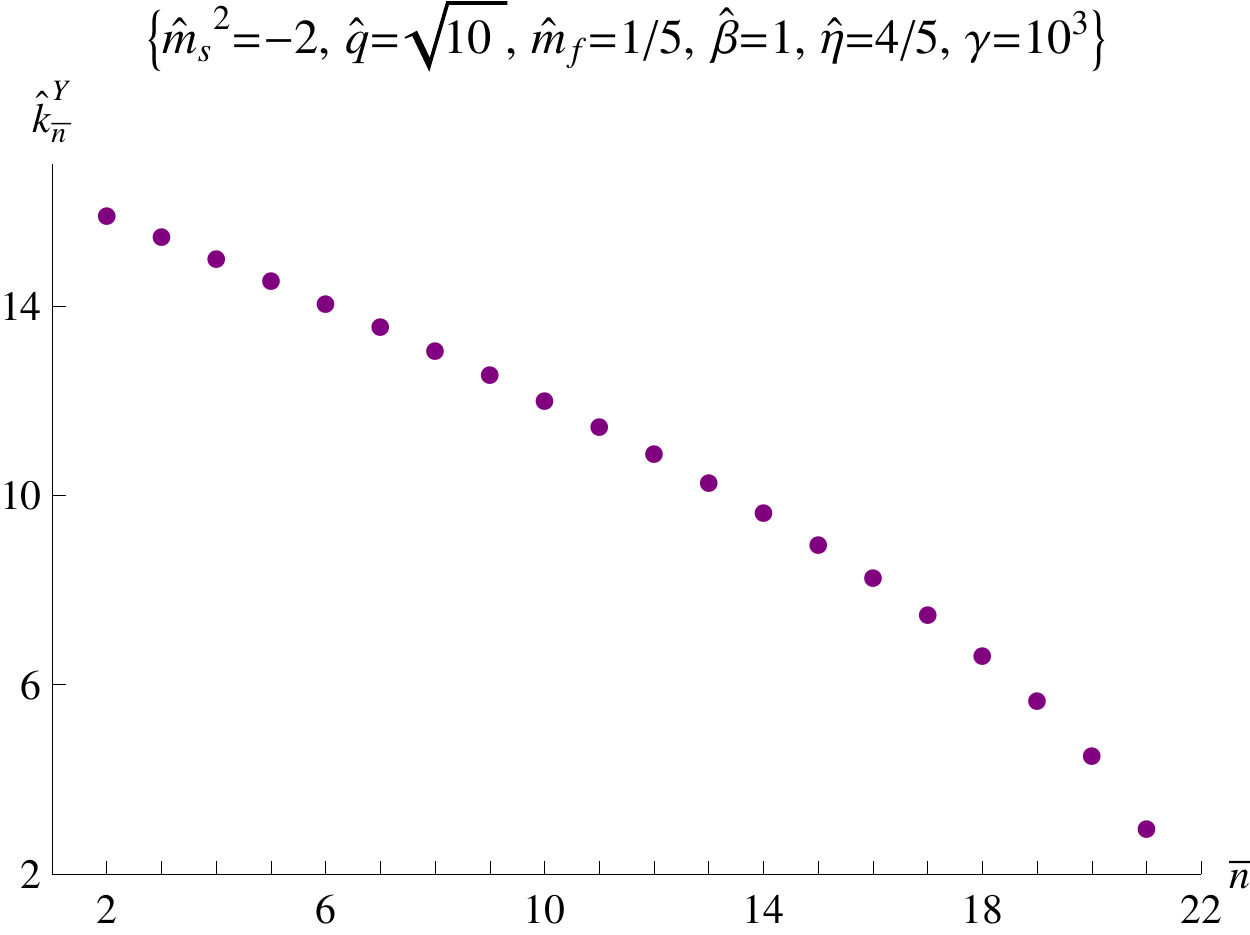}
\label{fig:poles-Y}}
\qquad
\subfloat[ ]{
\includegraphics[width=0.44\textwidth]{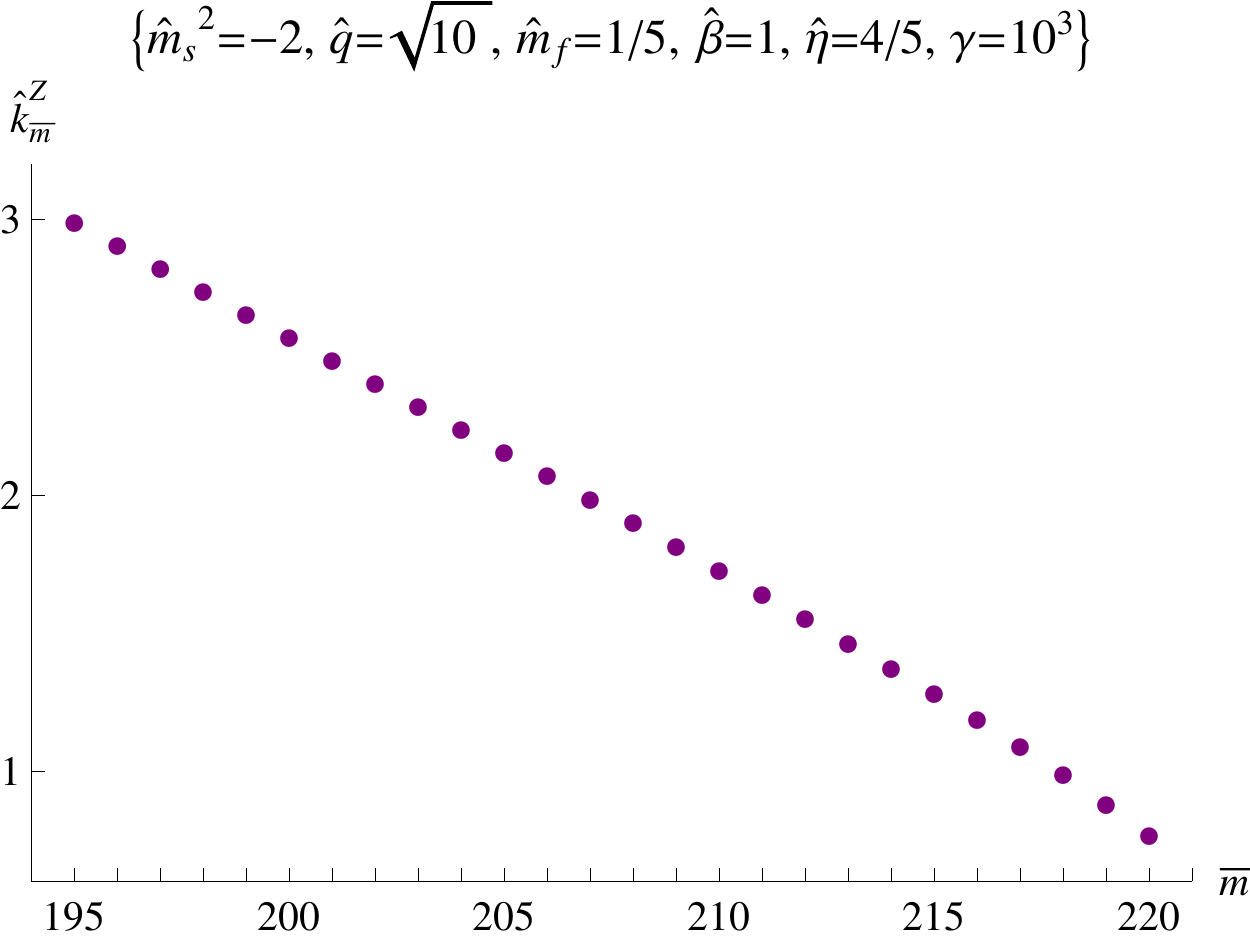}
\label{fig:poles-Z}}
\caption{Boundary Fermi momenta for a peCS solution. The $\hat{k}_{\bar{n}}^Y$'s and the
$\hat{k}_{\bar{m}}^Z$'s are the Fermi momenta of electron-like and hole-like Fermi surfaces in
the dual field theory, respectively.}
\label{fig:polesYZ}
\end{figure}

In Figure~\ref{fig:poles}, we display the poles of~(\ref{eq:mathcalG2}) in the
$(Y,Z)$-plan for constant values of $X$.
We see that in the large-$\gamma$ limit, the location of the poles of~(\ref{eq:mathcalG2}) are
well-approximated by 
\begin{align}
\label{eq:poles-sol-inf}
(Y,Z) \simeq \left(\frac{\pi}{2} + n\pi,Z\right) \quad \textnormal{and} \quad (Y,Z) \simeq
\left(Y,\frac{\pi}{2} + m\pi\right) \, ,\qquad n,m\in\mathbb{N} \, .
\end{align}
as suggested by Figure~\ref{fig:poles-4}.
We denote by $\hat{k}_{\bar{n}}^Y$ and $\hat{k}_{\bar{m}}^Z$ the boundary Fermi momenta
corresponding to the poles~(\ref{eq:poles-sol-inf}) and computed in the WKB approximation.
In Figure~\ref{fig:polesYZ} we give them for a peCS solution.
These poles match with a high accuracy to the normal modes of the Schr\"odinger-like
equation~(\ref{eq:wkb-eq}) that we have computed explicitly.
%
%

%%%%%%%%%%%%%%%%%%%%%%%%%%%%%%%%%%%%%%%%%%%%%%%%%%%%%%%%%%%%%%%%%%%%%%%%%%%%%%%%%%%%%%%%%%%%%%%
%%%%%%%%%%%%%%%%%%%%%%%%%%%%%%%%%%%%%%%%%%%%%%%%%%%%%%%%%%%%%%%%%%%%%%%%%%%%%%%%%%%%%%%%%%%%%%%
\subsection{The Luttinger count}
\label{sec:luttinger-count}
%%%%%%%%%%%%%%%%%%%%%%%%%%%%%%%%%%%%%%%%%%%%%%%%%%%%%%%%%%%%%%%%%%%%%%%%%%%%%%%%%%%%%%%%%%%%%%%
%%%%%%%%%%%%%%%%%%%%%%%%%%%%%%%%%%%%%%%%%%%%%%%%%%%%%%%%%%%%%%%%%%%%%%%%%%%%%%%%%%%%%%%%%%%%%%%

The Luttinger theorem relates the total charge of a Fermi liquid to the volume enclosed in the
Fermi surface.
For spin-$1/2$ particles with charge $|q_f|$ in two spatial dimensions, the Luttinger
count is~\cite{Oshikawa}
\begin{align}
\label{eq:lutt-count}
 Q_\textnormal{FS}= |q_f| \sum_i \frac{2}{(2\pi)^2} V_i \, ,
\end{align}
where $V_i$ are the volumes of the Fermi surfaces, given by $V_i=\pi p_i^2$ where $p_i$ is the
Fermi momentum of the $i$-th Fermi surface.
In the WKB approximation, one can approximate the discrete sum in~(\ref{eq:lutt-count}) by
an integral~\cite{Hartnoll:2011dm}; for eCS and pCS solutions we have
\begin{align}
 \sum_i \frac{1}{2\pi}p_i^2 = \sum_n \frac{1}{2\pi}k_n^2
= \frac{1}{2\pi}\int_0^{{k_F^\star}^2} \D E \, E \int_{r_1}^{r_2} \D r \, r\sqrt{g}
\frac{1}{\sqrt{k_F^2-E}}
\end{align}
where the Fermi momenta $k_n$ appear (with hats) in
the Green function~(\ref{eq:green-function-1region}).
This leads to the relations~\cite{Hartnoll:2011dm}
\begin{subequations}
\begin{align}
 |q_f|\sum_n \frac{1}{2\pi}k_n^2 = Q_\textnormal{e} \, , \\
 - |q_f|\sum_n \frac{1}{2\pi}k_n^2 = Q_\textnormal{p} \, ,
\end{align}
\end{subequations}
for eCS and pCS solutions, respectively, where the charges $Q_e$ and $Q_p$ are defined in
Section~\ref{sec:mixture}.
It means that as expected, the eCS and pCS solutions admit respectively boundary electronic
and positronic Fermi surfaces.
This is consistent with the fact that the particle excitations are electronic in the field
theory dual to the eCS solution and positronic in the field theory dual to the pCS solution.
The Luttinger relation is
\begin{align}
 |q_f|\sum_i \frac{1}{2\pi}p_i^2 = Q - Q_\textnormal{scalar} - Q_\textnormal{int,e}
\end{align}
for eCS solutions and
\begin{align}
 -|q_f|\sum_i \frac{1}{2\pi}p_i^2 = Q - Q_\textnormal{scalar} - Q_\textnormal{int,p}
\end{align}
for pCS solutions.
These situations are similar to the fractionalized phases of~\cite{Hartnoll:2011pp}; here the
bosonic field takes the role of the charged event horizon, and there is also screening of the
fermionic charge by the condensate.

For peCS solutions, the Luttinger count is
\begin{align}
 Q_\textnormal{FS}^\textnormal{e} = |q_f|\sum_i \frac{1}{2\pi}(p_i^e)^2 + |q_f|\sum_j
\frac{1}{2\pi}(q_j^e)^2
\end{align}
for electronic Fermi surfaces and
\begin{align}
 Q_\textnormal{FS}^\textnormal{p} = -|q_f|\sum_i \frac{1}{2\pi}(p_i^p)^2 - |q_f|\sum_j
\frac{1}{2\pi}(q_j^p)^2
\end{align}
for positronic Fermi surfaces.
Here, $p_i^e$ and $q_j^e$ denote the boundary Fermi momenta of electronic Fermi surfaces
corresponding to the cases where the potential $V$ is negative in one and two regions
respectively in the  bulk, and similarly for positrons.
For $\hat{k}_F^{\star,e}<\hat{k}_F^{\star,p}$, we have
\begin{subequations}
\begin{align}
|q_f| \sum_i \frac{1}{2\pi}(p_i^e)^2 &= 0 \, , \\
|q_f| \sum_j \frac{1}{2\pi}(q_j^e)^2
&= |q_f| \sum_{\bar{n}} \frac{1}{2\pi}(k_{\bar{n}}^Y)^2
= \frac{|q_f|}{2\pi}\int_0^{({k_F^{\star,e}})^2} \D E \, E \int_{r_1}^{r_2} \D r \,
r\sqrt{g}\frac{1}{\sqrt{k_F^2-E}} \, , \\
-|q_f| \sum_i \frac{1}{2\pi}(p_i^p)^2
&= -|q_f| \sum_{n} \frac{1}{2\pi} (k_{n}^Z)^2
= -\frac{|q_f|}{2\pi}\int_{{(k_F^{\star,e})}^2}^{{(k_F^{\star,p})}^2} \D E \, E
\int_{r_3}^{r_4} \D r \, r\sqrt{g}\frac{1}{\sqrt{k_F^2-E}} \, , \\
-|q_f| \sum_j \frac{1}{2\pi}(q_j^p)^2
&= -|q_f| \sum_{\bar{n}} \frac{1}{2\pi} (k_{\bar{n}}^Z)^2
= -\frac{|q_f|}{2\pi}\int_0^{({k_F^{\star,e}})^2} \D E \,
E \int_{r_3}^{r_4} \D r
\,
r\sqrt{g}\frac{1}{\sqrt{k_F^2-E}} \, .
\end{align}
\end{subequations}
We conclude that
\begin{align}
  Q_\textnormal{FS}^\textnormal{e} = Q_\textnormal{e} \qquad \textnormal{and} \qquad Q_\textnormal{FS}^\textnormal{p} =
Q_\textnormal{p} \, .
\end{align}
This result is also valid when $\hat{k}_F^{\star,p}<\hat{k}_F^{\star,e}$.
So we have
\begin{align}
 Q_\textnormal{FS}^\textnormal{e} + Q_\textnormal{FS}^\textnormal{p} = Q - Q_\textnormal{scalar} - Q_\textnormal{int,e} -
Q_\textnormal{int,p} \, .
\end{align}

Thus, we have shown that the charge that one can assign to fermionic fluid components (which does not include the effect of screening due to the scalar field) is reproduced by the total volume of particle-like and hole-like Fermi surfaces via the Luttinger count.  
% The formation of hole-like Fermi surfaces is allowed by the screening of the fermion charge by
% the charged condensate.

%%%%%%%%%%%%%%%%%%%%%%%%%%%%%%%%%%%%%%%%%%%%%%%%%%%%%%%%%%%%%%%%%%%%%%%%%%%%%%%%%%%%%%%%%%%%%%%
%%%%%%%%%%%%%%%%%%%%%%%%%%%%%%%%%%%%%%%%%%%%%%%%%%%%%%%%%%%%%%%%%%%%%%%%%%%%%%%%%%%%%%%%%%%%%%%
\subsection{Fermi surfaces and phase transitions}
\label{sec:FS-PT}
%%%%%%%%%%%%%%%%%%%%%%%%%%%%%%%%%%%%%%%%%%%%%%%%%%%%%%%%%%%%%%%%%%%%%%%%%%%%%%%%%%%%%%%%%%%%%%%
%%%%%%%%%%%%%%%%%%%%%%%%%%%%%%%%%%%%%%%%%%%%%%%%%%%%%%%%%%%%%%%%%%%%%%%%%%%%%%%%%%%%%%%%%%%%%%%

We have shown in the previous sections that the compact star solutions exhibit a large number of
boundary Fermi surfaces when the WKB approximation is applicable.
Since the field theory is rotationally invariant, they are circular and each of them is
specified by a Fermi momentum $\hat{k}_n$, which lie between zero and the maximal value $\hat{k}^\star_F$.

In the WKB approximation, the number of Fermi surfaces with Fermi momentum in the interval 
$(\hat{k},\hat{k}_F^\star)$ of a solution with one star -- eCS, pCS or ES -- is given by the
integral~\cite{Landau1977quantum}:
\begin{align}
\label{eq:number-surfaces-k-kFs}
N(\hat{k},\hat{k}_F^\star) \propto \gamma \int_{y_1}^{y_2} \D y \,
\sqrt{\hat{k}_F^2(y)-\hat{k}^2}
\end{align}
where $y_1$ and $y_2$ are boundaries of the region where $\hat{k}_F^2(y)>\hat{k}^2$.
The total number of Fermi surfaces $N$ is then
\begin{align}
\label{eq:number-surfaces-total}
N \propto \gamma \int_{y_{s,1}}^{y_{s,2}} \D y \, \hat{k}_F(y)
\end{align}
where $y_{s,1}$ and $y_{s,2}$ are the star boundaries.
This formula is in fact not exact, because for $\hat{k}$ close to $\hat{k}_F^\star$, the
WKB approximation is not valid.
However, the contribution of such momenta to the total number of levels is small, and non-vanishing, for eCS, pCS and ES  solutions. 
For the eCS and pCS solutions, the star boundaries $y_{s,1}$ and $y_{s,2}$ are finite, so the
total number of Fermi surfaces is finite; this also applies to the peCS solutions.
The corresponding Fermi momenta are bounded by zero and the extremal local Fermi momenta.

For the (unbounded) electron star solution, the total number of Fermi
surfaces~(\ref{eq:number-surfaces-total})
can be written as
\begin{align}
N \propto \gamma \int_{y_s}^{y_0} \D y \, \hat{k}_F(y) + \gamma \int_{y_0}^\infty \D y \,
\frac{\sqrt{g_\infty(h_\infty^2-\hat{m}_f^2)}}{y}
\end{align}
where $y_s$ is the star boundary and $y_0$ is an arbitrary point which belongs to the Lifshitz
region.
The first term is finite since the boundaries are finite and the integrand is a regular
function.
However, the second term is logarithmically divergent.
Then the electron star solution is dual to a field theory state admitting an infinite number
of Fermi surfaces $N=\infty$.
Indeed, it can be deduced from~(\ref{eq:number-surfaces-k-kFs}) that the density of levels at
small momentum $\hat{k}$ is
\begin{align}
 \rho(\hat{k}) \sim \frac{\gamma}{\hat{k}} \, , \qquad \hat{k}\to0 \, ,
\end{align}
so the number of Fermi surfaces in the interval $(\hat{k},\hat{k}_0)$ is
\begin{align}
 N(\hat{k},\hat{k}_0) \sim \gamma \log \frac{\hat{k}_0}{\hat{k}} \, , \qquad \hat{k} \to 0
\, ,
\end{align}
where $\hat{k}_0\ll1$ is a cutoff such that $\hat{k}<\hat{k}_0$.
It means that the Fermi surfaces are accumulating exponentially at small momentum,
\begin{align}
 \hat{k} \sim \hat{k}_0 \, e^{-N(\hat{k},\hat{k}_0)/\gamma} \, .
\end{align}
Notice that even if this computation of the total number of Fermi surfaces for the compact
star solutions and the electron star applies only in the WKB approximation, the result is also
valid when this approximation is not valid, as discussed in Section~\ref{sec:Schro-true-form}.

The infiniteness of fermionic modes in the electron star phase is removed at
finite frequency.
For $\hat{\omega}>0$, the number of resonances has to be counted between the extremal local
Fermi momentum and $\hat{k}\sim\hat{\omega}^{1/z}$.
For smaller momentum, the modes are unstable.
The total number of resonances is in this case
\begin{align}
N(\hat{\omega}>0) &\sim \gamma \int_{y_1}^{y_\star} \D y \,
\sqrt{\hat{k}_F^2(y)-\hat{\omega}^{2/z}} \, ,
\end{align}
where $y_1$ is the first turning point of $\tilde{V}$ and $\tilde{V}'(y_\star)=0$ at
$y=y_\star$, which belongs to the Lifshitz region.
Since $y_\star\sim\hat{\omega}^{-1/z}$, we conclude that the total number of resonances for
small and positive frequency is $N(\hat{\omega}>0) \sim \gamma \log \hat{\omega}^{-1/z}$.
An analogous computation can be done for $\hat{\omega}<0$ where the integration is taken up to
$\hat{k}=0$.
The result is similar and we conclude that at small non-zero frequency, the
total number of resonances is
\begin{align}
\label{eq:number-surfaces-omega-ES}
N(\hat{\omega}\neq0) &\sim \gamma \log |\hat{\omega}|^{-1/z} \, , \qquad \hat{\omega}\to0 \, .
\end{align}
This can be seen as the  number of Fermi surfaces that admit quasi-particle  excitations with frequency up to $\hat\omega$. This increases indefinitely as $\hat\omega$ is decreased, and as $\hat\omega \to 0$, we recover the infinite number of Fermi surfaces of the unbounded Electron star. 
 
On the other hand, for the compact star solutions, the number of resonances does not depend on
the frequency because the effects of finite and small $\hat\omega$ do not modify the two first
turning points of the potential $\tilde{V}$, which are still well approximated by the edges of the star: one finds the same finite number of stable excitations for $\hat\omega=0$ and small $\hat\omega$. More precisely, a small frequency $|\hat\omega| < \hat k_N$ (where $\hat k_N$ is the smallest eigenstate of the $\hat\omega=0$ potential)  does not change the number of bound states.  

A Fermi surface is defined when a system of fermions exhibits gapless low energy excitations
around a Fermi momentum $\hat{k}_n$.
A Fermi surface is not only defined by the existence of a Fermi momentum but also by a
dispersion relation for the low-energy excitations around it.
For this reason, the result that the number of Fermi surfaces is infinite in the field theory
state dual to the electron star needs to be clarified.
This result was obtained by setting $\hat{\omega}=0$.
From (\ref{eq:number-surfaces-omega-ES}), what we should rather say is that in the field
theory state dual to the electron star, the number of fermionic excitations at fixed energy and
fixed momentum diverges when the energy goes to zero.
In other words, at fixed energy the number of fermionic excitations of the system is
\textit{arbitrarily large} when the energy tends to zero.

The above discussion suggests that some of the Fermi surfaces disappear between
the electron star phase where it is infinite and the compact stars where it is finite.
Let us consider the case where the coupling constant $\hat{\eta}$ vanishes, corresponding to
the compact star solutions found in~\cite{Nitti:2013xaa}.
By increasing the elementary charge of the scalar field, one is expected to move from the
electron star phase to the compact star phase\footnote{In fact, although it is expected, a
phase transition between the electron star and the compact star was not found explicitly in~\cite{Nitti:2013xaa},  due to the difficulties in solving the system numerically for parameter values  close to the
possible transition.
Thus, we cannot exclude the  possibility  of the existence of another phase  between the electron star and the compact
star phases.}.
Some of the Fermi surfaces are destroyed, they are the Fermi surfaces of the flavors of
fermions which become superconducting.
These Fermi surfaces have small Fermi momentum, which goes in the opposite direction of what
was found in~\cite{Hartman:2010fk} for the Cooper pair creation in the Reissner-Nordstr\"om black
hole.
This suggests that the mechanism leading to superconductivity is here far from being described
by the BCS theory and the formation of Cooper pairs.
The scalar condensate can rather be though as being a very complicated operator made out of
many fermions, which does not have integer charge in elementary fermion charge unit.
Indeed, in the bulk we have in general $q\neq 2q_f$.

%%%%%%%%%%%%%%%%%%%%%%%%%%%%%%%%%%%%%%%%%%%%%%%%%%%%%%%%%%%%%%%%%%%%%%%%%%%%%%%%%%%%%%%%%%%%%%%
%%%%%%%%%%%%%%%%%%%%%%%%%%%%%%%%%%%%%%%%%%%%%%%%%%%%%%%%%%%%%%%%%%%%%%%%%%%%%%%%%%%%%%%%%%%%%%%
\section{Conclusion}
\label{sec:conclusion}
%%%%%%%%%%%%%%%%%%%%%%%%%%%%%%%%%%%%%%%%%%%%%%%%%%%%%%%%%%%%%%%%%%%%%%%%%%%%%%%%%%%%%%%%%%%%%%%
%%%%%%%%%%%%%%%%%%%%%%%%%%%%%%%%%%%%%%%%%%%%%%%%%%%%%%%%%%%%%%%%%%%%%%%%%%%%%%%%%%%%%%%%%%%%%%%

% \begin{itemize}
% \item Spectral weight of the fermionic excitations (see ES) + formation of condensate
% \item Number of states at finite $\hat{\omega}$ : in the ES, should be different for positive
% and negative $\hat{\omega}$, states dissipate for positive $\hat{\omega}$.
% \item No FS for sufficiently small CS, but for ES always a star
% \item Shall we talk about the semiconductors interpretation ?
% \item Why are the CS solutions stable ? Especially the peCS solutions
% \item Lower bound for the formation of the first FS in compact star solutions : can have bulk
% fermions but not    boundary Fermi surfaces (not possible in the ES, $N=\infty$ always in this
% case)
% \item Generalized free fields : infinite sum over $M$ for the ES, finite sum for the CS. What
% is the interpretation of the mass $M$ and the \lq smeared Fermi surface' (hypersurface in the
% space generated by the momentum $k$ and $M$) ?
% \end{itemize}

In this paper, we have shown that by coupling a fluid of charged fermions to a charged scalar
field in asymptotically AdS Einstein-Maxwell theory, both electron and hole-like Fermi surfaces
can form in the dual field theory. The current-current interaction causes the condensate to
effectively screen the electric charge of the fluid: the particle-type fluid charge is lowered,
and the hole-type fluid is overscreened and acquires a net positive charge. As a consequence,
beside the compact stars that were already found in these systems \cite{Nitti:2013xaa} there
exist static solutions with separate shells of electron-like and positron-like fluids. These
solutions, in the region of parameter space where they exist,  are thermodynamically favored
over both the compact stars made out entirely of particles or antiparticles, and the
holographic superconductor with no fermion fluid. This seems to confirm the trend, already
observed in \cite{Nitti:2013xaa}, that in holographic systems the solutions that dominate the
ensemble are those which contain the largest variety of charged components.  

This work has to be related to the fractionalized Fermi systems where the formation of Fermi
surfaces is controlled by a neutral scalar operator.
In our work, the scalar field plays the role of the charged black hole and is responsible for
the fact that not all the charge can be accounted for by the presence of Fermi surfaces. In our
solutions there is no horizon, hence no charge associated with non-gauge invariant degrees of
freedom. But as we argued the boson that condenses can be seen as a bound state of fermions,
although not of the fundamental fermions dual to the fermionic bulk operators; one can make the
hypothesis that it is a bound state of fractionalized fermions.  
%one could still  think of the fermions as fractionalized and forming bosonic bound states 
% \sout{fractionalization of the fermions} ({\bf wrong ! The fermions are not fractionalized,
% they form bound states ! Maybe not pairs, but...}).  {\em  why are they not fractionalized? does fractionalization in holography necessarily require they are behind a horizon?} \\
Moreover, in our system there is an additional ingredient since this also allows the formation of hole-like Fermi surfaces.

The Luttinger count computation shows that the charged Fermi fluids account for only a
part of the total charge, which gets contribution also from the scalar condensate and a
``binding charge'' which arises due to the interaction and encodes the screening effect.

The formation of the boundary Fermi surfaces is controlled by the bulk  current-current
interaction, as well as the scalar condensate. We have shown that, with respect to the case of
the non-compact electron star,  the  scalar condensate causes the disappearance of most Fermi
surfaces: an infinite number of them becomes gapped, and only a finite number remains. This
suggests that the operator which condenses couples to a large number (but not all) of the
constituent fermions. 

When the condensate is turned on, the disappearing Fermi surfaces are those with  Fermi
momentum smaller than a certain critical $k_F^*$  that depends on the microscopic parameters of
the model. This may seem puzzling, as one might have thought that ``inner'' Fermi surfaces
should be more stable. On the other hand, this agrees with the suggestion
\cite{Cubrovic:2011xm} that the order in which the Fermi surfaces are filled in  holographic
systems is characterized by a IR/UV duality: states with lowest $k_F$ are filled later than the
ones with higher $k_F$, thus they are the first to disappear when the condensate is turned on.

Another new feature of  compact star solutions is that  for small frequency, the surviving Fermi surfaces have stable particle-like excitations. This is unlike the case  of the non-compact electron star, where the quasi-particle states can decay into the critical bosonic  Lifshitz modes in the IR: in the compact case, dissipation of excitations around a given Fermi surface occurs only after an energy threshold is reached, meaning that the Fermi system does not interact with the bosons at arbitrarily low energy. 

As was shown in \cite{Cubrovic:2011xm}, the fact that the number of constituent fermions is
infinite in the non-compact electron star can be seen as a consequence of working in the regime
where the constituent charge $e q_f$ is very small compared to the total charge of the system.
As its value is increased, the number of Fermi surfaces decreases until finally one arrives at
the Dirac hair solution with only one species of fermions. It would be interesting to study the
effect of turning on a scalar condensate in the finite $q_f$ regime, where already for zero
condensate there is a finite number of Fermi surfaces.

% The dual field theory state can be interpreted as made of a number of fermions of different
% species coupled to a charged scalar condensate. 
% The bulk coupling constant plays the role of doping : it controls the chemical potential for
% each species and the formation of hole-like and electron-like Fermi surfaces.
% Each flavor of fermions behaves as a p-type or n-type semiconductor.
% For negative coupling constant, the system is a pure n-type semiconductor : only electron-like
% Fermi surfaces are present.
% In the peCS phase, the system admits n-type and p-type semiconductors.
%In the HSC phase, no Fermi surfaces form.

Notice that the WKB analysis we applied is valid when a large number of hole-like and/or
electron-like Fermi surfaces is present.
One can study the formation and destruction of Fermi surfaces in these regions of the phase
diagram which are far from the critical lines.
Following the work done to characterize the dual field theory to the electron star model, we
also expect Fermi surfaces to form for momentum close to the bulk extremal Fermi momenta.
We leave this for later considerations.

It would be interesting to characterize better the properties of the excitations around the
Fermi surfaces, determining the Fermi velocity 
and the residues as done for the electron star in \cite{Hartnoll:2011dm}. 
It would also be interesting to compute the electrical conductivity and especially the Hall
conductivity, which would determine the global nature of the fermionic system.
Another extension would be to consider a direct coupling interaction in holographic models with
emergent Lifshitz symmetry, in which case we expect a dispersion of the fermionic excitations
in the critical Lifshitz sector.
Finally, one could also study the temperature dependence of our system.

% \begin{itemize}
%  \item Contribution of the scalar field to the Green function ?
%  \item Poles in the e-fluid have larger spectral weight
%  \item Velocity of excitations ? Small, large ?
%  \item Fermi surfaces are expected to be present when the global $U(1)$ symmetry is not broken.
% Make this point clear.
% \end{itemize}

%%%%%%%%%%%%%%%%%%%%%%%%%%%%%%%%%%%%%%%%%%%%%%%%%%%%%%%%%%%%%%%%%%%%%%%%%%%%%%%%%%%%%%%%%%%%%%%
\subsection*{Acknowledgements}
%%%%%%%%%%%%%%%%%%%%%%%%%%%%%%%%%%%%%%%%%%%%%%%%%%%%%%%%%%%%%%%%%%%%%%%%%%%%%%%%%%%%%%%%%%%%%%%

We would like to thank Beno\^it Dou\c{c}ot and Konstantina Kontoudi for useful discussions.
This work made in the ILP LABEX (under reference ANR-10-LABX-63) was supported by French state
funds managed by the ANR within the Investissements d'Avenir programme under reference
ANR-11-IDEX-0004-02.

%%%%%%%%%%%%%%%%%%%%%%%%%%%%%%%%%%%%%%%%%%%%%%%%%%%%%%%%%%%%%%%%%%%%%%%%%%%%%%%%%%%%%%%%%%%%%%%
%%%%%%%%%%%%%%%%%%%%%%%%%%%%%%%%%%%%%%%%%%%%%%%%%%%%%%%%%%%%%%%%%%%%%%%%%%%%%%%%%%%%%%%%%%%%%%%
%\newpage 
\addcontentsline{toc}{section}{Appendices}
\appendix
\renewcommand{\theequation}{\Alph{section}.\arabic{equation}}
%%%%%%%%%%%%%%%%%%%%%%%%%%%%%%%%%%%%%%%%%%%%%%%%%%%%%%%%%%%%%%%%%%%%%%%%%%%%%%%%%%%%%%%%%%%%%%%
%%%%%%%%%%%%%%%%%%%%%%%%%%%%%%%%%%%%%%%%%%%%%%%%%%%%%%%%%%%%%%%%%%%%%%%%%%%%%%%%%%%%%%%%%%%%%%%

%%%%%%%%%%%%%%%%%%%%%%%%%%%%%%%%%%%%%%%%%%%%%%%%%%%%%%%%%%%%%%%%%%%%%%%%%%%%%%%%%%%%%%%%%%%%%%%
%%%%%%%%%%%%%%%%%%%%%%%%%%%%%%%%%%%%%%%%%%%%%%%%%%%%%%%%%%%%%%%%%%%%%%%%%%%%%%%%%%%%%%%%%%%%%%%
\section{Action and field equations}
\label{app:action}
%%%%%%%%%%%%%%%%%%%%%%%%%%%%%%%%%%%%%%%%%%%%%%%%%%%%%%%%%%%%%%%%%%%%%%%%%%%%%%%%%%%%%%%%%%%%%%%
%%%%%%%%%%%%%%%%%%%%%%%%%%%%%%%%%%%%%%%%%%%%%%%%%%%%%%%%%%%%%%%%%%%%%%%%%%%%%%%%%%%%%%%%%%%%%%%

We give in this appendix the details about the action from which we derive the field equations.
The metric has signature $(-,+,+,+)$ and we take the conventions of~\cite{Nitti:2013xaa}.

The action is
\begin{equation}
\label{eq:action-app}
S = \int \D^4x \sqrt{-g}\left(\mathcal{L}_{\textnormal{Eins.}}+\mathcal{L}_{\textnormal{Mxwl.}}
+ \mathcal{L}_{\textnormal{scalar}} + \mathcal{L}_{\textnormal{fluid}} +
\mathcal{L}_\textnormal{int} \right)
\end{equation}
where the two first terms define Maxwell-Einstein theory with
\begin{equation}
\mathcal{L}_{\textnormal{Eins.}}=\frac{1}{2\kappa^2}\left(R+\frac{6}{L^2}\right) \, , \qquad
\mathcal{L}_\textnormal{Mxwl.}=-\frac{1}{4e^2}F_{ab}F^{ab} \, ,
\end{equation}
where $F=\D A$ is the field strength of the gauge field $A$, $\kappa$ is Newton's constant,
$L$ is the asymptotic $AdS_4$ length and $e$ is the $U(1)$ coupling.

The matter content consists of a charged scalar field and a perfect fluid of charged fermions
which interact directly through the interaction Lagrangian $\mathcal{L}_\textnormal{int}$.
The Lagrangian density of the scalar is
\begin{equation}
\mathcal{L}_\textnormal{scalar}=-\frac{1}{2} \left( \left| \nabla\psi-iqA\psi
\right|^2 + m_s^2 |\psi|^2 \right) \, ,
\end{equation}
where $q$ is the coupling to the $U(1)$ gauge field and $m_s$ the mass.
It contributes to Einstein and Maxwell equations through its stress tensor
\begin{align}
T_{ab}^\textnormal{scalar} &= \frac{1}{2}\left( g_a^c g_b^d + g_b^c g_a^d -g_{ab}g^{cd}
\right) \left( \nabla_c\psi-iqA_c\psi\right)\left(\nabla_d\psi^*+iqA_d\psi^*\right) -
\frac{1}{2} m_s^2g_{ab}\psi\psi^* 
\end{align}
and its electromagnetic current
\begin{align}
\label{eq:current-psi}
J^a_\textnormal{scalar} &= -i \frac{q}{2}g^{ab}\left[
\psi^*\left(\nabla_a-iqA_a\right)\psi - \psi\left(\nabla_a+iqA_a\right)\psi^*
\right] \, .
\end{align}
In the Thomas-Fermi treatment of the fermion $\chi$, the electromagnetic current of the
fluid is
\begin{align}
\label{eq:current-fluid-app}
  J^a_\textnormal{fluid} = - q_f \langle \bar{\chi}\Gamma^a \chi \rangle = q_f n u^a
\end{align}
where $n>0$ is the particle number density, $u^a$ the fluid velocity and $q_f$ the fermion
charge.

To keep the fluid approximation valid for the fermions, the direct interaction between the
fluid and the scalar field is chosen to be the current-current interaction
\begin{align}
\label{eq:int-lagrangian}
 \mathcal{L}_\textnormal{int} = \eta J_a^\textnormal{fluid}J^a_\textnormal{scalar} \, ,
\end{align}
where $\eta$ is a coupling constant.
To compute the fluid quantities, we derive the field equations of the
Lagrangian
\begin{align}
\label{eq:tot-fluid-lagrangian}
\mathcal{L}_\textnormal{fluid} + \mathcal{L}_\textnormal{int} = - \rho(n) + n
u^a\left(\partial_a \phi
+ q_f A_a+\eta
q_f J_a^\textnormal{scalar}\right)+\lambda(u^a u_a+1) \, ,
\end{align}
where $\rho$ is the energy density, $\phi$ a Clebsch coefficient and $\lambda$ a Lagrange
multiplier, for $n$ and $\lambda$~\cite{Hartnoll:2010gu}.
One obtains the local chemical potential for particle number
\begin{align}
 \widetilde{\mu}_l(n) \equiv \rho'(n) = q_f u^a(A_a + \eta J_a^\textnormal{scalar}) \, ,
\end{align}
where we have set\footnote{We could in principle allow for a non-zero \lq\lq intrinsic''
chemical potential for particle number not related to the fermion charge $q_f$. However our
choice avoids possible singularities~\cite{Hartnoll:2010gu}.}
$\partial_a\phi=0$, and the normalization condition
\begin{align}
\label{eq:norm-condition-u}
 u^a u_a = -1 \, .
\end{align}
We will assume that the fluid is made of non-interacting fermionic particles.
The energy density $\rho(n)$ is then an increasing function of the particle number
density and the chemical potential for particle number $\widetilde{\mu}_l$ is positive for all
$q_f$ (positive or negative).
We also define the pressure $p$ through the thermodynamical relation
\begin{align}
 p(n) \equiv -\rho(n) + n \, \widetilde{\mu}_l(n) \, .
\end{align}
The energy and particle number densities are functions of the local chemical potential for
particle number,
\begin{align}
\rho = \beta \int_{m_f}^{\widetilde{\mu}_l}\D \epsilon \,
\Theta\left(\epsilon-m_f\right) \epsilon^2 \, \sqrt{\epsilon^2-m_f^2}
\, ,
\qquad n = \beta \int_{m_f}^{\widetilde{\mu}_l} \D \epsilon \,
\Theta\left(\epsilon-m_f\right) \epsilon \,
\sqrt{\epsilon^2-m_f^2}\, ,
\end{align}
where $m_f$ is the mass of the fermions, $\beta$ a parameter related to the spin of the
fermions and $\Theta$ the Heaviside step function.
One can equivalently work with the charge density
\begin{align}
 \sigma = q_f n
\end{align}
and the chemical potential for charge density
\begin{align}
\mu_l = \frac{\widetilde{\mu}_l}{q_f}
\end{align}
instead of $n$ and $\widetilde{\mu}_l$.
Notice that the chemical potential for charge density $\mu_l$ has the same sign as $q_f$ since $\widetilde{\mu}_l>0$.
%In our conventions, electrons have $q_f>0$ while positrons have $q_f<0$.
%It follows that the electron and positron fluid quantities are non-vanishing when $\mu_l>m_f>0$
%and $\mu_l<-m_f<0$, respectively.
The stress tensor of the fluid is
\begin{align}
T_{ab}^\textnormal{fluid} &= (\rho+p)u_a u_b + p g_{ab}
\end{align}
and its electromagnetic current is given by~(\ref{eq:current-fluid-app}).
The electromagnetic current of the interaction~(\ref{eq:int-lagrangian}) is
\begin{align}
 J_\textnormal{int}^a = -\eta q^2|\psi|^2\sigma u^a \, .
\end{align}
Finally, Einstein equations are
\begin{equation}
 R_{ab}-\frac{1}{2}g_{ab}R-\frac{3}{L^2}g_{ab} =
\kappa^2\left(T_{ab}^\textnormal{Mxwl.}+T_{ab}^\textnormal{fluid}+T_{ab}^\textnormal{scalar}
\right)
\end{equation}
where
\begin{align}
 T_{ab}^\textnormal{Mxwl.} = \frac{1}{e^2}\left(F_{ac}F_b^{\ c}-\frac{1}{4} g_{ab}
F_{cd}F^{cd}\right) \, ,
\end{align}
the equation of motion for $\psi$ is
\begin{equation}
\label{eq:eom-psi-coupling}
-\left(\nabla_a -iqA_a-2iq\eta J_a^\textnormal{fluid}\right)\left(\nabla^a -iqA^a\right)\psi +
\left(m_s^2 + iq\eta\nabla_a J^a_\textnormal{fluid}\right)\psi = 0
\end{equation}
and Maxwell equations are
\begin{equation}
 \nabla_a F^{ba}=e^2 \left(J^b_\textnormal{fluid} + J^b_\textnormal{scalar} +
J^b_\textnormal{int}\right)
\, .
\end{equation}
%

%%%%%%%%%%%%%%%%%%%%%%%%%%%%%%%%%%%
\subsection{Ansatz, physical parameters and solution-generating symmetries}
%%%%%%%%%%%%%%%%%%%%%%%%%%%%%%%%%%%
\label{app:ansatz}
We will restrict to the homogeneous and isotropic ansatz
\begin{equation}
\begin{aligned}
\label{eq:ansatz-full-app}
\D s^2 &= L^2\left[-f(r)\D t^2 + g(r)\D r^2 + \frac{1}{r^2} \left(\D x^2 + \D
y^2\right)\right] \, , \\
A &= \frac{eL}{\kappa} h(r) \D t \ , \qquad \psi = \psi(r) \, , \qquad u^a = (u^t,0,0,0) \, .
\end{aligned}
\end{equation}
From~(\ref{eq:norm-condition-u}), the fluid velocity has non-zero component
$u^t=1/(L\sqrt{f})$.
The local chemical potential is now given by
\begin{align}
\label{eq:local-mu-app}
 \mu_l = \frac{e}{\kappa}\frac{h}{\sqrt{f}} \left(1 - \eta q^2\psi^2 \right) \, .
\end{align}

One can eliminate $e$, $\kappa$, $L$ from the field equations by performing the parameter and
field redefinitions
\begin{equation}
\begin{aligned}
\label{eq:rescalings}
A_a = \frac{eL}{\kappa} \hat{A}_a \, , \qquad \psi = \frac{1}{\kappa}\hat{\psi} \, ,
\qquad m_s &= \frac{1}{L}\hat{m}_s \, , \qquad q=\frac{\kappa}{eL}\hat{q} \, , \\
\eta=e^2L^2\hat{\eta} \, , \qquad \mu_l=\frac{e}{\kappa}\hat{\mu}_l \, , \qquad
\frac{m_f}{|q_f|}&=\frac{e}{\kappa}\hat{m}_f \, , \qquad \beta q_f^4 =
\frac{\kappa^2}{e^4L^2}\hat{\beta} \, , \\
\rho = \frac{1}{\kappa^2L^2}\hat{\rho} \, , \qquad p &= \frac{1}{\kappa^2L^2}\hat{p} \, ,
\qquad \sigma = \frac{1}{e\kappa L^2}\hat{\sigma} \, .
\end{aligned}
\end{equation}

The $r$-component of Maxwell equations implies that when the scalar condenses, the phase of
$\psi$ is constant, and we can fix it to zero in the whole solution by a global $U(1)$
transformation.
$\psi(r)$ can be now considered as a real field.
With the rescalings~(\ref{eq:rescalings}), the field equations~(\ref{eq:field-equations})
depend only on the hatted parameters.
Let us obtain the expressions for the rescaled energy density $\hat{\rho}$ and charge density $\hat{\sigma}$ explicitly.
From~(\ref{eq:rescalings}) we have
\begin{equation}
\begin{aligned}
\rho &= \beta \int_{\frac{e}{\kappa}|q_f|\hat{m}_f}^{\frac{e}{\kappa}q_f\hat{\mu}_l}\D \epsilon \
\Theta\left(\epsilon-\frac{e}{\kappa}q_f\hat{m}_f\right) \epsilon^2 \, \sqrt{\epsilon^2-\frac{e^2}{\kappa^2}q_f^2\hat{m}_f^2}
\, , \\
\qquad \sigma &= q_f \beta \int_{\frac{e}{\kappa}|q_f|\hat{m}_f}^{\frac{e}{\kappa}q_f\hat{\mu}_l}\D \epsilon \
\Theta\left(\epsilon-\frac{e}{\kappa}q_f\hat{m}_f\right) \epsilon \, \sqrt{\epsilon^2-\frac{e^2}{\kappa^2}q_f^2\hat{m}_f^2}
\, .
\end{aligned}
\end{equation}
Since $q_f\mu_l=\widetilde{\mu}_l>0$, we can replace in the upper bound of the integrals $q_f\hat{\mu}_l$ by $|q_f||\hat{\mu}_l|$.
Replacing $m_f$, $\beta$ and $\mu_l$ by the rescaled quantities~(\ref{eq:rescalings}) and performing a simple change of variable, one obtains
\begin{equation}
\begin{aligned}
\rho &= \frac{1}{\kappa^2L^2} \hat{\beta} \int_{\hat{m}_f}^{|\hat{\mu}_l|}\D \epsilon \
\Theta\left(\epsilon-\hat{m}_f\right) \epsilon^2 \, \sqrt{\epsilon^2-\hat{m}_f^2}
\, , \\
\qquad \sigma &= \frac{1}{e\kappa L^2} \frac{|q_f|}{q_f} \hat{\beta} \int_{\hat{m}_f}^{|\hat{\mu}_l|}\D \epsilon \
\Theta\left(\epsilon-\hat{m}_f\right) \epsilon \, \sqrt{\epsilon^2-\hat{m}_f^2} \, ,
\end{aligned}
\end{equation}
where $|q_f|$ arises from the square root.
Since $q_f\hat{\mu}_l>0$, $|q_f|/q_f=\textnormal{sign}(q_f)=\textnormal{sign}(\hat{\mu}_l)$, using the last line of~(\ref{eq:rescalings}) we obtain the final expressions for the rescaled energy and charge densities,
\begin{equation}
\begin{aligned}
\hat{\rho} &=  \hat{\beta} \int_{\hat{m}_f}^{|\hat{\mu}_l|}\D \epsilon \
\Theta\left(\epsilon-\hat{m}_f\right) \epsilon^2 \, \sqrt{\epsilon^2-\hat{m}_f^2}
\, , \\
\qquad \hat{\sigma} &= \textnormal{sign}(\hat{\mu}_l) \, \hat{\beta} \int_{\hat{m}_f}^{|\hat{\mu}_l|}\D \epsilon \
\Theta\left(\epsilon-\hat{m}_f\right) \epsilon \, \sqrt{\epsilon^2-\hat{m}_f^2} \, .
\end{aligned}
\end{equation}
These expressions are non-zero for $0<\hat{m}_f<|\hat{\mu}_l|$.
The charge density $\hat{\sigma}$ is positive for $0<\hat{m}_f<\hat{\mu}_l$ and negative for $\hat{\mu}_l<-\hat{m}_f<0$.
The energy density is positive in both cases.

The field equations~(\ref{eq:field-equations}) are invariant under the transformations
\begin{subequations}
\begin{align}
 (r,x,y)\to a(r,x,y) \, , \qquad & f\to a^{-2}f \, \qquad g\to a^{-2} g \, , \qquad h \to
a^{-1} h \, ; \\
\nonumber \\
& f\to b^{-2}f \, , \qquad h\to b^{-2} h \, .
\end{align}
\end{subequations}
These transformations do not leave the ansatz~(\ref{eq:ansatz-full-app}) invariant; they are
solution-generating symmetries which take one solution to a physically different one with
generically different boundary chemical potential and charge for example.

In the applied classical gravity regime and local flat space treatment of the fermions, the
parameters must satisfy
\begin{align}
 e^2\sim\frac{\kappa}{L}\ll1 \, .
\end{align}
\section{The Dirac equation}
\label{app:dirac-eq}
%%%%%%%%%%%%%%%%%%%%%%%%%%%%%%%%%%%%%%%%%%%%%%%%%%%%%%%%%%%%%%%%%%%%%%%%%%%%%%%%%%%%%%%%%%%%%%%
%%%%%%%%%%%%%%%%%%%%%%%%%%%%%%%%%%%%%%%%%%%%%%%%%%%%%%%%%%%%%%%%%%%%%%%%%%%%%%%%%%%%%%%%%%%%%%%

Consider the action
\begin{align}
 S_\chi = \int\D^4x
\sqrt{-g}\left[-i\left(\bar{\chi}\Gamma^a\mathcal{D}_a\chi-m_f\bar{\chi}\chi\right) + \eta
J_a^\textnormal{ferm}J^a_\textnormal{scal}\right]
\end{align}
of a probe spinor field $\chi$ representing an electronic excitation of charge
$|q_f|$, on top of a background solution of the
form~(\ref{eq:ansatz-full-app}) of the theory~(\ref{eq:action-app}), where
\begin{subequations}
\label{eq:currents-probe}
\begin{align}
 J^a_\textnormal{scal} &=
-i\frac{q}{2}g^{ab}\left[\bar{\psi}(\partial_b-iqA_b)\psi-\psi(\partial_b+iqA_b)\bar{\psi}
\right ] \, , \\
J_a^\textnormal{ferm} &= - |q_f|\bar{\chi}\Gamma_a\chi \, ,
\end{align}
\end{subequations}
and
\begin{align}
 \mathcal{D}_a &= \partial_a+\frac{1}{4}\omega_{ija}\Gamma^{ij}-i|q_f|A_a \, , \\
 \bar{\chi} &= \chi^\dagger\Gamma^0 \, .
\end{align}
Notice that the field $\chi$ has positive electric coupling $|q_f|$ consistent with our
background conventions.
The Dirac equation is
\begin{align}
 i\,\Gamma^a\mathcal{D}_a\chi - im_f\chi + \eta |q_f|J^a_\textnormal{scal}\Gamma_a\chi =
0
\, .
\end{align}
By setting $\chi=rf^{1/4}\xi(r)e^{ikx-i\omega t}$, it becomes
\begin{align}
 r^{-1}g^{-1/2}\left(\Gamma^1\partial_r- L m_f g^{1/2}\right)\xi(r) + i K_i\Gamma^i
\xi(r) = 0
\end{align}
where
\begin{align}
 K_0 = - r^{-1}f^{-1/2}\left[\omega+\frac{eL}{\kappa}|q_f| h\left(1 - \eta
q^2\bar{\psi}\psi\right)\right] \, , \qquad K_2 = k \, , \qquad K_1 = K_3=0 \, .
\end{align}
The Dirac equation does not depend on $\Gamma^3$.
We choose the following basis for Gamma-matrices,
\begin{align}
 \Gamma^0 =
\begin{pmatrix}
i \sigma^1 & 0 \\
0 & i\sigma^1
\end{pmatrix} \, , \quad
 \Gamma^1 =
\begin{pmatrix}
-\sigma^3 & 0 \\
0 & -\sigma^3
\end{pmatrix} \, , \quad
 \Gamma^2 =
\begin{pmatrix}
-\sigma^2 & 0 \\
0 & \sigma^2
\end{pmatrix} \, , \quad
\Gamma^3 =
\begin{pmatrix}
0 & \sigma^2 \\
\sigma^2 & 0
\end{pmatrix} \, ,
\end{align}
where $\sigma^l$, $l=1,2,3$, are Pauli matrices.
By writing $\xi=(\Phi,\tilde{\Phi})$, we obtain two decoupled first order equations for the
Dirac spinors $\Phi$ and $\tilde{\Phi}$ which differ only by the momentum $k\rightarrow-k$.
The equation for $\Phi$ is~\cite{Faulkner:2009wj}
\begin{align}
\left(\partial_r + \gamma\hat{m}_f g^{1/2}\sigma^3\right)\Phi =
g^{1/2}\left\{i\gamma\sigma^2 \left[\hat{\omega}f^{-1/2}+\hat{\mu}_l
\right] - \gamma\hat{k}r\sigma^1\right\} \Phi \, ,
\end{align}
or in components,
\begin{subequations}
\label{eq:dirac-coupled-eqs}
\begin{align}
 \frac{1}{\sqrt{g}}\partial_r\Phi_1 + \gamma \hat{m}_f\Phi_1 -
\gamma\left(\frac{\hat{\omega}}{\sqrt{f}}+\hat{\mu}_l - \hat{k}r\right)\Phi_2 &= 0 \, ,
\\
 \frac{1}{\sqrt{g}}\partial_r\Phi_2 - \gamma \hat{m}_f\Phi_2 +
\gamma\left(\frac{\hat{\omega}}{\sqrt{f}}+\hat{\mu}_l + \hat{k}r \right)\Phi_1 &= 0 \, .
\end{align}
\end{subequations}
In the above equations, we have rescaled the momentum and the frequency by
\begin{align}
 \omega = \gamma \hat{\omega} \, , \qquad k = \gamma \hat{k} \, ,
\end{align}
where $\gamma$ is given by~(\ref{eq:def-gamma}).
The equations~(\ref{eq:dirac-coupled-eqs}) can be written in the form
\begin{subequations}
\label{eq:phi-eq}
\begin{align}
\label{eq:phi2-eq}
 \Phi_2'' = &
\frac{\left[\sqrt{g}\left(\frac{\hat{\omega}}{\sqrt{f}}+\hat{\mu}_l+\hat{k}r\right)\right]'}{
\sqrt{g}\left(\frac{\hat{\omega}}{\sqrt{f}}+\hat{\mu}_l+\hat{k}r\right)}\Phi_2' \nonumber \\
&+
\left\{\gamma^2 g\left[\hat{m}_f^2 + \hat{k}^2r^2 -
\left(\frac{\hat{\omega}}{\sqrt{f}} +
\hat{\mu}_l\right)^2\right] -
\gamma\hat{m}_f\sqrt{g}
\frac{\left(\frac{\hat{\omega}}{\sqrt{f}}+\hat{\mu}_l+\hat{k}r\right)'}{\frac{\hat
{\omega}}{\sqrt{f}}+\hat{\mu}_l+\hat{k}r}\right\}\Phi_2 \, , \\
\Phi_1 =& \frac{1}{\frac{\hat{\omega}}{\sqrt{f}}+\hat{\mu}_l+\hat{k}r}
\left(\hat{m}_f\Phi_2-\frac{1}{\gamma}\frac{1}{\sqrt{g}}\Phi_2'\right) \, ,
\end{align}
\end{subequations}
where primes denote derivatives with respect to the radial coordinate $r$.
In the large-$\gamma$ limit, it can be shown that the $\Phi_2'$ term in~(\ref{eq:phi2-eq}) is
negligible by putting it in a Schr\"odinger form. In this limit, the system~(\ref{eq:phi-eq})
can
be written as
\begin{align}
 \Phi_2'' &=
\left\{\gamma^2 g\left[\hat{m}_f^2 + \hat{k}^2r^2 -
\left(\frac{\hat{\omega}}{\sqrt{f}} +
\hat{\mu}_l\right)^2\right] -
\gamma\hat{m}_f\sqrt{g}
\frac{\left(\frac{\hat{\omega}}{\sqrt{f}}+\hat{\mu}_l+\hat{k}r\right)'}{\frac{\hat
{\omega}}{\sqrt{f}}+\hat{\mu}_l+\hat{k}r}\right\}\Phi_2 \, , \\
\Phi_1 &= \frac{1}{\frac{\hat{\omega}}{\sqrt{f}}+\hat{\mu}_l-\hat{k}r}
\left(\hat{m}_f\Phi_2-\frac{1}{\gamma}\frac{1}{\sqrt{g}}\Phi_2'\right) \, .
\end{align}
%

%%%%%%%%%%%%%%%%%%%%%%%%%%%%%%%%%%%%%%%%%%%%%%%%%%%%%%
\section{Solving the Schr\"odinger-like equation}
\label{app:wkb}
%%%%%%%%%%%%%%%%%%%%%%%%%%%%%%%%%%%%%%%%%%%%%%%%%%%%%%

We will solve the Schr\"odinger equation~(\ref{eq:wkb-eq}) in the WKB approximation.
We focus here on the case where there is a region of the bulk where the Schr\"odinger potential
is negative.
This region is bounded by the turning points $r=r_1$ and $r=r_2$, with $r_1<r_2$, where the
potential vanishes.
We recall that in the WKB approximation, the formal solution to the equation~(\ref{eq:wkb-eq})
is
\begin{align}
\Phi_2 \simeq& C_+ \exp\left[
\gamma\int_{r_0}^r\D s\sqrt{V(s)} - \int_{r_0}^r\D s \frac{V'(s)}{4V(s)}  \right]
\nonumber \\
+& C_- \exp\left[
-\gamma\int_{r_0}^r\D s \sqrt{V(s)} - \int_{r_0}^r\D s \frac{V'(s)}{4V(s)} \right]
\end{align}
where $r_0$ is an arbitrary (fixed) point.

Close to a turning point $r=r_\star$, we have
\begin{align}
 \gamma\int_{r_\star}^r \D s \sqrt{|V(s)|} &\sim
\pm\varphi(r) \, , \qquad r\to r_\star^\pm \, ,
\end{align}
where
\begin{align}
 \varphi(r) =& \frac{2}{3}\gamma\sqrt{\left|V'(r_\star)\right|}\cdot|r-r_\star|^{3/2} \, .
\end{align}
The matching conditions around a turning point where $V(r)$ vanishes linearly are
\begin{subequations}
\label{eq:matching-conditions}
\begin{align}
 \frac{1}{2}e^{-\varphi} \leftrightarrow \sin(\varphi+\pi/4) \, , \\
 e^{\varphi} \leftrightarrow \cos(\varphi+\pi/4) \, .
\end{align}
\end{subequations}
%

%%%%%%%%%%%%%%%%%%%%%%%%%%%%%%%%%%%%%%%%%%%%%%%%%%%%%
\subsection{The WKB solution for one $V<0$ region}
\label{sec:wkb-1}
%%%%%%%%%%%%%%%%%%%%%%%%%%%%%%%%%%%%%%%%%%%%%%%%%%%%%

We consider first the case where the potential $V$ is negative in one region bounded by
$r_1$ and $r_2$ with $r_1<r_2$.
By imposing normalizability on the wave function, in the inner region $r>r_2$ we have
\begin{align}
\label{eq:phi-in}
 \Phi_2^{in} \sim C^{in} \, e^{-\gamma\int_{r_2}^r\D s \sqrt{V(s)}} \, .
\end{align}
In the intermediate region $r_1<r<r_2$, the wave function is
\begin{align}
 \Phi_2^{inter} &\sim C_+^{inter}  \, e^{i\gamma\int_{r_1}^r\D
s\sqrt{|V(s)|}} + C_-^{inter}  \, e^{-i\gamma\int_{r_1}^r\D
s\sqrt{|V(s)|}} \\
&\sim C_+^{inter} \, e^{\theta_{12}}\,e^{i\gamma\int_{r_2}^r\D
s\sqrt{|V(s)|}} + C_-^{inter} \,
e^{-\theta_{12}}\,e^{-i\gamma\int_{r_2}^r\D
s\sqrt{|V(s)|}}
\end{align}
where
\begin{align}
 \theta_{ij} = \gamma\int_{r_i}^{r_j}\D r\sqrt{|V(r)|} \, .
\end{align}
Finally, in the UV region, we have
\begin{align}
 \Phi_2^{out} &\sim C_+^{out}  \, e^{\gamma\int_{r_1}^r\D
s\sqrt{V(s)}} + C_-^{out}  \, e^{-\gamma\int_{r_1}^r\D
s\sqrt{V(s)}} \\
&\sim C_+^{out} \, e^{-\theta_{\epsilon}}\,e^{\gamma\int_{\epsilon}^r\D
s\sqrt{V(s)}} + C_-^{out} \,
e^{\theta_{\epsilon}}\,e^{-\gamma\int_{\epsilon}^r\D s\sqrt{V(s)}}
\end{align}
where
\begin{align}
 \theta_{\epsilon} = \gamma\int_{\epsilon}^{r_1}\D r\sqrt{V(r)}
\end{align}
and $\epsilon\ll1$ is a UV cutoff.
Close to the UV boundary, the WKB solution is
\begin{align}
\label{eq:phi-uv-expansion}
\Phi_2^{UV} \sim C_+^{UV} e^{-\theta_{\epsilon}}
\left(\frac{r}{\epsilon}\right)^{\gamma\hat{m}_f} + C_-^{UV} e^{\theta_{\epsilon}}
\left(\frac{r}{\epsilon}\right)^{-\gamma\hat{m}_f+1} \, .
\end{align}
Notice that we have taken into account the $\mathcal{O}(\gamma^{-1})$-correction
\begin{align}
- \gamma^{-1}\hat{m}_f\sqrt{g}
\frac{\left(\frac{\hat{\omega}}{\sqrt{f}}+\hat{\mu}_l+\hat{k}r\right)'}{\frac{\hat
{\omega}}{\sqrt{f}}+\hat{\mu}_l+\hat{k}r}
\end{align}
to the Schr\"odinger-like potential~(\ref{eq:potential}) to obtain the subleading power $r^1$
in
the second term in~(\ref{eq:phi-uv-expansion}).
Applying the matching conditions~(\ref{eq:matching-conditions}) and
using~(\ref{eq:asympt-spinor1}) and (\ref{eq:Green1}), we obtain the Green
function~(\ref{eq:green-function}) with
\begin{align}
 \mathcal{G} = \frac{1}{2}\tan W
\end{align}
where
\begin{align}
 W = \gamma\int_{r_1}^{r_2}\D r\sqrt{|V(r)|} \, .
\end{align}
%

%%%%%%%%%%%%%%%%%%%%%%%%%%%%%%%%%%%%%%%%%%%%%%%%%%%%%
\subsection{The WKB solution for two $V<0$ region}
\label{sec:wkb-2}
%%%%%%%%%%%%%%%%%%%%%%%%%%%%%%%%%%%%%%%%%%%%%%%%%%%%%

Now we consider the case where $V$ is negative in two regions bounded by
$r_1$, $r_2$ and $r_3$, $r_4$ respectively, with $r_1<r_2<r_3<r_4$.
We impose regularity on the wave function at infinity, so in the inner region $r>r_4$, $\Phi_2$
is given by~(\ref{eq:phi-in}) where $r_2$ is replaced by $r_4$. The UV expansion is again
given by~(\ref{eq:phi-uv-expansion}). By matching the solution in the different regions at the
turning points, the constant $\mathcal{G}$ in the Green function~(\ref{eq:green-function}) is
now given by
\begin{align}
 \mathcal{G} =
\frac{4 e^{2X}\sin Y \cos Z + \cos Y \sin Z}{8e^{2X}\cos Y \cos Z - 2\sin Y\sin Z}
\end{align}
where
\begin{align}
X = \gamma\int_{r_2}^{r_3} \D r \sqrt{V(r)} \, , \qquad
Y = \gamma\int_{r_1}^{r_2} \D r \sqrt{|V(r)|} \, , \qquad
Z = \gamma\int_{r_3}^{r_4} \D r \sqrt{|V(r)|} \, .
\end{align}

\bibliographystyle{JHEP}
%\bibliography{/Users/thomasvanel/Dropbox/BibTex/myrefs}
%\bibliography{/ada2/etudiants/vanel/Dropbox/BibTex/myrefs}
%\bibliography{\string~/Dropbox/ES-SC/PAPER-2014/PAPER-2014-Draft/myrefs}
\bibliography{myrefs}
\end{document}